\documentclass[fleqn,usenatbib]{mnras}

\usepackage{newtxtext,newtxmath}

\usepackage[T1]{fontenc}

\DeclareRobustCommand{\VAN}[3]{#2}
\let\VANthebibliography\thebibliography
\def\thebibliography{\DeclareRobustCommand{\VAN}[3]{##3}\VANthebibliography}

\usepackage{graphicx}	
\usepackage{amsmath}	
\usepackage{multicol}
\usepackage{physics}
\usepackage{bm}
\usepackage{ulem}
\usepackage{comment} 
\usepackage{xcolor}
\usepackage{mathrsfs}
\usepackage{tabularx}
\usepackage{threeparttable}
\usepackage{hyperref}
\usepackage[title]{appendix}
\allowdisplaybreaks

\newcommand{\sign}[1]{\mathrm{sgn}\qty(#1)}

\graphicspath{{./}{figures/}}

\title{The Impact of Cosmic Rays on Thermal and Hydrostatic Stability in Galactic Halos} 

\author[Tsung et al.]{
Tsun Hin Navin Tsung,$^{1}$\thanks{E-mail: ttsung@ucsb.edu}
S. Peng Oh,$^{1}$
Chad Bustard$^{2}$
\\
$^{1}$Dept. of Physics, University of California, Santa Barbara, CA 93106, USA\\
$^{2}$Kavli Institute for Theoretical Physics, University of California at Santa Barbara, Kohn Hall, Santa Barbara, CA 93107, USA
}

\date{Accepted XXX. Received YYY; in original form ZZZ}

\pubyear{2015}

\begin{document}
\label{firstpage}
\pagerange{\pageref{firstpage}--\pageref{lastpage}}
\maketitle

\begin{abstract}
We investigate how cosmic rays (CRs) affect thermal and hydrostatic stability of circumgalactic (CGM) gas, in simulations with both CR streaming and diffusion. 
Local thermal instability can be suppressed by CR-driven entropy mode propagation, in accordance with previous analytic work. However, there is only a narrow parameter regime 
where this operates, before CRs overheat the background gas. As mass dropout from thermal instability causes the background density and hence plasma $\beta \equiv P_g/P_{\rm B}$ to fall, the CGM becomes globally unstable. At the cool disk to hot halo interface, a sharp drop in density boosts Alfven speeds and CR gradients, driving a transition from diffusive to streaming transport. CR forces and heating  strengthen, while countervailing gravitational forces and radiative cooling weaken, resulting in a loss of both hydrostatic and thermal equilibrium. In lower $\beta$ halos, CR heating drives a hot, single-phase diffuse wind with velocities $v \propto (t_\mathrm{heat}/t_\mathrm{ff})^{-1}$, which exceeds the escape velocity when $t_\mathrm{heat}/t_\mathrm{ff} \lesssim 0.4$. In higher $\beta$ halos, where the Alfven Mach number is higher, CR forces drive multi-phase winds with cool, dense fountain flows and significant turbulence. These flows are CR dominated due to `trapping' of CRs by weak transverse B-fields, and have the highest mass loading factors. Thus, local thermal instability can result in winds or fountain flows where either the heat or momentum input of CRs dominates. 

\end{abstract}




\section{Introduction}

In recent years, there has been a surge of interest in how cosmic rays (CRs) affect feedback in the circumgalactic medium (CGM) and intracluster medium (ICM), in particular on how they can drive a wind and provide thermal support. Unlike thermal gas, CRs do not suffer radiative losses and have smaller adiabatic index, making them able to sustain their pressure far away from their sources. Simulations have indeed shown that CRs can drive winds in the CGM \citep{jubelgas08_cr_feedback,uhlig12_cr_wind,booth13_cr_wind,hanasz13_cr_wind,salem14_cr_wind,girichidis16_cr_wind,simpson16_cr_wind,ruszkowski17_cr_wind,hopkins21_cr_outflows,chan22_disk_halo_interface} and heat the ICM gas sufficiently to prevent a cooling catastrophe \citep{guo08-CR,ruszkowski17_icm,jacob17a,wang20_cr_heating_feedback}, but these are apparently dependent on the model for CR transport and the gas properties, for which results can differ by orders of magnitude \citep{pakmor16_cr_wind,buck20_cr_cgm,hopkins21_cr_transport,hopkins22_cr_transport_not_consistent}

To date, there is still considerable uncertainty about CR effects on galaxy evolution, despite herculean efforts to simulate CR-modulated feedback and compare to observations. To help dissect the influence of CRs, we take a distinctly different but complementary approach to that of usual feedback simulations. Our main science questions are as follows. For a hydrostatic atmosphere in initial thermal equilibrium supported by thermal gas, magnetic fields, and CRs, is this atmosphere thermally and dynamically stable? How do CRs affect local and global stability? Finally, in regimes where neither stability criterion holds, what is the nonlinear outcome? Our results confirm previous analytic expectations, illuminate connections between local and global instability, and reveal new insights into how CRs create and modify large-scale gas flows.

A key uncertainty in CR feedback models is the nature of CR transport. CRs scatter collisionlessly off magnetic turbulence, and the rapid scattering rate renders their behavior fluid-like. However the exact details of this process are still not yet fully understood, and sometimes at odds with observations in our Galaxy \citep{kempski22,hopkins21-CR-problems}. CR transport is often divided into 2 distinct modes: CR streaming, and CR diffusion. In the self-confinement picture of CR transport, CRs are scattered by magnetic turbulence they generate and can lock themselves with self-excited Alfven waves. They advect or {\it stream} down the CR pressure gradients at the Alfven speed \citep{kulsrud69_stream_instab,zweibel17_cr_review}. In addition, since the CR scattering rate is finite, CRs are not completely locked to the Alfven waves, but drift slowly with respect to the Alfven wave frame. This can be represented as a field-aligned diffusion term. More generally, CRs undergo a random walk due to small-scale tangled B-fields, known as 'Field Line Wandering', even if CRs stream along B-field. In addition, CRs can random walk due to scattering by extrinsic turbulence\footnote{Scattering by extrinsic turbulence is thought to be important only at higher energies, $E > 100$GeV, with lower energy CRs -- where the bulk of the energy resides-- predominantly self-confined.}. This random walk renders CR transport diffusive, or even super-diffusive \citep{yan04,mertsch20,sampson22}. In reality, CRs are likely to both stream and diffuse; these processes must be considered in parallel.   

There are two aspects of CR streaming which are germane to this paper. Streaming CRs locked with the Alfven waves transfer energy to the thermal gas at the rate $v_A\cdot\nabla P_c$, reflecting the work done by the CRs to excite magnetic waves, which then damp and heat the gas. This only takes place with CR streaming; there is no collisionless CR heating with CR diffusion. CR heating could plausibly explain elevated heating-- as inferred from line ratios-- in the Reynolds layer of our Galaxy \citep{wiener13}, reflecting its potential importance in the disk halo interface and CGM. It can drive acoustic waves unstable in sufficiently magnetized environments ($\beta \lesssim 0.5$, \citet{begelman94_acoustic_instab}) and may potentially have significant effect on wind driving in the CGM \citep{tsung22-staircase,quataert22_cr_wind,huang22_cr_wind}. In another context, the ICM, CR heating has been shown able to balance radiative cooling and suppress cooling flows \citep{guo08_global_stab,wang20_cr_heating_feedback}. Recently, \citet{kempski20_thermal_instability} explored, through linear analysis, the effect CR heating has on local thermal instability, finding it can cause thermal entropy modes to propagate and suppress the instability in certain parameter regimes. The nonlinear effects have not been explored yet.

Secondly, for the streaming instability to be excited, the drift speed must exceed the local Alfv\'en velocity $v_{\rm A}$. In regions where the CRs are isotropic ($\nabla P_{\rm c}=0 $), or have small drift speed, $v_{\rm D} < v_{\rm A}$, CRs will not scatter; they decouple from the gas and free stream out of these `optically thin' regions at the speed of light. This leads to the `CR bottleneck effect' \citep{skilling71,begelman95,wiener17-cold-clouds}, which can significantly modulate CR transport in a multi-phase medium. Since $v_{\rm D} \sim v_A \propto \rho^{-1/2}$, a cloud of warm ($T\sim 10^{4}$K) ionized gas embedded in hot ($T\sim 10^{6}$K) gas results in a minimum in drift speed. This produces a `bottleneck' for the CRs: CR density is enhanced as CRs are forced to slow down, akin to a traffic jam. Since CRs cannot stream up a gradient, the system readjusts to a state where the CR profile is flat up to the minimum in $v_A$; thereafter the CR pressure falls again. If there are multiple bottlenecks, this produces a staircase structure in the CR profile \citep{tsung22-staircase}. Importantly, since $\nabla P_c = 0$ in the plateaus, CRs there are no longer coupled to the gas, and can no longer exert pressure forces or heat the gas. Instead, momentum and energy deposition is focused at the CR `steps'. Small-scale density contrasts can thus have global influence on CR driving and heating.

The impact of CRs on halo gas is by now well-trodden ground; there is a vast and rapidly expanding literature on this topic. Nonetheless, as hinted above, there are several key aspects which motivate this study. Firstly, the influence of collisionless CR heating $v_{\rm A} \cdot \nabla P_c$ on thermal instability and the development of winds, which is a key prediction of the self-confinement theory of CR transport, is often neglected. To date thermal instability simulations with CR streaming do not have background CR heating--they either have horizontal B-fields, so that there is no CR streaming in the background profile \citep{butsky20_ti_cr}, or take place in an unstratified medium \citep{huang22_cr_instability}. Wind simulations are also often run in limits (e.g., ignoring streaming, isothermal winds, or considering high $\beta$ winds) where only the momentum input of CRs drive the wind, while CR thermal driving is negligible. To date, there is only an analytic linear analysis \citep{kempski20_thermal_instability} and 1D CR wind models \citep{ipavich75, modak23} where CR heating plays an important role in thermal instability and CR winds respectively. We suggest that CR heating could play a more crucial role than previously thought. 

Secondly, we take care to consider the {\it combined} effects of CR streaming and diffusion, operating simultaneously. Until $\sim 5$ years ago, due to numerical challenges (see \S\ref{sec:governing_equations}), CR streaming was either ignored or treated in limits where the Alfven speed changes only on large lengthscales\footnote{This is not possible in a multi-phase medium, where $v_{\rm A} \sim \rho^{-1/2}$ change on the short lengthscale of the interface width between cold and hot phases.}. These difficulties have since been overcome with the two moment method \citep{jiang18_cr_numerical_method,thomas19,chan19}. Nonetheless, wind studies often consider effective limits where either CR streaming {\it or} diffusion are dominant\footnote{Of course, there are exceptions, such as the FIRE simulations \citep{chan19,hopkins21-CR-test}, which incorporate simultaneous streaming and diffusion with the two moment method. However, they run fully self-consistent simulations from cosmological initial conditions; the B-field strength and plasma $\beta$ is not an adjustable parameter, as in our idealized simulations, but a simulation output. Thus, we have greater flexibility to survey parameter space. See additional discussion in \S\ref{sec:discussion}.}. We shall see that the combined effects of diffusion and streaming can be non-trivial, as each can dominate in different regimes. For instance, diffusion can dominate in the disk, allowing CRs to escape without strong heating losses, while streaming dominates in the halo, which provides strong CR heating in a low density regime where radiative cooling is weak. By contrast, streaming-only simulations lead to strong CR losses at the disk-halo interface, while diffusion-only simulations ignore the effects of CR heating. 

Finally, the impact of {\it local} thermal instability on {\it global} hydrostatic and thermal stability have not been sufficiently studied. CR winds are often studied in models where conditions in the wind base (e.g., star formation rate) change, leading to a higher CR momentum flux which drives an outflow. Our models consider the opposite case where conditions at the base are fixed, but conditions in the halo gas change. Local thermal instability reduces the background gas density, thereby reducing plasma $\beta$ and radiative cooling rates and increasing Alfven speeds. It also introduces CR `bottlenecks' in a multi-phase medium. These changes can lead to a loss of global hydrostatic and thermal stability, and the emergence of phenomena such as CR heated winds and fountain flows. We find it is particularly important to include and resolve the disk/halo interface, where sharp density gradients drive sharp gradients in Alfven speed and hence CR pressure. Winds and fountain flows are generally launched at this interface, and are qualitatively different if this phase transition is not modelled.

This paper is organized as follows. In \S\ref{sec:simulation} we review the governing equations, and describe the simulation setup used for this study. In \S\ref{sec:TI}, we study the effect of CRs on linear thermal instability. In \S\ref{subsec:nonlinear_outcome}, we discusses the nonlinear results of the simulations, in particular the emergence of galactic fountain flows and winds. We discuss some implications in \S\ref{sec:discussion}, and conclude in \S\ref{sec:conclusions}.


\section{Methods} \label{sec:simulation}

\subsection{Governing equations} \label{sec:governing_equations}

We utilize the two-moment method \citep{jiang18_cr_numerical_method}, which has been tested in stringent conditions (e.g. CR-modified shocks, \citet{tsung21_shock}). A merit of this method is its ability to handle CR pressure extrema self-consistently and efficiently, which previously resulted in grid-scale numerical instabilities which rapidly swamped the true solution. Previous remedies relied on ad-hoc regularization \citep{sharma09_regularize}, where the timestep scales quadratically with resolution; this becomes prohibitively expensively at high resolution. The ability to resolve sharp gradients in Alfven speed is important in simulating the CR bottleneck effect \citep{wiener17_cold_clouds,bustard21_bottleneck,tsung22-staircase}, a crucial feature of CR streaming transport in multi-phase media where large volumes of zero CR pressure gradient are found. 

Assuming the gas is fully ionized, the gas flow is non-relativistic and the gyroradii of the CRs are much smaller than any macro scale of interest, the two-moment equations governing the dynamics of a CR-MHD fluid are given by \citep{jiang18_cr_numerical_method}:
\begin{gather}
    \pdv{\rho}{t} + \nabla\cdot\qty(\rho\vb{v}) = 0, \label{eqn:continuity} \\
    \pdv{\qty(\rho\vb{v})}{t} + \nabla\cdot\qty(\rho\vb{v}\vb{v} - \vb{B}\vb{B} + P^*\vb{I}) = \bm{\sigma}_c\cdot\qty[\vb{F}_c - \qty(E_c + P_c)\vb{v}] + \rho\vb{g}, \label{eqn:momentum} \\
    \pdv{E}{t} + \nabla\cdot\qty[\qty(E + P^*)\vb{v} - \vb{B}\qty(\vb{B}\cdot\vb{v})] = \qty(\vb{v} + \vb{v}_s)\cdot\bm{\sigma}_c\cdot \nonumber \\
    \qquad\qty[\vb{F}_c - \qty(E_c + P_c)\vb{v}] + \rho\vb{g}\cdot\vb{v} + \mathcal{L}, \label{eqn:energy} \\
    \pdv{\vb{B}}{t} = \nabla\cross\qty(\vb{v}\cross\vb{B}), \label{eqn:induction} \\
    \pdv{E_c}{t} + \nabla\cdot\vb{F}_c = -\qty(\vb{v} + \vb{v}_s)\cdot\bm{\sigma}_c\cdot\qty[\vb{F}_c - \qty(E_c + P_c)\vb{v}] + \mathcal{Q}, \label{eqn:cr_energy} \\
    \frac{1}{c_\mathrm{red}^2}\pdv{\vb{F}_c}{t} + \nabla P_c = -\bm{\sigma}_c\cdot\qty[\vb{F}_c - \qty(E_c + P_c)\vb{v}], \label{eqn:cr_flux}
\end{gather}
where $c_\mathrm{red}$ is the reduced speed of light, $\mathcal{L} = \mathcal{H} - \mathcal{C}$ is the net heating, defined by source heating minus cooling, $\mathcal{Q}$ is the CR source/sink term, $\vb{v}_s = -\vb{v}_A\sign{\vb{B}\cdot\nabla P_c}$ is the streaming velocity, where $\vb{v}_A = \vb{B}/\sqrt{\rho}$ is the Alfven velocity, $\vb{g}$ is the gravitational acceleration, $P^* = P_g + B^2/2$ is the total pressure, equal to the sum of thermal gas pressure and magnetic pressure, $E = \rho v^2/2 + P_g/\qty(\gamma_g - 1) + B^2/2$ is the total energy density, equal to the sum of kinetic, thermal and magnetic energy densities and $\bm{\sigma}_c$ is the interaction coefficient defined by 
\begin{gather}
    \bm{\sigma}_c^{-1} = \bm{\sigma}_d^{-1} + \bm{\sigma}_s^{-1}, \label{eqn:interaction_coef} \nonumber \\
    \bm{\sigma}_d^{-1} = \frac{\bm{\kappa}}{\gamma_c - 1}, \quad \bm{\sigma}_s^{-1} = \frac{\vb{B}}{\abs{\vb{B}\cdot\nabla P_c}}\vb{v}_A\qty(E_c + P_c), \label{eqn:diffusion_coef}
\end{gather}
where $\bm{\kappa}$ is the CR diffusion tensor. The interaction coefficient $\bm{\sigma}_c$ links the thermal gas with CRs and acts as a bridge for momentum and energy transfer (through the source terms $\bm{\sigma}_c\cdot\qty[\vb{F}_c - \qty(E_c + P_c)\vb{v}]$ and $\vb{v}_s\cdot\bm{\sigma}_c\cdot\qty[\vb{F}_c - \qty(E_c + P_c)\vb{v}]$). It describes the strength of the CR-gas coupling and consists of two parts: a streaming part $\bm{\sigma}_s$ and a diffusive part $\bm{\sigma}_d$, to model different modes of CR transport. The reduced speed of light $c_\mathrm{red}$, in combination with $\bm{\sigma}_c$, set the timescale for CRs to couple with thermal gas. $c_\mathrm{red}$ is designed to capture the speed of the free-streaming CRs, which in reality is close to the speed of light, though in practice it is always set much lower to allow for a longer Courant time-step; it has been shown that results are converged with respect to $c_\mathrm{red}$ as long as it is much greater than any other velocity in the system \citep{jiang18_cr_numerical_method}. 
Note that if $\sigma_c c_\mathrm{red}^2 \Delta t \gg 1$ (where $\Delta t$ is the time step), the time derivative $\pdv*{\vb{F}_c}{t}$ will be negligible and one would recover the steady-state CR flux 
\begin{equation}
    \vb{F}_\mathrm{c,steady} = \qty(E_c + P_c)\qty(\vb{v} + \vb{v}_s) - \frac{\bm{\kappa}}{\gamma_c - 1}\cdot\nabla P_c. \label{eqn:cr_flux_steady_state}
\end{equation}
We see two components of CR transport in eqn.\ref{eqn:cr_flux_steady_state}, the first term showing CR energy advecting at the combined velocity $\vb{v} + \vb{v}_s$ and the second term depicting diffusion. Note that from eqn.\ref{eqn:diffusion_coef} $\sigma_c c_\mathrm{red}^2 \Delta t \gg 1$ is not possible if $\nabla P_c\approx 0$. In this case $\sigma_c\approx 0$, $\pdv*{\vb{F}_c}{t}$ is not negligible and no closed form expression for $\vb{F}_c$ exists. CR momentum and energy transfer $\approx 0$. In this regime CRs are said to be uncoupled from the thermal gas and free streaming.  On the other hand, if $\nabla P_c$ is finite and $c_\mathrm{red}$ is sufficiently large, the CR flux would be in steady-state (eqn.\ref{eqn:cr_flux_steady_state}) and CR-gas are said to be coupled. In this regime CRs transfer momentum and energy to the gas at the rates
\begin{gather}
    \bm{\sigma}_c\cdot\qty[\vb{F}_c - \qty(E_c + P_c)\vb{v}] \rightarrow -\nabla P_c, \label{eqn:coupled_momentum_transfer}\\
    \vb{v}_s\cdot\bm{\sigma}_c\cdot\qty[\vb{F}_c - \qty(E_c + P_c)\vb{v}] \rightarrow -\vb{v}_s\cdot\nabla P_c. \label{eqn:coupled_energy_transfer}
\end{gather}
Note that there is no heat transfer if $\vb{v}_s$ (or the magnetic field) is perpendicular to $\nabla P_c$. Since $\vb{v}_s$ always points down the $P_c$ gradient, CRs always heat the gas instead of the other way around.

The diffusion tensor can be expressed in general as $\bm{\kappa} = \kappa_\parallel\vu{b}\vu{b} + \kappa_\perp (\vb{I} - \vu{b}\vu{b})$, where $\kappa_\parallel$ and $\kappa_\perp$ are the field-aligned and cross-field diffusion coefficients. Cross-field diffusion is ignored in this study (i.e. $\kappa_\perp\approx 0$). We also ignore any CR collisional losses due to Coulomb collisions and hadronic interactions. In this context, $\bm{\kappa}$ accounts for the slippage from perfect wave locking due to damping. If damping is weak, slippage is small and $\kappa_\parallel$ will be small. In principle, $\kappa_\parallel$ is a function of various plasma parameters (e.g., \citealt{wiener_13_cr_stream_clusters,jiang18_cr_numerical_method}), but to date the exact contributions from wave damping are unclear, so in this study unless otherwise stated we shall consider damping to be weak, and we will set $\kappa_\parallel$ to be a small constant. For an implementation of $\kappa_\parallel$ with non-negligible ion-neutral damping, for example, see \citet{bustard21_bottleneck}.

\subsection{Simulation Setup} \label{subsec:setup}


In this section we describe the simulation setup that is used throughout the study. The simulations were performed with Athena++ \citep{stone20_athena}, an Eulerian grid-based MHD code using a directionally unsplit, high-order Godunov scheme with the constrained transport (CT) technique. CR streaming was implemented with the two-moment method \citep{jiang18_cr_numerical_method}, which solves eqn.\ref{eqn:continuity}-\ref{eqn:cr_flux}. Cartesian geometry is used throughout.

We run our setup in 2D and 3D, 2D for high resolution and 3D for full dimensional coverage. The setup consists of a set of initial profiles, source terms and appropriate boundary conditions. Gravity defines the direction of stratification, which is taken to be in the $x$-direction ($\vb{g} = -g\qty(x)\vu{x}$). We sometimes use `vertical' and `horizontal' to denote stratification ($x$) and perpendicular ($y, z$) directions respectively. Both CR transport modes are present (streaming and diffusion). 

\subsubsection{Initial Profiles} \label{subsubsec:initial}

\begin{figure}
    \centering
    \includegraphics[width=0.48\textwidth]{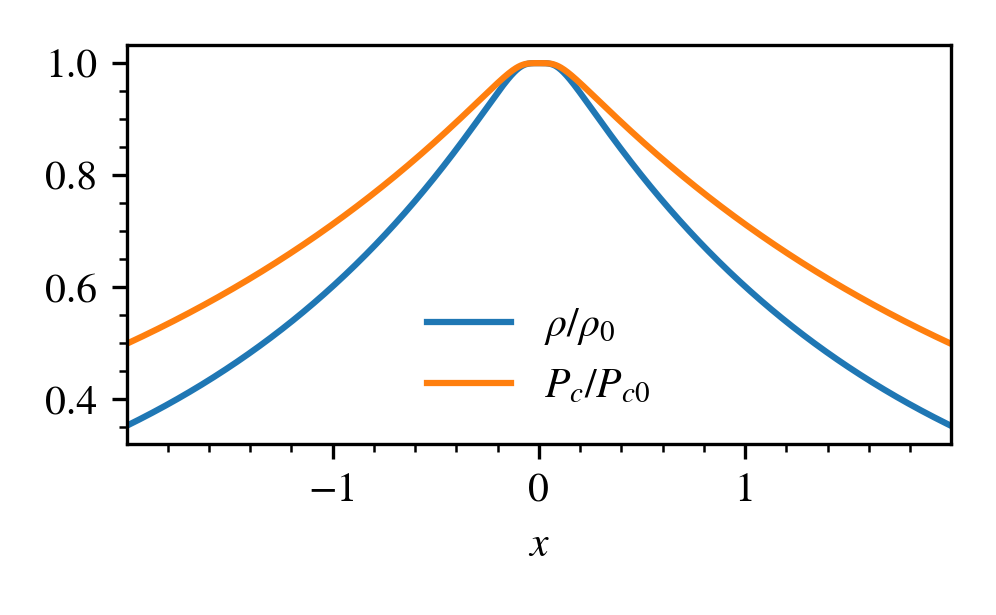}
    \includegraphics[width=0.48\textwidth]{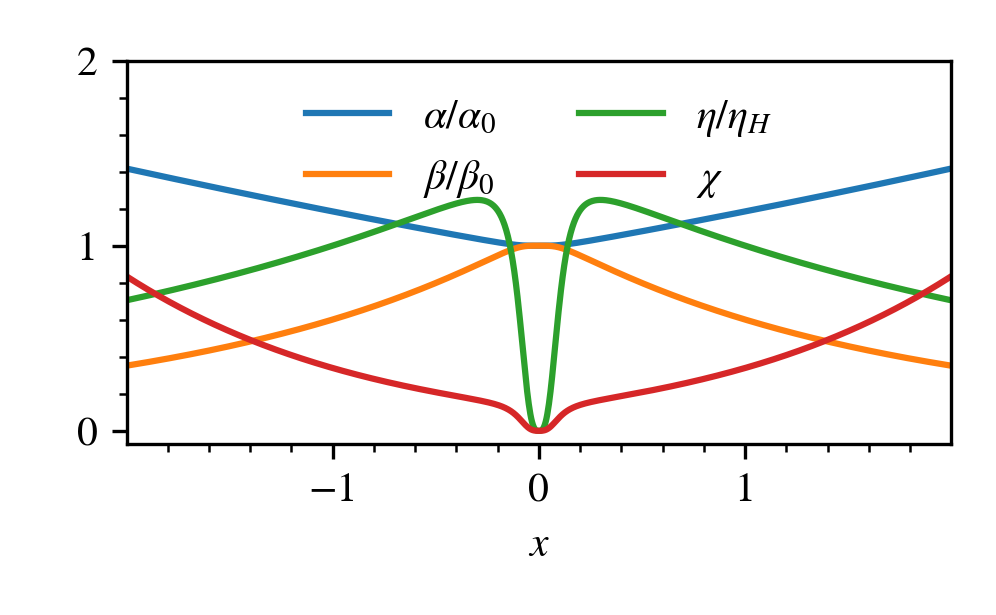}
    \caption{\emph{Top}: Example of the initial density (blue) and $P_c$ (orange) profiles. Found from solving eqn.\ref{eqn:hydrostatic_algebraic} numerically. Gravity is tapered to zero at $x=0$, resulting in zero derivatives for $\rho$ and $P_c$ there. \emph{Bottom}: Example of $\alpha, \beta, \eta, \chi$ for the same initial profile, normalized by their reference values. Note that $\alpha_0 = 1, \beta_0 = 5, \eta_H = 0.01$ and $\delta_H = 1$.}
    \label{fig:init_profile}
\end{figure}

The initial profiles are calculated by solving a set of ODEs assuming hydrostatic and thermal equilibrium. In the absence of any instability, the initial profiles will remain time-steady. We align the magnetic field with the direction of stratification for background CR heating. It is initially spatially constant ($\vb{B} = B\vu{x}$)\footnote{By symmetry, the magnetic field can only vary along $x$, the direction of stratification, i.e. $\vb{B} = B\qty(x)\vu{x}$. To satisfy $\div{B}=0$, $\dv*{B}{x} = 0$, i.e. the magnetic field is constant.}. Gravity is taken to be
\begin{equation}
    g\qty(x) = g_0 \frac{\qty(x/a)^3}{1 + \abs{x/a}^3}.  \label{eqn:gravity}
\end{equation}
Thus, $g\qty(x)$ approaches $g_0$, a constant, as $x\rightarrow\infty$ and approaches $g_0\qty(x/a)^3$ when $x\ll a$. The smoothing parameter $a$ tapers the gravitational field to zero as $x\rightarrow 0$ so as to avoid discontinuities in $\nabla P_g$ and $\nabla P_c$. 
We found that this functional form maintains hydrostatic equilibrium better than the gravitational softening employed by \citet{mccourt_ti_stratified} and thereafter. 
In hydrostatic equilibrium,
\begin{equation}
    \dv{P_g}{x} + \dv{P_c}{x} = -\rho g. \label{eqn:hydrostatic}
\end{equation}
We take the initial profile to be isothermal with temperature $T_0$ such that $\dv*{P_g}{x} = T_0\dv*{\rho}{x}$. If CR transport is streaming dominated, $P_c = P_{c0}\qty(\rho/\rho_0)^{\gamma_c/2}$, where $\rho_0$ and $P_{c0}$ are some reference density and CR pressure. Substituting into eqn.\ref{eqn:hydrostatic},
\begin{equation}
    \qty[T_0 + \frac{\gamma_c P_{c0}}{2\rho_0}\qty(\frac{\rho}{\rho_0})^{\gamma_c/2 - 1}]\dv{\rho}{x} = -\rho g. \label{eqn:hydrostatic_substitute}
\end{equation}
Integrating both sides,
\begin{equation}
    \ln(\frac{\rho}{\rho_0}) + \frac{\gamma_c}{2 - \gamma_c}\frac{P_{c0}}{P_{g0}}\qty[1 - \qty(\frac{\rho}{\rho_0})^{\gamma_c/2 - 1}] = -\frac{1}{T_0}\int_0^x g\qty(x')\dd{x'}, \label{eqn:hydrostatic_algebraic}
\end{equation}
where $P_{g0} = \rho_0 T_0$ is some reference gas pressure. The density profile $\rho\qty(x)$ is then found numerically from eqn.\ref{eqn:hydrostatic_algebraic} using a numerical integrator and root-finder. The gas and CR pressure, CR flux profiles are then found from $P_g = \rho T_0$, $P_c = P_{c0}\qty(\rho/\rho_0)^{\gamma_c/2}$ and 
\begin{equation}
    F_c\qty(x) = \frac{\gamma_c}{\gamma_c - 1} P_c v_A  - \frac{\kappa_\parallel}{\gamma_c - 1}\dv{P_c}{x}, \label{eqn:initial_crflux}
\end{equation}
where we have used eqn.\ref{eqn:cr_flux_steady_state}. See the top panel of fig.\ref{fig:init_profile} for an example of the density and $P_c$ profile. Here we discuss several important ratios characterizing our initial profiles:
\begin{gather}
    \alpha = \frac{P_c}{P_g}, \ \beta = \frac{2 P_g}{B^2}, \ \eta = \frac{\kappa_\parallel}{\gamma_c v_A L_c}, \ \delta = \frac{t_\mathrm{cool}}{t_\mathrm{ff}}, \ \chi = \frac{v_A\abs{\nabla P_c}}{\rho^2\Lambda}, \label{eqn:important_ratios} 
\end{gather}
which determine the CR to gas ($\alpha$), magnetic to gas ($\beta$) pressure ratios, diffusive to streaming flux ratio ($\eta$) and the ratio of cooling to free-fall time ($\delta$) and CR heating to radiative cooling ($\chi$). $L_c = P_c/\abs{\nabla P_c}$ is the CR scale height. $\eta$, the ratio of the diffusive to streaming flux, is small if streaming transport dominates. As $\rho, P_c, P_g$ in the initial profiles are functions of $x$, the ratios in eqn.\ref{eqn:important_ratios} in general also vary with $x$. 

The density profile can be fully determined given $g_0,a,\rho_0, T_0, \alpha_0$. 
The reference values $\rho_0, \alpha_0$ are set at the base $x=0$. Note that $T_0$ is a constant in our isothermal profile. With $\rho\qty(x)$ determined, $P_g\qty(x), P_c\qty(x)$ can be obtained easily from the ideal gas law and $P_c\propto\rho^{\gamma_c/2}$.  The latter is true for steady-state, static streaming dominated flows\footnote{We ignore CR diffusion in our initial profiles. Thus, our background profiles are not exactly in steady state, particularly in profiles where diffusion is comparable to streaming, $\eta_{\rm H} \sim 1$. In practice, we have found that our results are not sensitive to the initial deviation from perfect equilibrium. We also find that the global background profile eventually always evolves significantly, once thermal instability triggers mass dropout.} \citep{breitschwerdt91,wiener17_cold_clouds}. The magnetic field can be obtained by specifying $\beta_0$, i.e. $\beta$ at $x=0$. Note that in our setup the field is aligned with gravity. The diffusion coefficient $\kappa_\parallel$ is found by setting $\eta$ \textit{not} at $x = 0$ as $L_c$ is infinite there but at a thermal scale height $x=H=T_0/g_0$ (we shall denote this by $\eta_H$, with subscript $H$ meaning it is set at $x=H$). Without loss of generality, we shall set $g_0, T_0, \rho_0$ all to 1 and $a = 0.1 H$. In fig.\ref{fig:init_profile} we show an example of the how these profiles (top panel) and the respective ratios $\alpha,\beta,\eta$ (bottom panel) vary in space. Since CR pressure  declines more weakly with density ($P_c \propto \rho^{2/3}$) than isothermal gas pressure ($P_{\rm g} \propto \rho$), the profile becomes slightly more CR dominated the further out. Plasma $\beta$ decreases with height as the $B$-field is spatially constant. Apart from the peak at $x\sim a$ where $\abs{\nabla P_c}$ is maximized, $\eta$ generally decreases with increasing $x$. 

\subsubsection{Source Terms} \label{subsubsec:source_terms}

We adopt a power law cooling function
\begin{equation}
    \Lambda = \Lambda_0 (T/T_0)^{\Lambda_T}.
\end{equation}
Using the density and $P_c$ profiles found from \S\ref{subsubsec:initial}, the cooling strength $\Lambda_0$ is determined by $\delta$ or $\chi$. The cooling index $\Lambda_T$ can be adjusted to mimic the cooling curve in cluster ($\Lambda_T = 0.5$) and galaxy halo ($\Lambda_T = -2/3$; we use this exclusively in this paper) contexts.

When $\delta_H$ (i.e. $\delta$ at a thermal scale-height) is specified, the cooling strength is given by
\begin{equation}
    \Lambda_0 = \frac{T_0}{\qty(\gamma_g - 1)\delta_H\rho_H t_\mathrm{ff, H}}, \label{eqn:cooling_strength_delta}
\end{equation}
where $t_\mathrm{ff,H}$ is the free-fall time at $x=H$ defined by 
\begin{equation}
    t_\mathrm{ff} = \sqrt{\frac{2 x}{g_0}}, \label{eqn:ff_time}
\end{equation}
with $x$ replaced by $H$. The subscript $H$ again denotes quantity evaluated at $x=H$. When $\chi_H$ is specified, the cooling strength is: 
\begin{equation}
    \Lambda_0 = \frac{\abs{v_{A,H}\nabla P_{c,H}}}{\chi_H\rho_H^2}. \label{eqn:cooling_strength_chi}
\end{equation}
Thus, we can specify either $\delta$ or $\chi$ to our desired value for the purpose of the study. 

Unless $\chi = 1$ everywhere, CR heating cannot fully balance radiative cooling. The residual heating needed to attain thermal balance is provided by `feedback heating' (or `magic heating') $\mathcal{H}$ \citep{mccourt_ti_stratified,sharma12}, a phenomenological heating model where global thermal equilibrium is enforced by fiat. At each time-step, uniform heating is input at a rate given by the spatially averaged cooling rate at a given height (or radius), so that the average net cooling is zero. However, the fluctuations in the net cooling rate can give rise to thermal instability. In our system, this means that the heating rate is given by:
\begin{equation}
    \mathcal{H}\qty(x, t) = \langle \rho^2\Lambda + v_A\nabla P_c \rangle, \label{eqn:feedback_heating}
\end{equation}
where $\langle\cdot\rangle$ denotes spatial average over the $y,z$-slice at any particular $x$. This term is activated only when the RHS of eqn.\ref{eqn:feedback_heating} is greater than zero; $\mathcal{H}$ will be set to zero if the RHS is negative. $\mathcal{H}$ can be interpreted physically as other sources of heating, such as thermal star formation or AGN feedback. 

In the absence of a CR source at the base ($x = 0$), the CR profiles will not maintain steady state as CRs stream away, causing the $P_c$ profile to flatten (see fig.1 of \citet{jiang18_cr_numerical_method} for an example). We supply CRs at the base by fixing the CR pressure $P_c$ as
\begin{gather}
    P_c\qty(x, t) = \alpha_0 P_g\qty(x, t)  \label{eqn:fix_pc}
\end{gather}
for $\abs{x} < 0.7\Delta x$, where $\Delta x$ is the grid size. Physically this represents sources of CRs from the galactic disk (e.g., due to supernovae or AGN). An alternative is to fix the CR flux $F_c$ at the base. We show some results from this in Appendix \ref{app:fix_fc}. There is no qualitative change in our conclusions.  


\subsubsection{Simulation Box and Boundary Conditions} \label{subsubsec:boundary_condition}

The simulation box extends symmetrically in all directions about the origin for 2 thermal scale-heights $H$, employing hydrostatic boundary conditions in the $x$-direction and periodic boundaries otherwise. Hydrostatic boundaries mandate
\begin{equation}
    \eval{\dv{P_g}{x}}_{\mathrm{bond}} + \eval{\dv{P_c}{x}}_{\mathrm{bond}} = -\eval{\rho g}_{\mathrm{bond}} \label{eqn:bond_hydro}
\end{equation}
at the ghost zone cell faces. We provide details of our boundary implementation in Appendix \ref{app:hydro_boundary_implement}. In \S\ref{subsec:nonlinear_outcome} we shall see that in some cases the flow could become non-hydrostatic, but no significant difference is seen when we adopt an outflow type boundary condition (see Appendix \ref{app:hydro_boundary_implement}). We refrain from extending the box beyond $2 H$ to prevent plasma $\beta$ from dropping below $\sim 0.5$ and exciting acoustic instabilities right from the beginning of the simulation \citep{begelman94_acoustic_instab,tsung22-staircase}. It is also to prevent $\chi > 1$ at large $x$ (see fig.\ref{fig:init_profile}) for which overheating occurs and the gas would be out of thermal equilibrium.

To prevent spurious numerical behavior, we apply buffers with thickness $a$ near the $x$ boundaries and the base. 
There is no cooling or CR heating within these buffers.

\subsubsection{Resolution, Reduced Speed of Light and Temperature Floors} \label{subsubsec:resolution_rsol}

We run our simulations in 2D with $1024\times 512$ grids (higher resolution along the $x$-axis). In Appendix \ref{app:res_3d} we run a selected subset of cases in higher resolution ($2048\times 512$) and in 3D with $256\times 128\times 128$ grids and show that our conclusions remain unchanged. We use a reduced speed of light $c_\mathrm{red} = 200$, which is much greater than any other velocity scale in the problem (in most cases this should be sufficient, though in some cases with particularly strong magnetic field and low density, for which $v_A$ is large, we increase $c_\mathrm{red}$ accordingly). The temperature floor is set to $T_0/100$ while the ceiling is set to $5 T_0$. In general, in a multi-phase medium, cooling is dominated by the cool gas, so enforcing global thermal equilibrium means that the hot gas, where cooling is inefficient, could be heated up to even higher temperatures. It is customary, in thermal instability studies, to set a temperature ceiling to prevent the time-step from becoming extremely small. However, given the possibility that CR heating can potentially heat the gas to very high temperatures in the nonlinear evolution, we remove the ceiling for simulations in \S\ref{subsec:nonlinear_outcome}.



\subsubsection{Simulation Runs}

\begin{table*}
    \centering
    \begin{tabular}{c|c|c|c|c|c|c|c|c|c}
        Identifier & $\alpha_0$ & $\beta_0$ & $\eta_H$ & $\chi_H$ & $\delta_H$ & $\Lambda_T$ & Resolution & $c_\mathrm{red}$ & Remarks\\
        \hline
        Test cases for \S\ref{subsec:propagation} & & & & & & & & & \\
        \hline
        $\mathrm{a1b3k.01c.4in.67res1024c200single}$ & 1 & 3 & 0.01 & 0.4 & - & -2/3 & $1024\times 512$ & 200 & - \\
        $\mathrm{a.5b3k.01c.3in.67res1024c200single}$ & 0.5 & 3 & 0.01 & 0.3 & - & -2/3 & $1024\times 512$ & 200 & - \\
        $\mathrm{a5b100k.01c.7in.67res1024c200single}$ & 5 & 100 & 0.01 & 0.7 & - & -2/3 & $1024\times 512$ & 200 & - \\
        $\mathrm{a1b3k.01c.4in.67res1024c200single-nocrh}$ & 1 & 3 & 0.01 & 0.4 & - & -2/3 & $1024\times 512$ & 200 & No CR heating \\
        $\mathrm{a.5b3k.01c.3in.67res1024c200single-nocrh}$ & 0.5 & 3 & 0.01 & 0.3 & - & -2/3 & $1024\times 512$ & 200 & No CR heating \\
        $\mathrm{a5b100k.01c.7in.67res1024c200single-nocrh}$ & 5 & 3 & 0.01 & 0.7 & - & -2/3 & $1024\times 512$ & 200 & No CR heating \\
        $\mathrm{a1b3k1c.4in.67res1024c200single}$ & 1 & 3 & 1 & 0.4 & - & -2/3 & $1024\times 512$ & 200 & - \\
        \hline
        Test cases for \S\ref{subsec:suppress_propagation} & & & & & & & & & \\
        \hline
        $\mathrm{a1b3k.01c.4in.67res1024c200}$ & 1 & 3 & 0.01 & 0.4 & - & -2/3 & $1024\times 512$ & 200 & - \\
        $\mathrm{a1b3k.01c.4in.67res1024c200-nocrh}$ & 1 & 3 & 0.01 & 0.4 & - & -2/3 & $1024\times 512$ & 200 & No CR heating \\
        $\mathrm{a1b3k1c.4in.67res1024c200}$ & 1 & 3 & 1 & 0.4 & - & -2/3 & $1024\times 512$ & 200 & - \\
        $\mathrm{a1b3k.01c1in.67res1024c200}$ & 1 & 3 & 0.01 & 1 & - & -2/3 & $1024\times 512$ & 200 & - \\
        $\mathrm{a1b3k.01c1in.67res1024c200-nocrh}$ & 1 & 3 & 0.01 & 1 & - & -2/3 & $1024\times 512$ & 200 & No CR heating \\
        $\mathrm{a1b3k1c1in.67res1024c200}$ & 1 & 3 & 1 & 1 & - & -2/3 & $1024\times 512$ & 200 & - \\
        $\mathrm{a1b3k.01c2.5in.67res1024c200}$ & 1 & 3 & 0.01 & 2.5 & - & -2/3 & $1024\times 512$ & 200 & - \\
        $\mathrm{a1b3k.01c2.5in.67res1024c200-nocrh}$ & 1 & 3 & 0.01 & 2.5 & - & -2/3 & $1024\times 512$ & 200 & No CR heating \\
        $\mathrm{a1b3k1c2.5in.67res1024c200}$ & 1 & 3 & 1 & 2.5 & - & -2/3 & $1024\times 512$ & 200 & - \\
        \hline
        Test cases for \S\ref{subsec:nonlinear_outcome} & & & & & & & & & \\
        \hline
        $\mathrm{a1b5k.0001d1in.67res1024c200}$ & 1 & 5 & 0.0001 & - & 1 & -2/3 & $1024\times 512$ & 200 & - \\
        $\mathrm{a1b5k.001d1in.67res1024c200}$ & 1 & 5 & 0.001 & - & 1 & -2/3 & $1024\times 512$ & 200 & - \\
        $\mathrm{a1b5k.01d1in.67res1024c200}$ & 1 & 5 & 0.01 & - & 1 & -2/3 & $1024\times 512$ & 200 & `slow wind' \\
        $\mathrm{a1b5k.01d1in.67res1024c200-nocrh}$ & 1 & 5 & 0.01 & - & 1 & -2/3 & $1024\times 512$ & 200 & `slow wind' (nocrh) \\
        $\mathrm{a1b5k.1d1in.67res1024c200}$ & 1 & 5 & 0.1 & - & 1 & -2/3 & $1024\times 512$ & 200 & - \\
        $\mathrm{a1b5k.5d1in.67res1024c200}$ & 1 & 5 & 0.5 & - & 1 & -2/3 & $1024\times 512$ & 200 & - \\
        $\mathrm{a1b5k1d1in.67res1024c200}$ & 1 & 5 & 1 & - & 1 & -2/3 & $1024\times 512$ & 200 & `fast wind' \\
        $\mathrm{a1b5k1d1in.67res1024c200-nocrh}$ & 1 & 5 & 1 & - & 1 & -2/3 & $1024\times 512$ & 200 & `fast wind' (nocrh) \\
        $\mathrm{a1b5k5d1in.67res1024c1000}$ & 1 & 5 & 5 & - & 1 & -2/3 & $1024\times 512$ & 1000 & - \\
        $\mathrm{a5b5k1d1in.67res1024c3000}$ & 10 & 5 & 1 & - & 1 & -2/3 & $1024\times 512$ & 3000 & - \\
        $\mathrm{a10b5k.01d1in.67res1024c1000}$ & 10 & 5 & 0.01 & - & 1 & -2/3 & $1024\times 512$ & 1000 & - \\
        $\mathrm{a1b10k1d1in.67res1024c200}$ & 1 & 10 & 1 & - & 1 & -2/3 & $1024\times 512$ & 200 & - \\
        $\mathrm{a1b30k1d1in.67res1024c200}$ & 1 & 30 & 1 & - & 1 & -2/3 & $1024\times 512$ & 200 & - \\
        $\mathrm{a1b50k1d1in.67res1024c200}$ & 1 & 50 & 1 & - & 1 & -2/3 & $1024\times 512$ & 200 & - \\
        $\mathrm{a1b100k1d1in.67res1024c200}$ & 1 & 100 & 1 & - & 1 & -2/3 & $1024\times 512$ & 200 & - \\
        $\mathrm{a1b300k.01d1in.67res1024c200}$ & 1 & 300 & 0.01 & - & 1 & -2/3 & $1024\times 512$ & 200 & - \\
        $\mathrm{a1b300k1d1in.67res1024c200}$ & 1 & 300 & 1 & - & 1 & -2/3 & $1024\times 512$ & 200 & `fountain' \\
        $\mathrm{a1b300k1d1in.67res1024c200-nocrh}$ & 1 & 300 & 1 & - & 1 & -2/3 & $1024\times 512$ & 200 & `fountain' (nocrh) \\
        $\mathrm{a1b300k10d1in.67res1024c200}$ & 1 & 300 & 10 & - & 1 & -2/3 & $1024\times 512$ & 200 & - \\
        $\mathrm{a1b1000k10d1in.67res1024c200}$ & 1 & 1000 & 1 & - & 1 & -2/3 & $1024\times 512$ & 200 & - \\
        $\mathrm{a1b10000k10d1in.67res1024c200}$ & 1 & 10000 & 1 & - & 1 & -2/3 & $1024\times 512$ & 200 & - \\
        $\mathrm{a10b300k.01d1in.67res1024c1000}$ & 10 & 300 & 0.01 & - & 1 & -2/3 & $1024\times 512$ & 1000 & - \\
        $\mathrm{a10b300k1d1in.67res1024c1000}$ & 10 & 300 & 1 & - & 1 & -2/3 & $1024\times 512$ & 1000 & - \\
        $\mathrm{a.3b300k.01d1in.67res1024c200}$ & 0.3 & 300 & 0.01 & - & 1 & -2/3 & $1024\times 512$ & 200 & - \\
        $\mathrm{a.1b300k1d1in.67res1024c200}$ & 0.1 & 300 & 1 & - & 1 & -2/3 & $1024\times 512$ & 200 & - \\
        $\mathrm{a.1b5k1d1in.67res1024c200-nocrh}$ & 0.1 & 5 & 1 & - & 1 & -2/3 & $1024\times 512$ & 200 & No CR heating \\
        $\mathrm{a.01b5k1d1in.67res1024c200-nocrh}$ & 0.01 & 5 & 1 & - & 1 & -2/3 & $1024\times 512$ & 200 & No CR heating \\
        \hline
        Test cases for Appendix \ref{app:res_3d} & & & & & & & & & \\
        \hline
        $\mathrm{a1b5k.01d1in.67res2048c200}$ & 1 & 5 & 0.01 & - & 1 & -2/3 & $2048\times 512$ & 200 & - \\
        $\mathrm{a1b5k1d1in.67res2048c200}$ & 1 & 5 & 1 & - & 1 & -2/3 & $2048\times 512$ & 200 & - \\
        $\mathrm{a1b300k1d1in.67res2048c200}$ & 1 & 300 & 1 & - & 1 & -2/3 & $2048\times 512$ & 200 & - \\
        $\mathrm{a1b5k.01d1in.67res256c2003d}$ & 1 & 5 & 0.01 & - & 1 & -2/3 & $256\times 128\times 128$ & 200 & - \\
        $\mathrm{a1b5k1d1in.67res256c2003d}$ & 1 & 5 & 1 & - & 1 & -2/3 & $256\times 128\times 128$ & 200 & - \\
        $\mathrm{a1b300k1d1in.67res256c2003d}$ & 1 & 300 & 1 & - & 1 & -2/3 & $256\times 128\times 128$ & 200 & - \\
        \hline
    \end{tabular}
    \caption{Test cases used in this study. Each test case has an identifier, listed in column 1. Identifiers suffixed with `single' have a single density bump as initial perturbation while `nocrh' denotes no CR heating. Column 2 and 3 list the profile parameters $\alpha_0,\beta_0$ used in determining the initial profiles $\rho,P_g,P_c$ ($\alpha_0$ and $\beta_0$ are the initial ratio of CR to gas pressure and magnetic to gas pressure at $x=0$, respectively). With the initial $\rho,P_g,P_c$, column 4 to 7 are parameters used to determine the CR diffusivity $\kappa$, cooling normalization $\Lambda_0$ and cooling index $\Lambda_T$, which are constants throughout the simulation. $\eta_H$ refers to the initial ratio of CR diffusive to streaming flux at a scaleheight $H$, and similarly for $\chi_H$ and $\delta_H$. Please refer to \S\ref{subsubsec:initial} and \S\ref{subsubsec:source_terms} for complete description. The resolution, reduced speed of light and box domain are listed in column 8 to 10.}
    \label{tab:cases}
\end{table*}

Table \ref{tab:cases} summarizes the test cases used to produce the results shown in this study. As previously discussed, the initial profile is characterized by $g_0, \rho_0, T_0, a, \alpha_0, \beta_0, \eta_H$ while the cooling term is determined by the cooling index $\Lambda_T$, and $\chi_H$ or $\delta_H$. These parameters are defined therein. Without loss of generality, we set $g_0,\rho_0,T_0 = 1$ in all our simulations. The scale height $H$ should always be understood as the {\it initial} gas scale height $x=H=T_0/g_0=1$, which is therefore a constant. Although we report all our results in code units, we translate our results into physical units scaled to the Milky Way in \S\ref{sec:discussion-code}. 

\section{Linear Evolution: Thermal Instability} \label{sec:TI}

\subsection{Previous Work; Analytic Expectations} \label{subsec:linear_analytics}


Local thermal instability is caused by runaway radiative cooling, i.e. hot gas that has been cooled slightly becomes denser, causing it to cool faster. 
In gravitationally stratified media, however, buoyant oscillations can damp local thermal instability \citep{mccourt_ti_stratified,sharma12}. Stability is determined by the ratio of two timescales: the cooling time $t_\mathrm{cool}$ and the free fall $t_\mathrm{ff}$, 
where $t_{\rm ff} \approx \sqrt{2h/g}$ for the constant gravity setup in this paper. 
If cooling acts faster than buoyant damping, the instability can proceed, otherwise it is damped. This idea has been pursued by many others in various geometries and background profiles, generally leading to an instability condition of $t_\mathrm{cool}/t_\mathrm{ff}\lesssim 10$, although in our particular setup, where we evaluate $t_\mathrm{cool}/t_\mathrm{ff}$ at a scale-height, the condition is $t_\mathrm{cool}/t_\mathrm{ff}\lesssim 1$\footnote{This result only holds for small linear perturbations. \citet{choudhury19} have shown that buoyant oscillations cannot suppress large amplitude perturbations $\delta \rho/\rho \sim \mathcal{O}(1)$, where thermal instability is independent of $t_{\rm cool} /t_{\rm ff}$ and only depends on $t_{\rm cool}$.}. Observationally, this threshold has been quite successful in flagging clusters which host substantial cold gas \citep{donahue22}, though the applicability to galaxy halos, which are not in hydrostatic or thermal equilibrium, and where large amplitude density perturbations are present, is less clear \citep{nelson20,esmerian21}.

Non-thermal forces can modify thermal instability. Magnetic tension suppresses buoyant oscillations, leading to instability even when the threshold is exceeded \citep{ji18_ti}. This effect takes hold even for high plasma $\beta\sim 300$ and is independent of field orientation relative to the gravitational field. Magnetic fields can also provide pressure support for the cold clouds, so that they can be vastly out of thermal pressure balance with the surrounding. While CRs can similarly provide non-thermal pressure support \citep{huang22-clouds}, their impact {\it does} depend on the orientation of magnetic fields. \citet{butsky20_ti_cr} include CR streaming transport in their stratified simulations. However, the magnetic field is oriented perpendicular to gravity in the study, making CR streaming heating, $-v_s\cdot\nabla P_c$, non-existent in the background and only a second order effect in the evolution of the instability. In this case, the main influence of CRs is via their pressure support. Cold gas is underdense relative to the purely thermal case (reducing net density fluctuations $\delta \rho/\rho$ in the atmosphere), and thus more buoyant; they can levitate for longer and TI saturation is less sensitive to $t_\mathrm{cool}/t_\mathrm{ff}$.  However, if the magnetic field is aligned with gravity (as should be the case if there are outflows), the background heating will change the nature of thermal instability. This has been studied in a linear stability analysis by \citet{kempski20_thermal_instability}, although it has not yet been simulated.


We now review analytic expectations for the CR-modified thermal instability when the magnetic field is aligned with gravity \citep{kempski20_thermal_instability}. 
The relevant dimensionless parameters are the cooling index $\Lambda_T \equiv {\partial \ln{\Lambda}}/{\partial \ln{T}}$, 
the ratio of CR pressure to gas pressure
$\alpha = P_c/P_g$, and
\begin{equation}
    \xi = \frac{\kappa_\parallel}{\alpha t_\mathrm{cool} v_A^2} \sim \left( \frac{F_{\rm c,diff}}{F_{\rm c,st}} \right) \left( \frac{t_{\rm heat}}{t_{\rm cool}} \right) \sim \frac{\eta}{\chi} \label{eqn:stream_dominance}
\end{equation}
where in the last equality of equation \ref{eqn:stream_dominance}, we have used the diffusive flux $F_\mathrm{c,diff}\sim\kappa_\parallel\nabla P_c$, the streaming flux $F_\mathrm{c,st}\sim P_c v_A$, and the heating time $t_{\rm heat} \sim P_g/v_A \cdot \nabla P_c$; the symbols $\eta, \chi$ are defined in equation \ref{eqn:important_ratios}. In the limit where background cooling is balanced by CR heating, $t_{\rm cool} \sim t_{\rm heat}$, $\xi$ is simply the ratio of diffusive to streaming flux. For fiducial values in galaxies, it is of order unity: 
\begin{equation}
    \xi
     \sim 1 \frac{(\kappa_\parallel/10^{28}\ \mathrm{cm}^2\mathrm{s}^{-1})\qty(\beta/10)}{\qty(\alpha/1)\qty(t_\mathrm{cool}/30 \ \mathrm{Myr})\qty(c_s/100\ \mathrm{km}\mathrm{s}^{-1})^2}. \label{eqn:xi_estimate}
\end{equation}


The CR to gas pressure $\alpha$ determines how gas density changes with cooling. 
If $\alpha\gg 1$, CR pressure dominates and cooling is isochoric ($\Delta P_g/P_g \gg \delta \rho/\rho$), while if $\alpha\ll 1$, cooling is isobaric ($\Delta P_g/P_g \ll \delta \rho/\rho$) \footnote{Assuming that the perturbation $l \ll c_s t_{\rm cool}$, where $c_s t_{\rm cool}$ is evaluated at the background temperature, so that it is in sonic contact with its surroundings. Note that as the perturbation cools to lower temperatures and $c_{\rm s} t_{\rm cool}$ falls, it can fall out of pressure balance and be subject to fragmentation by `shattering' \citep{mccourt18_shattering}, but this is immaterial in the linear evolution.}. 
The cooling index $\Lambda_T$ determines if the gas is isobarically ($\Lambda_T < 2$) and/or isochorically ($\Lambda_T < 0$) thermally unstable \citep{field65}. Ignoring the influence of cosmic ray transport for now, this means that gas will be thermally unstable for $\Lambda_T < 2$ when $\alpha \ll 1$ (and cooling is isobaric), and it will be thermally unstable for $\Lambda_T < 0$ when $\alpha \gg 1$ (and cooling is isochoric). In between, there is a critical cooling index $0 \lesssim \Lambda_{\rm T,c}(\alpha) \lesssim 2$, for which gas with $\Lambda_T < \Lambda_{\rm T,c}(\alpha)$ will be thermally unstable. As ambient hot galaxy halo gas in the temperature range $10^5 < T < 10^7$K always has $\Lambda_T < 0$ (we adopt $\Lambda_{\rm T} = -2/3$ in our simulations), it will always be thermally unstable. It turns out that inclusion of CR transport changes some details, but does not change the conclusion that CRs do not generally suppress thermal instability (except in specific conditions described below) \citep{kempski20_thermal_instability}. 

 The perturbed CR heating rate due to CR streaming has two potential effects \citep{kempski20_thermal_instability}. If it is in phase with the perturbed cooling rate, and also is sufficiently strong, it can suppress thermal instability. If CR heating is out of phase with the perturbed cooling rate, which is more generally the case, the associated gas pressure fluctuations will drive an acoustic mode\footnote{For adiabatic sound waves, this can drive an acoustic instability, where sound waves grow in amplitude and steepen into shocks \citep{begelman94_acoustic_instab,tsung22-staircase}.}. In this case, thermally unstable modes result in overstable oscillations which propagates at the characteristic velocity of the heating front, i.e. the Alfven velocity. 

We can gain some intuition from the perturbation equations. In Appendix \ref{app:linear}, we present a fuller analysis, but below we outline the main elements. The perturbed CR heating is
\begin{equation}
    \delta\qty(-\vb{v}_A\cdot\nabla P_c) = -i\omega_A\delta P_c, \label{eqn:pert_cr_heat} 
\end{equation}
while the perturbed cooling is
\begin{equation}
    \delta\qty(-\rho^2\Lambda) = -\rho^2\Lambda\qty(2 - \Lambda_T)\frac{\delta\rho}{\rho} - \rho^2\Lambda\Lambda_T\frac{\delta P_g}{P_g}. \label{eqn:pert_cool}
\end{equation}
If the background is in thermal equilibrium, local thermal stability is then determined by the perturbed CR heating and gas cooling rates. Their ratio in the isobaric ($\delta P_g/P_g\ll\delta\rho/\rho$) and isochoric ($\delta P_g/P_g\gg\delta\rho/\rho$) cases is: 
\begin{equation}
    \frac{\delta\qty(\textrm{CR Heating})}{\delta\qty(\mathrm{Cooling})} =
    \begin{cases}
        \frac{i\omega_A}{\qty(2 - \Lambda_T)\omega_c}\frac{\delta P_c/P_g}{\delta\rho/\rho} & \textrm{Isobaric} \\
        \frac{i\omega_A}{\omega_c\Lambda_T} & \textrm{Isochoric},
    \end{cases}
    \label{eqn:ratio_cr_heat_cool}
\end{equation}
where $\omega_c = \rho^2\Lambda/P_c$ is the cooling rate and we have used $\delta P_c \sim - \delta P_g$ in the isochoric case. In the isochoric case, the perturbed CR heating is always $\pi/2$ out of phase with cooling. Thus, they cannot cancel. The effect of CR heating in this case is to cause the modes to oscillate and propagate at frequency $\propto\omega_A$ up the CR pressure gradient, as \citet{kempski20_thermal_instability} has shown. 
In the isobaric case, CR heating can suppress cooling if there is an out of phase component between $\delta P_c$ and $\delta\rho$. CR diffusion can provide this phase shift. In the strong diffusion limit ($k \kappa_\parallel/v_A\gg 1$), $\delta P_c$ will scale as $\delta P_c/P_g\sim i (\alpha\omega_A/\omega_d) (\delta\rho/\rho)$, i.e. shifted by a phase of $\pi/2$ from $\delta\rho$. Substituting this into eqn.\ref{eqn:ratio_cr_heat_cool} gives
\begin{equation}
    \frac{\delta\qty(\textrm{CR Heating})}{\delta\qty(\mathrm{Cooling})} \sim -\frac{\alpha v_A^2 t_\mathrm{cool}}{\kappa_\parallel} = -\xi^{-1}, \label{eqn:ratio_fast_diff}
\end{equation}
where the minus sign indicates the opposite nature of CR heating and cooling. The perturbed CR heating suppresses cooling only if diffusion is subdominant in the background ($\xi < 1$). Then, on small scales where CR diffusion across the perturbation dominates ($k \kappa/v_A \gg 1$), there is a CR 'Field length' $\lambda_{CRF}$ below which the perturbed heating balances cooling \citep{kempski20_thermal_instability}. Thermal instability is suppressed for $\lambda < \lambda_{\rm CRF}$, where 
\begin{equation} 
\lambda_{\rm CRF} \sim {\rm min}(\alpha^{1/2}, \alpha^{-1/2}) \sqrt{\kappa t_{\rm cool}}. 
\end{equation}
Note the close analogy to the Field length $\lambda_{\rm F} \sim \sqrt{\kappa_{\rm cond} t_{\rm cool}}$ set by thermal conduction, where $\kappa_{\rm cond}$ is the heat diffusion coefficient associated with thermal electrons. 

To summarize, isochoric modes ($\Lambda_{\rm T} < 0$) are always thermally unstable. In addition, if $\xi > 1$, all isobaric modes are unstable. If $\xi < 1$, then small scale modes are stabilized, but large scale modes $\lambda > \lambda_{\rm CRF}$ are still unstable. In general, CRs are unable to directly quench thermal instability in galaxy halos, where $\Lambda_{\rm T} < 0$. We shall see this is consistent with our simulations.

However, the phase velocity of thermal modes up the CR pressure gradient, which can be approximated as \citep{kempski20_thermal_instability}:
\begin{equation}
    v_\mathrm{ph} \sim {\rm min} (\frac{2}{3}, \frac{4}{15} \alpha) \ \ v_A
    \label{eqn:phase_velocity}
\end{equation}
is potentially of more interest. The ratio of the crossing time $t_{\rm cross} \sim L/v_{\rm ph}$ to the cooling time $t_{\rm cool}$ is: 
\begin{equation} 
\theta \equiv \frac{t_{\rm cool}}{t_{\rm cross}} \sim \left(\frac{t_{\rm cool}}{t_{\rm heat}} \right) \left(\frac{t_{\rm heat}}{t_{\rm cross}} \right) \sim 
\chi \frac{L_c}{L} \ \ {\rm min} (\frac{2}{3}\alpha^{-1}, \frac{4}{15}), \label{eqn:theta_model}
\end{equation} 
where $L_c = P_c/\nabla P_c$, and $t_{\rm heat}  \sim P_g/(v_A \cdot \nabla P_c)$, and we have used equation \ref{eqn:phase_velocity}. If $\theta > 1$, then thermal modes will propagate out of the system before cooling significantly. In general, we expect $\theta < 1$, since $\chi = t_{\rm cool}/t_{\rm heat} < 1$ (otherwise the background will be overheated), and we also expect $L_c/L \lesssim 1$. We can also see this from the parametrization:
\begin{equation} 
\theta \equiv \frac{t_{\rm cool}}{t_{\rm cross}} \sim \left(\frac{t_{\rm cool}}{t_{\rm ff}} \right) \left(\frac{t_{\rm ff}}{t_{\rm cross}} \right) \sim \frac{\delta}{\beta^{1/2}} \ \ {\rm min} (\frac{2}{3}, \frac{4}{15} \alpha)
\end{equation} 
so that when there is thermal instability in our setup, $\delta = t_{\rm cool}/t_{\rm ff} < 1$, the fact that $\beta \gtrsim 1$ means that $\theta < 1$.

However, there is sufficient uncertainty that it is worth investigating numerically how the propagation of thermal modes affect thermal instability. In addition, \citet{kempski20_thermal_instability} suggest that the oscillations induced by mode propagation could also potentially damp thermal instability and change the threshold for thermal instability, particularly if the oscillation frequency $\omega_{\rm A}$ is higher than the free-fall frequency. We now address this in our simulations.

\subsection{Propagation of modes} \label{subsec:propagation}

\begin{figure*}
    \centering
    \includegraphics[width=\textwidth]{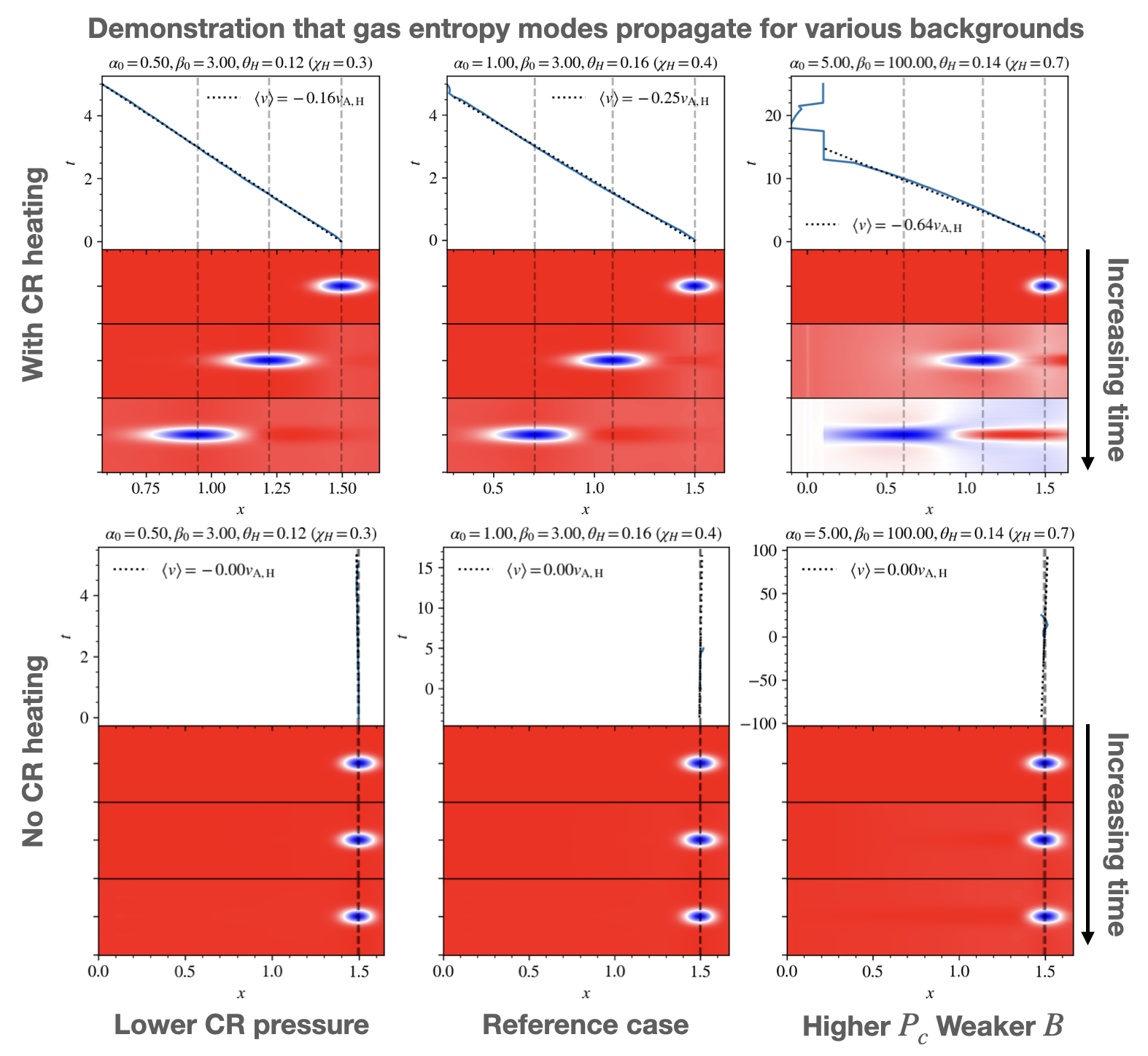}
    \caption{Propagation of thermal entropy modes for different values of $\alpha_0, \beta_0, \theta_H = t_\mathrm{cool,H}/t_\mathrm{cross,H}$ with (top row) and without (bottom row) CR heating (note that $\theta_H$ is related to $\chi_H$, the ratio of CR heating to radiative cooling through eqn.\ref{eqn:theta_model}). Note that subscripts $0$ and $H$ means quantities evaluated at the base $x=0$ and at the initial gas scale-height $x=H=T_0/g_0=1$ respectively. A Gaussian density bump is placed at $x=1.5$ initially and its trajectory followed. In each panel we trace the location of the bump with time, showing snapshots of the temperature field. With CR heating, the bump is clearly moving. No propagation is seen without CR heating. We fit the slope of the $x-t$ plot while the bump is in its linear phase (i.e. $\delta\rho/\rho < 1$) to extract its propagation velocity $\langle v\rangle$ (in units of $v_{A,H}$). The propagation speeds are consistent with that predicted from linear theory eqn.\ref{eqn:phase_velocity}. The propagating mode appear stretched out due to slight differences in $v_\mathrm{ph}$ across its width.}
    \label{fig:xshow}
\end{figure*}

We begin by verifying that thermal entropy modes do propagate at the expected velocity given by linear analysis. To do this we insert a Gaussian density bump of form
\begin{equation}
    \delta\rho = \mathcal{A}\rho\qty(x_b) e^{-\qty(\frac{\qty(x - x_b)^2 + \qty(y - y_b)^2}{\Delta})}, \label{eqn:gaussian_bump}
\end{equation}
where $\mathcal{A}$ is the amplitude (in units of $\rho (x_b)$, the local background density), $(x_b, y_b)$ is the location of the bump and $\Delta$ is the width. We place the bump at $(1.5 H, 0)$, with amplitude $\mathcal{A} = 0.01$ and width $0.1 H$. The other parameters used are listed on top of each panel in fig.\ref{fig:xshow} and the resulting evolution is shown. In each panel we plot the $x-t$ trajectory of the perturbation and display snapshots of the temperature field at different times. The perturbation is tracked by finding the location with minimum temperature. CR heating to the thermal gas $-v_A\cdot\nabla P_c$ is switched off for the bottom row as a control to illustrate the effect of CR heating. To ensure the background is in steady state while the mode is propagating (for the sake of clarity in our demonstration), we impose the CR source term $\mathcal{Q}$ ({\it only} for the studies of mode propagation in  \S\ref{sec:TI}; we do not include such source terms in \S\ref{subsec:nonlinear_outcome}) on the right hand side of equation \ref{eqn:cr_energy}: 
\begin{equation}
   \mathcal{Q}\qty(x) = \dv{F_c}{x} - v_A\dv{P_c}{x}. \label{eqn:cr_source}
\end{equation}
Since $P_c\propto\rho^{\gamma_c/2}$ initially, this simplifies to
\begin{equation}
   \mathcal{Q}\qty(x) = -\frac{\kappa}{\gamma_c - 1}\dv[2]{P_c}{x} = -\frac{\gamma_c}{\gamma_c - 1}\eta L_c v_A\dv[2]{P_c}{x}. \label{eqn:cr_source_simplify}
\end{equation} 

In fig.\ref{fig:xshow} we present three test cases, corresponding (from left to right) to parameters $(\alpha_0,\beta_0,\eta_H,\theta_H)=$ $(0.5,3,0.01,0.12)$, $(1,3,0.01,0.16)$ and $(5,100,0.01,0.14)$. CR streaming dominate the transport as $\eta_H\ll 1$. Note that $\theta_H$ is adjusted through $\chi_H$ via eqn.\ref{eqn:theta_model}. Changing $\beta_0$ changes the Alfven speed $v_A$ while changing $\alpha_0$ affects the fraction of the Alfven speed the modes propagate at. Using the $(1,3,0.01,0.16)$ (middle column) as a reference case, halving the CR pressure roughly halves the propagation speed while increasing it by 5 times boosts the propagation to the asymptotic limit of $v\sim -2 v_A/3\sim -0.67 v_A$. The propagation speed of the reference case, $\langle v\rangle = -0.25 v_{A,H}$, is consistent with $-4\alpha v_A/15$ predicted from linear theory (eqn.\ref{eqn:phase_velocity}). Note that the mode velocity displayed in the figure $\langle v\rangle$ is given in terms of $v_{A,H}$, the Alfven velocity at a thermal scale-height. $v_A$ varies with density along the profile, so the slight difference found in our test cases from that predicted from linear theory is to be expected. 
The minus sign in the propagation speeds indicate the modes are propagating up the $P_c$ gradient.

\begin{figure}
    \centering
    \includegraphics[width=0.40\textwidth]{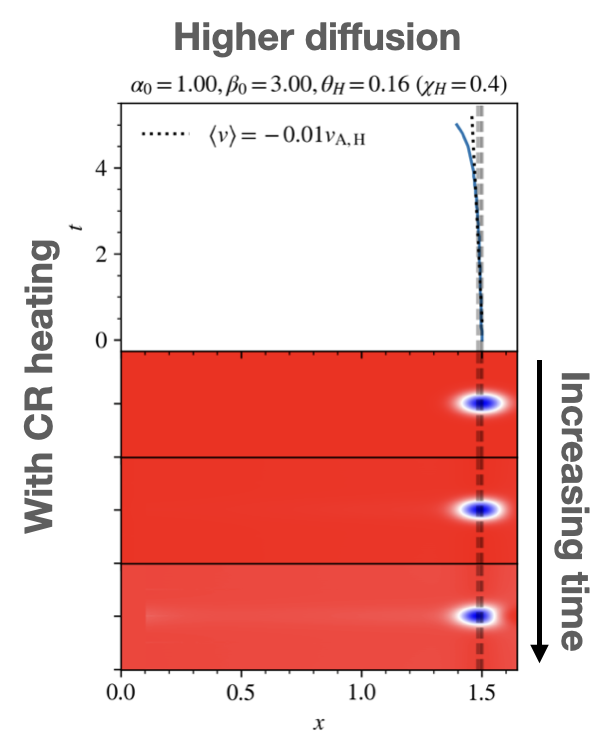}
    \caption{Same as the middle column of fig.\ref{fig:xshow} but with enhanced CR diffusion ($\eta_H = 1$). There is no propagation in this case even when CR heating is present.}
    \label{fig:xshow_diff}
\end{figure}

The bottom row of fig.\ref{fig:xshow}, with CR heating switched off, shows no propagation of the modes and reflects clearly that propagation is completely due to CR heating. In fig.\ref{fig:xshow_diff} we show again the middle column case ($\alpha_0=1,\beta_0=3,\theta_H=0.16$) but with increased diffusion ($\eta_H = 1$, i.e. the diffusive flux is equal to the streaming flux at a thermal scale-height). There is no mode propagation in this case. The reason is diffusion causes CRs to slip out of the perturbation before they can heat the gas, thus removing the effects of CR heating.

From these test cases we have shown that CR streaming, through streaming heating, can cause thermal entropy modes to propagate at some fraction of the Alfven velocity consistent with linear theory. If one removes the effect of CR heating, either by switching off the source term $-v_A\cdot\nabla P_c$ or by increasing diffusion, the modes do not propagate.

\subsection{Does propagation suppress thermal instability?} \label{subsec:suppress_propagation}

Having shown in \S\ref{subsec:propagation} that thermal entropy modes propagate in a CR streaming dominated flow under the effect of CR heating, we consider whether thermal instability can be suppressed as proposed in \S\ref{subsec:linear_analytics}, that is, if mode propagation sets a time limit $t_\mathrm{cross}$ on how much the perturbations can grow before moving out of the cooling region. If $t_\mathrm{cross}\ll t_\mathrm{cool}$, perturbations can hardly grow before they propagate out of the cooling region, effectively suppressing the instability. 

In this section, we initiate a stratified profile in hydrostatic and thermal balance and seed random isobaric perturbations: 
\begin{equation}
    \frac{\delta\rho}{\rho} =
    \begin{cases}
        \sum_{mn}\frac{4 A_{mn}}{\sqrt{N}}\sin(\frac{2\pi n x}{L_x} + \phi_{x,n}) \sin(\frac{2\pi m y}{L_y} + \phi_{y,m}), & \mathrm{(2D)} \\
        \sum_{lmn}\frac{8 A_{lmn}}{\sqrt{N}}\sin(\frac{2\pi n x}{L_x} + \phi_{x,n}) \sin(\frac{2\pi m y}{L_y} + \phi_{y,m}) &  \mathrm{(3D)} \\
        \qquad \sin(\frac{2\pi l z}{L_z} + \phi_{z,l}), & 
    \end{cases}
    \label{eqn:rand_pert}
\end{equation}
where $1\leq n, m, l \leq 10$, $\phi_x, \phi_y, \phi_z$ are phase shifts selected randomly from $(0, 2\pi)$, $L_x, L_y, L_z$ are domain sizes in the $x, y, z$ directions, $A_{lmn}$ are mode amplitudes selected randomly from a Gaussian pdf with $(\mu, \sigma) = (0, 0.1)$ and $N$ is the total number of modes.

We then let the simulation evolve, and record the amount of cold gas formed near a thermal scale-height ($0.9 H < x< 1.1 H$). As a comparison, we also run simulations without CR heating and with higher CR diffusion. Switching off CR heating allows us to isolate the effect of CR heating on thermal instability, whereas increasing CR diffusion allows us to isolate the effect of mode propagation (see results from \S\ref{subsec:propagation}). In particular, recall that in our setup, $\theta_{\rm H} \propto \chi_{\rm H}$ (eqn.\ref{eqn:theta_model}). Thus, if thermal instability is suppressed, it is difficult to tell if this is due to mode propagatiion ($\theta > 1$) or CR overheating ($\chi > 1$). In order to break this degeneracy and isoloate the effects of mode propagation, we utilize the fact that increasing CR diffusion can suppress mode propagation. We saw this explicitly in \S\ref{subsec:propagation}, when $\eta_{\rm H}=1$. Our `enhanced diffusion' tests here use the same value of $\eta_{\rm H}$.

\begin{figure*}
    \centering 
    \includegraphics[width=\textwidth]{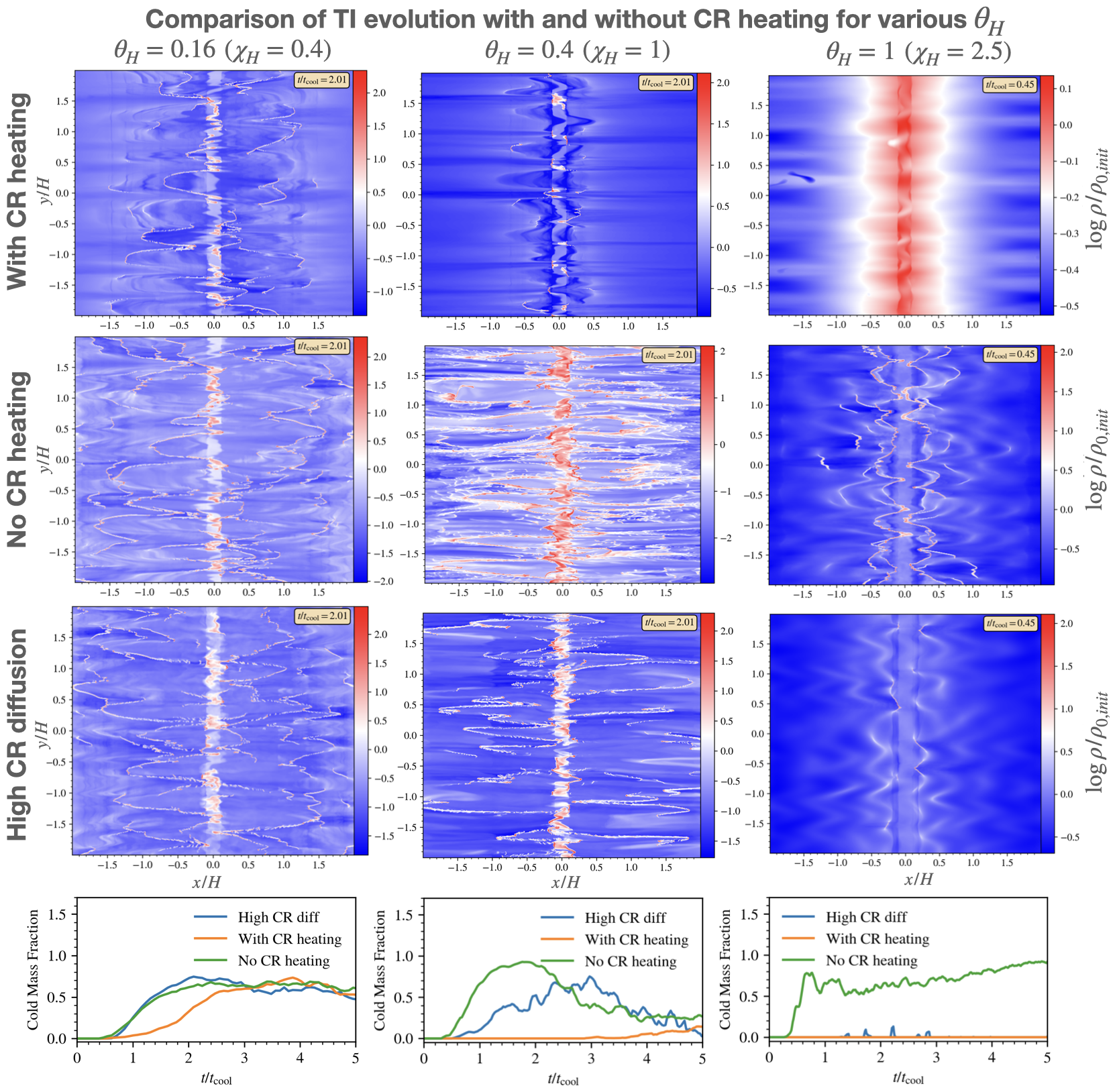} 
    \caption{Comparison of thermal instability evolution with (top row) and without (second row) CR heating for streaming dominated transport ($\eta_H=0.01$), and with higher diffusion (third row, $\eta_H=1$) for (from left to right) $\theta_H=t_\mathrm{cool,H}/t_\mathrm{cross,H} = 0.16,0.4,1$. The bottom row displays the cold mass fraction taken near a thermal scale-height ($0.9 H<x<1.1H$) as a function of time $t/t_\mathrm{cool,H}$. The density slices displayed are taken roughly at times where the cold mass fraction peaks. Note that quantities with subscript $H$ are taken at the initial gas scaleheight $x=T_0/g_0=1$. The remaining parameters required to uniquely determine the initial profiles for these test cases are $\alpha_0=1,\beta_0=3$. The conversion from $\theta_H$ to $\chi_H$ in our setup is given by eqn.\ref{eqn:theta_model}.}
    \label{fig:nonlinear_snapshots_a1b3k_01}
\end{figure*}

In fig.\ref{fig:nonlinear_snapshots_a1b3k_01} we compare the evolution of thermal instability with (top row) and without (second row) CR heating for streaming dominated transport ($\eta_H=0.01$), and with enhanced diffusion (third row, $\eta_H=1$), for (from left to right) $\theta_H=0.16,0.4,1$. The bottom row displays the cold mass fraction taken near a thermal scale-height ($0.9H < x < 1.1H$) as a function of time $t/t_\mathrm{cool,H}$. The density slices displayed are taken roughly at times where the cold mass fraction peaks. The initial profile parameters for these test cases are $\alpha_0=1,\beta_0=3$. Note that $\theta_H, \chi_H, t_\mathrm{cool,H}$, with the subscript $H$, are parameters evaluated at $x=H$ (where $H$ is the thermal scale-height of the initial profile). Note also that the initial density, $P_c$ and $P_g$ profiles and magnetic field strength of the test cases displayed in fig.\ref{fig:nonlinear_snapshots_a1b3k_01} are all the same. Thus, initial background CR forces $\nabla P_c$ and heating rates $v_A \cdot \nabla P_c$ are identical in all cases. This remains true even when we change the amplitude of CR diffusion, which ordinarily would change CR profiles and heating rates. However, the CR source terms implemented in equations \ref{eqn:cr_source} and \ref{eqn:cr_source_simplify} guarantee identical $P_c(x)$ profiles. We emphasize that this is a numerical convenience to isolate the impact of mode propagation by enabling the background $P_c(x)$ profile to be held fixed. We do {\it not} include source terms $Q(x)$ in our study of non-linear outcomes in \S\ref{subsec:nonlinear_outcome}.

Let us first compare the left ($\chi_{\rm H}=0.4)$) and rightmost ($\chi_{\rm H}=2.5$) columns in Fig \ref{fig:nonlinear_snapshots_a1b3k_01}, which correspond to the cases where CR heating provides only a fraction of the heating rate ($\chi_{\rm H}=0.4)$) and overheats the gas ($\chi_{\rm H}=2.5)$). As one might expect, when there is CR heating, there is ample cold gas in the former case (broadly comparable to the `no CR heating' case), and almost no cold gas in the latter case. The strong suppression in the overheated $\chi_{\rm H}=2.5$ case is still present when CR diffusion is included. This suggests that overheating, rather than mode propagation (which is absent once CR diffusion is included), suppresses thermal instability. On the other hand, when CR heating marginally balances cooling, for $\chi_{\rm H} = 1, \theta_{\rm H}  =0.4 $, the case including CR heating has almost no cold gas, but including CR diffusion allows ample cold gas to form. This suggests that mode propagation, rather than overheating, is responsible for the suppression of thermal instability. We have verified this directly by examining heating rates, as well as observing the propagation of cooling gas clouds.

\begin{figure}
    \centering
    \includegraphics[width=0.49\textwidth]{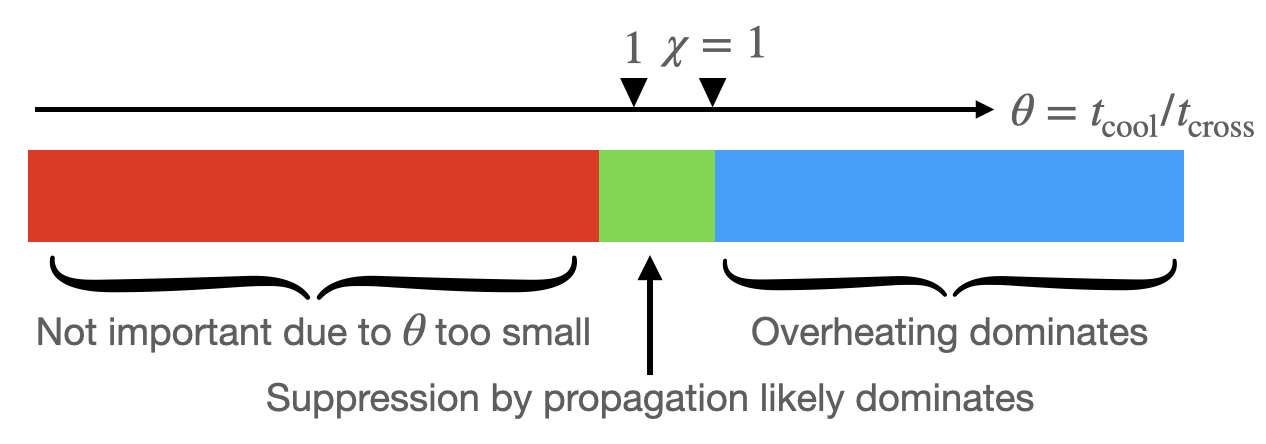}
    \caption{A color-bar illustrating the effect of mode propagation on TI for various regimes of $\theta=t_\mathrm{cool}/t_\mathrm{cross}$.}
    \label{fig:prop_effect_bar}
\end{figure}

Thus, suppression of TI by propagation effects can occur. However, it only occurs in a narrow range around $\theta \approx 1$, as illustrated in Fig \ref{fig:prop_effect_bar}: for low values of $\theta$, mode propagation is too slow, while for high values of $\theta$, overheating suppresses thermal instability. Thus, mode propagation is unlikely to play an important role in regulating the abundance of cold gas. In fact, overheating during the non-linear stages is much more interesting. We turn to this next.  

\section{Nonlinear Outcomes: Winds and fountain flows} 
\label{subsec:nonlinear_outcome}

Since suppression of TI by mode propagation is at best marginally important, thermal instability will likely develop in a system in global thermal equilibrium, i.e. when there is no overheating. What would be the nonlinear outcome of TI then, particularly when CR heating plays an important thermodynamic role in the system? Note that we started with a profile in both hydrostatic and thermal balance,
\begin{gather}
    \nabla P_g + \nabla P_c = -\rho g, \\
    \rho^2\Lambda = -v_A\cdot\nabla P_c + \mathcal{H}.
\end{gather}
As TI develops, it draws mass out of the atmosphere and causes the density to decrease. It is not immediately clear, in the subsequent evolution, that both hydrostatic and thermal balance will be maintained. In particular, since radiative cooling varies with density much more sensitively than CR heating, one could imagine in the nonlinear evolution, the energy budget is likely dominated by CR heating. This could eventually drive the system out of both hydrostatic and thermal balance. Indeed, we shall see that this is exactly what happens. We shall also see that the reduced gas density also reduces gas pressure, causing $\beta$ to decrease in the atmosphere. 


\begin{figure*}
    \centering
    \includegraphics[width=\textwidth]{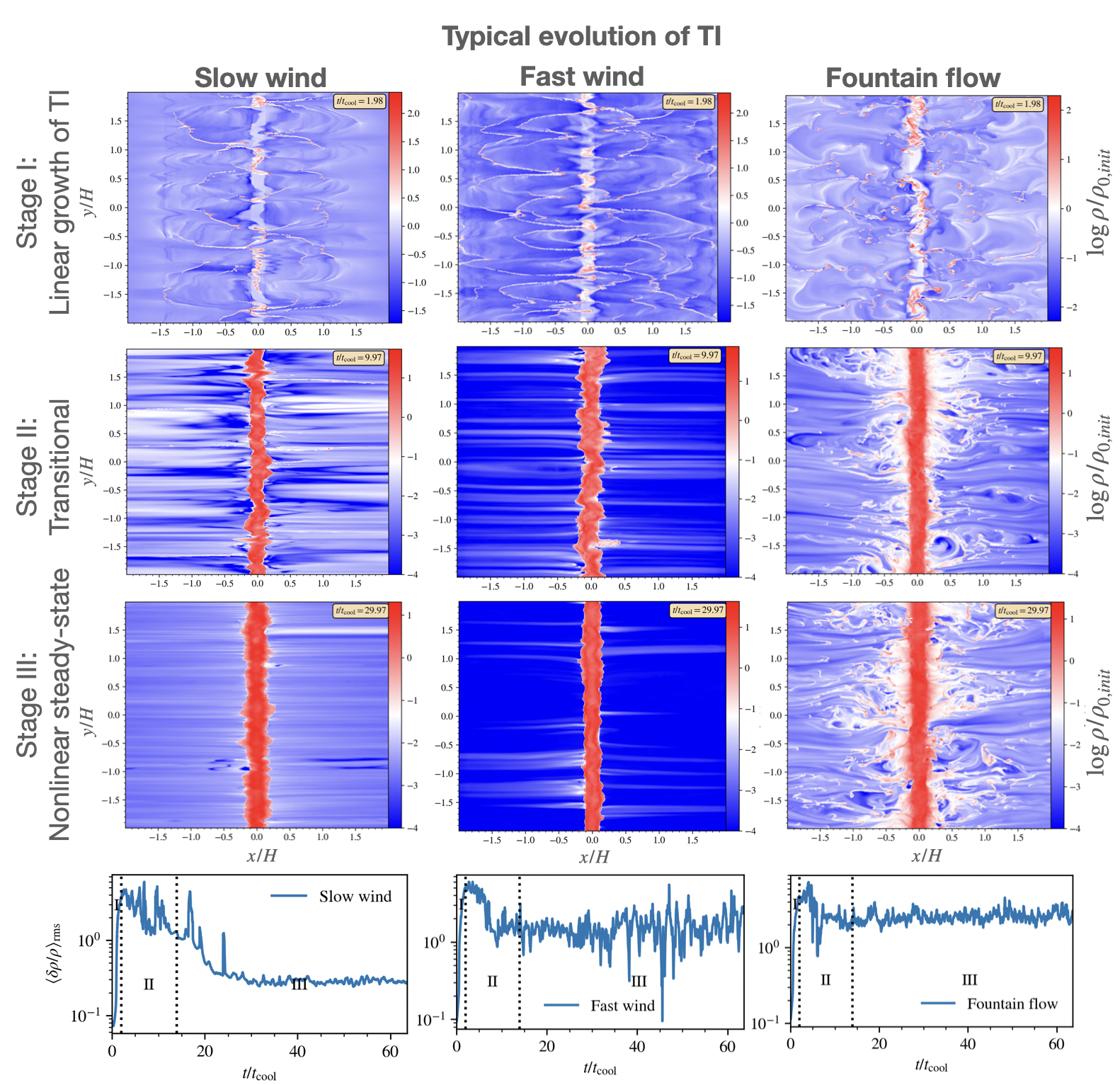}
    \caption{Typical evolution of TI with CR heating. Shown above are density slices at $t=2,10,30 t_\mathrm{cool}$ for the three categories of outcomes - slow wind, fast wind and fountain flow, with blue and red coloring indicating less dense and denser gas respectively. In this figure, $t_\mathrm{cool}$ refers to the initial cooling time at $x=H$. Shown at the bottom are $\langle\delta\rho/\rho\rangle_\mathrm{rms}$ (r.m.s. fractional density deviation from the mean) as a function of time taken at a strip near a scale height ($0.9 H<x<1.1H$) for the three cases shown, with the black dotted line demarcating different stages of evolution. 
    The case identifier for the three cases shown are: `slow wind' - a1b5k.01d1in.67res1024c200; `fast wind' - a1b5k1d1in.67res1024c200; `fountain flow' - a1b300k1d1in.67res1024c200. $\rho_{0,init}$ with the extra subscript $init$ is added for clarity and is synonymous to $\rho_0=1$, the density at the base of the initial profile.}
    \label{fig:typical_evo}
\end{figure*}

\begin{figure*}
    \centering
    \includegraphics[width=\textwidth]{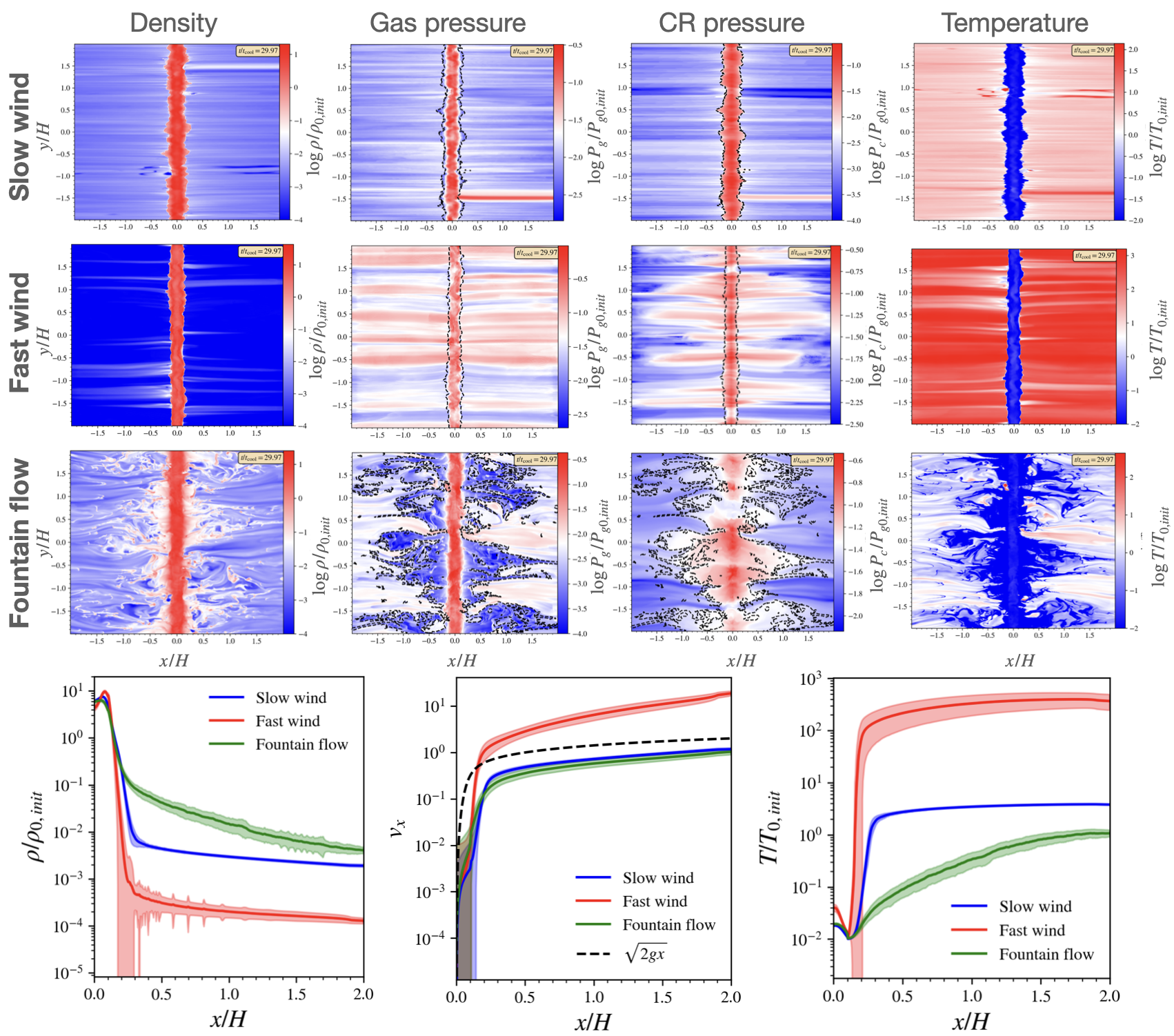}
    \caption{Nonlinear outcomes of TI with CR heating. Density, gas pressure, CR pressure and temperature slices at $t=30 t_\mathrm{cool}$. $t_\mathrm{cool}$ shown in this figure refers to the initial cooling time at $x=H$. The black dashed contours in the gas and CR pressure slices demarcate gas with temperature below and above $0.3 T_{0,init}$, where $T_{0,init}$ is the temperature of the initial profile and is synonymous to $T_0=1$ (similarly for $\rho_{0,init}=\rho_0=1$ and $P_{g0,init}=P_{g0} = 1$, the added subscript $init$ for added clarity). Shown at the bottom are the time averaged projection plots of the density, outflow velocity and temperature for the three cases shown. Time average projection refers to averaging from $t=31.8 t_\mathrm{cool}$ to $63.6 t_\mathrm{cool}$, when the flows have settled onto their nonlinear steady-states, and then spatial averaging over $y$. $t_\mathrm{cool}$ refers to the initial cooling time at a $x=H$. The shaded regions denote $1\sigma$ variations throughout the time averaging. The case identifiers are the same as fig.\ref{fig:typical_evo}.}
    \label{fig:three_outcomes}
\end{figure*}

\begin{figure*}
    \centering
    \includegraphics[width=\textwidth]{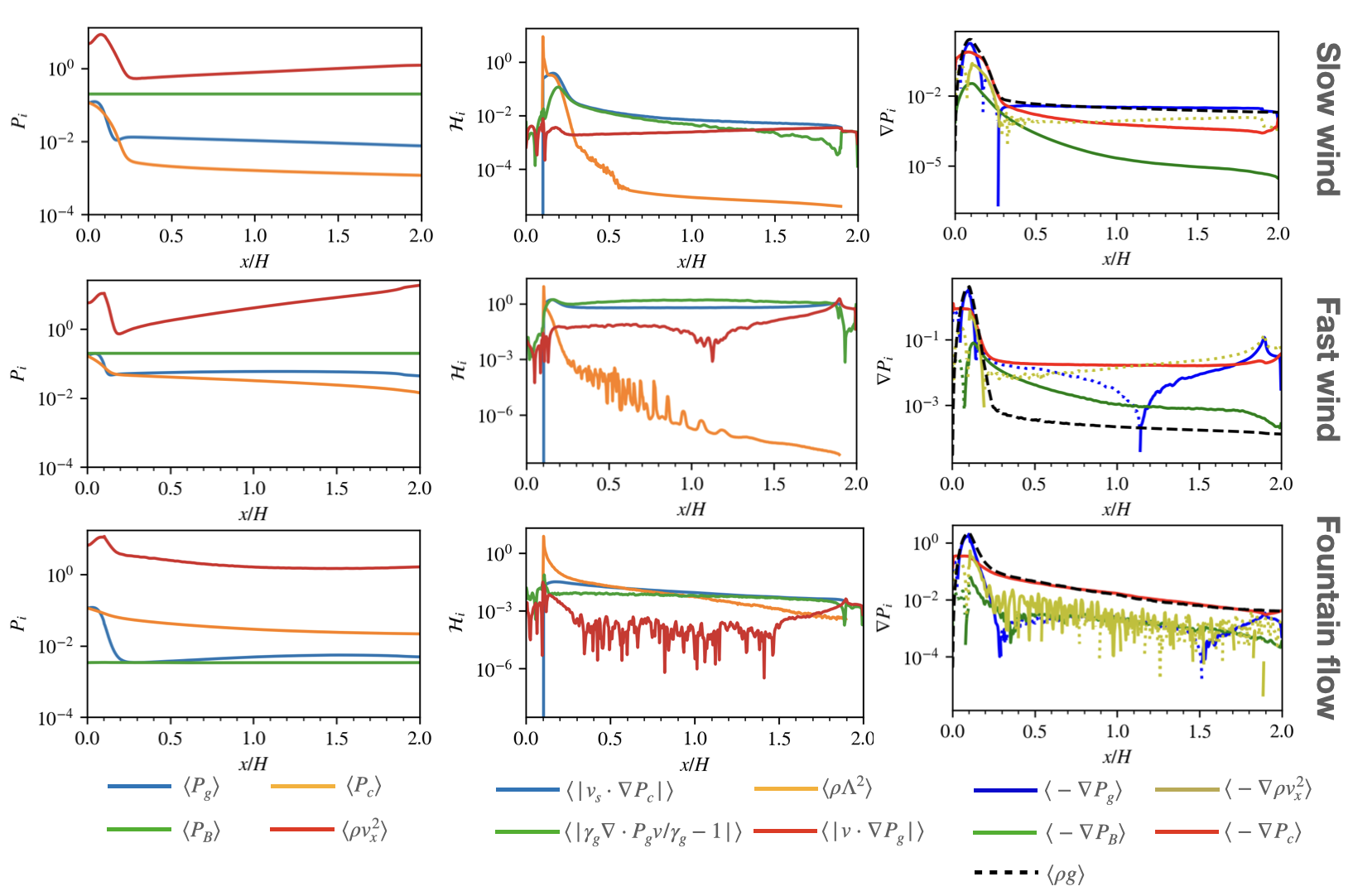}
    \caption{Time averaged projection plots of the pressures (left column), energy terms (center column) and pressure gradients (right column) for the three categories of outcomes. Time average projection refers to averaging from $t=31.8 t_\mathrm{cool}$ to $63.6 t_\mathrm{cool}$, when the flows have settled onto their nonlinear steady-states, and then spatial averaging over $y$. $t_\mathrm{cool}$ refers to the initial cooling time at $x=H$. The cases displayed are the same as that in fig.\ref{fig:three_outcomes}. In the pressure plots (left column), the pressures are represented by: gas pressure $\langle P_g\rangle$ - blue; CR pressure $\langle P_c\rangle$ - orange; magnetic pressure $\langle B^2/2\rangle$ - green; ram pressure $\langle \rho v^2_x\rangle$ - red. In the energy plots (center column): CR heating $\langle |\vb{v}_s\cdot\nabla P_c|\rangle$ - blue; cooling $\langle \rho^2 \Lambda\rangle$ - orange; enthalpy flux $\langle |\gamma_g\nabla\cdot P_g \vb{v}|/\gamma_g - 1\rangle$ - green; gas work done $\langle |\vb{v}\cdot\nabla P_g|\rangle$ - red; There is no cooling and CR heating within the buffer zones ($0<x<a$ and $2 H - a<x<2 H$). In the pressure gradient plots: $\langle -\nabla P_g\rangle$ - blue; $\langle -\nabla P_c\rangle$ - red; $\langle -\nabla P_B\rangle$ - green; $\langle -\nabla\rho v^2_x\rangle$ - yellow. Positive values are represented by solid curves, negative by dotted lines. $\langle \rho g\rangle$, the gravitational force, is denoted by a black dashed line. The angled brackets indicate the time average projection, they are omitted in the legend and in other plots to reduce clutter.}
    \label{fig:flow_properties}
\end{figure*}

\begin{figure}
    \centering
    \includegraphics[width=0.48\textwidth]{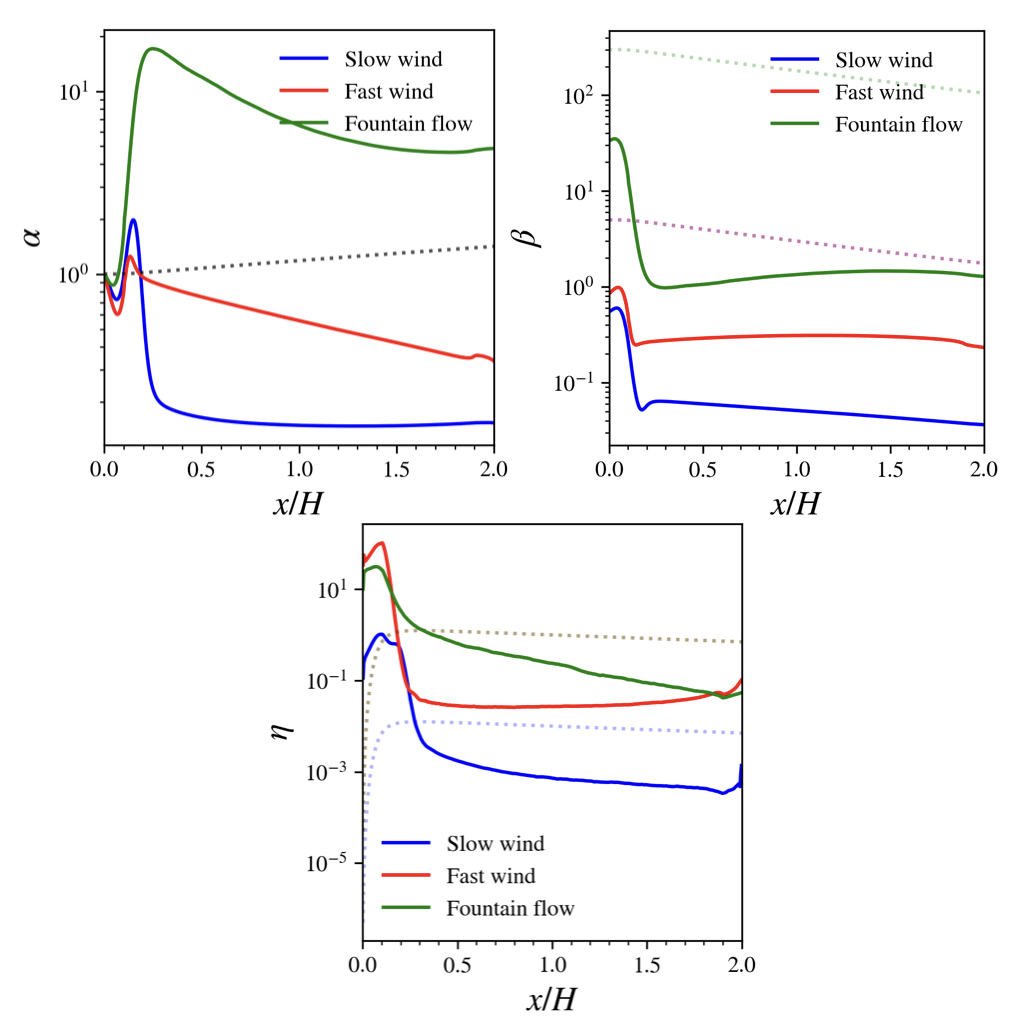}
    \caption{$\alpha,\beta,\eta$ are in general functions of position and time. The figure show the time average projection plots of $\alpha,\beta,\eta$ when steady state has been reached (solid lines) in comparison to their initial values (dotted line). Time average projection refers to averaging from $t=31.8 t_\mathrm{cool}$ to $63.6 t_\mathrm{cool}$, when the flows have settled onto their nonlinear steady-states, and then spatial averaging over $y$. $t_\mathrm{cool}$ refers to the initial cooling time at $x=H$. Note that due to mass draw-out from TI, the ending $\alpha,\beta,\eta$ could be very different from their starting values.}
    \label{fig:alpha_beta_eta}
\end{figure}

\begin{figure}
    \centering
    \includegraphics[width=0.48\textwidth]{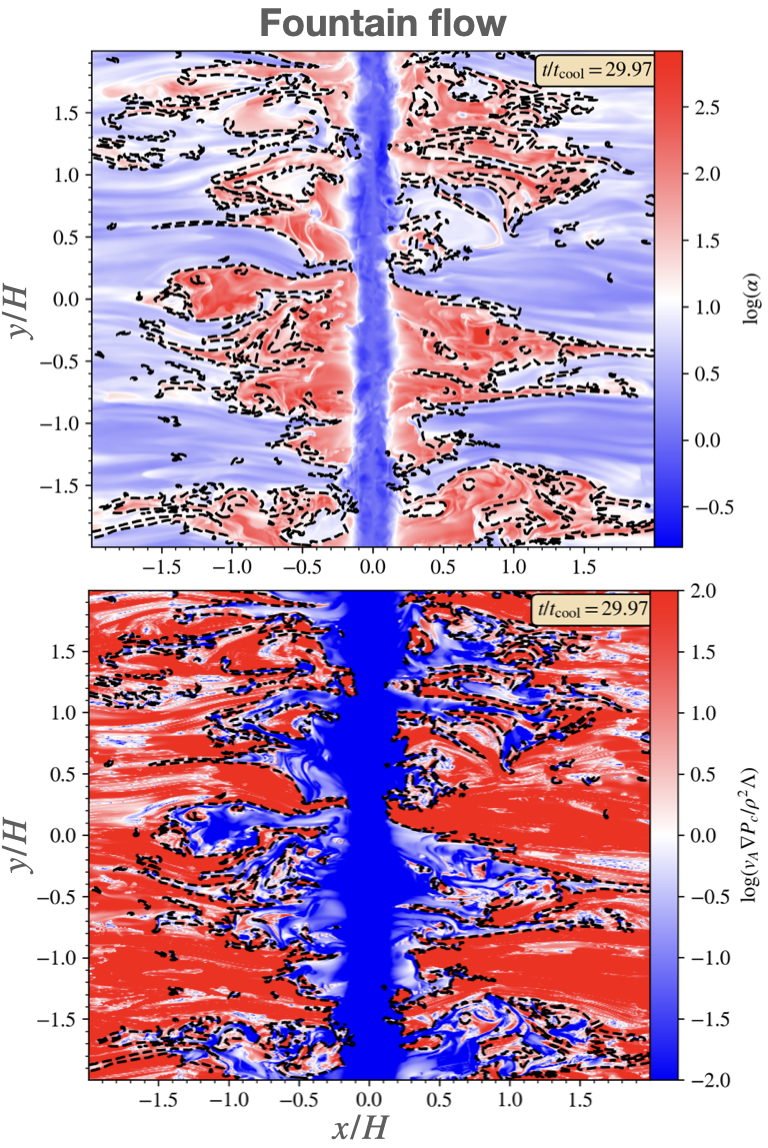} 
    \caption{Ratio of CR to gas pressure $\alpha=P_c/P_g$ (top) and CR heating to radiative cooling $\abs{v_A\nabla P_c}$ (bottom) for the fountain flow case at $t=30 t_\mathrm{cool}$. $t_\mathrm{cool}$ refers to the initial cooling time at $x=H$. Contours demarcating cold gas ($T<0.3 T_{0,init}$) from the hotter gas are marked by black dashed lines.}
    \label{fig:alpha_heat_cool_fountain}
\end{figure}

\begin{figure*}
    \centering
    \includegraphics[width=\textwidth]{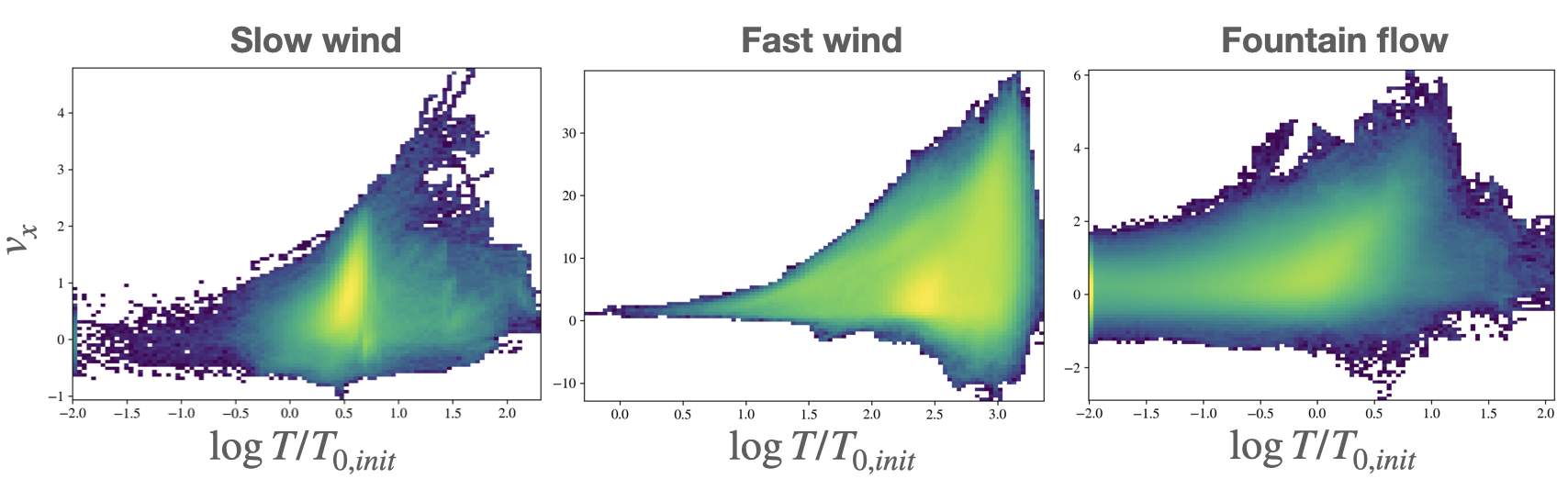}
    \caption{$v_x - T$ (velocity vs. temperature) phase plots for the three displayed cases in fig.\ref{fig:typical_evo}: slow wind case on the left; fast wind in the middle; fountain flow on the right. The $v_x - T$ plots are constructed from binning the gas cells by their $x$ velocity and temperature from $21.2-24.7 t_\mathrm{cool}$. $t_\mathrm{cool}$ refers to the initial cooling time at $x=H$. $T_{0,init}$ refers to the initial profile temperature, i.e. $T_{0,init}=T_0=1$, the extra subscript $init$ for added clarity.}
    \label{fig:vx_temp}
\end{figure*}

\subsection{Overview of simulation outcomes} 
\label{subsubsec:overview_outcome}

There are in general three categories of outcomes for nonlinear TI with CR heating\footnote{The reader can view videos pertaining to the discussion in this section at the following link: \url{https://www.youtube.com/playlist?list=PLQqhpX30dsYq2cD51L4M2pNQAlm0GSle9}.}. Here, we analyze 3 prototypical simulations which exemplify these outcomes: 'slow wind' ($\beta_0=5, \eta_{\rm H}=0.01$), 'fast wind' ($\beta_0 = 5, \eta_{\rm H}=1$), and 'fountain flow' ($\beta_0 = 300, \eta_{\rm H}=1$). All simulations have $\delta_{\rm H}=1,\alpha_0=1$. We run the simulations for up to $60 t_\mathrm{cool,H}$, long enough for the flow to settle onto a nonlinear steady state. Although we fix CR pressure at the base, in Appendix \ref{app:fix_fc} we show that similar outcomes arise if we fix the CR flux. 
As mentioned in \S\ref{subsubsec:resolution_rsol}, with the expectation that CR heating dominates the nonlinear evolution, we henceforth remove the temperature ceiling. 

The typical evolution of these simulations are shown in fig.\ref{fig:typical_evo}, showing the density slices at $t=2,10,30 t_\mathrm{cool}$, which mark the three stages of TI evolution: stage I, linear growth of TI; stage II, the transitional phase; stage III, the nonlinear steady-state. Shown at the bottom of the figure are the r.m.s density variations $\langle\delta\rho/\rho\rangle_\mathrm{rms}$ as a function of time for the three displayed cases, with black dotted lines demarcating the different stages of evolution. In stage I, seed density perturbations grow to nonlinear amplitudes over several $t_\mathrm{cool}$ (from $t=0$ to $\sim 2-3 t_\mathrm{cool}$ in our sims with $10\%$ seed amplitude). Over this time, $\langle\delta\rho/\rho\rangle_\mathrm{rms}$ grows exponentially. In stage II, cold, dense gas formed from TI collapses under gravity, forming a dense mid-plane disk. Stage II marks the transitional period where such a two-phase medium (a cold, dense mid-plane disk bounded by hot, rarefied halo gas) is formed. This is the typical end state of TI simulations \citep{mccourt_ti_stratified, ji18_ti, butsky20_ti_cr}. However, in the presence of CR heating, this two-phase hydrostatic disk-halo medium can be globally unstable, and TI generally veers towards one of three possible outcomes in Stage III, the non-linear steady state. 

The first outcome is a slow wind, where the disk-halo structure is well maintained and the interface clearly defined. A wind with velocity less than the local escape velocity ($\sqrt{2 g x}$) develops. The second outcome is a fast wind, where again the disk-halo structure is well maintained with a distinctive interface, but the wind has a velocity greater than the local escape velocity, so that much of the gas in the halo are blown away, leaving the halo much more rarefied compared to the weak-wind case. The third outcome is a fountain flow, characterized by filaments of cold, dense gas rising and falling from the central disk. The warmer gas is generally outflowing, leaking through the space between the cold tendrils, occasionally entraining several tiny cold clouds out. The halo gas flow is turbulent. 

The transition from linear TI to these outcomes through the formation of a disk-halo structure takes around $10 t_\mathrm{cool}$, marked by high values of $\langle\delta\rho/\rho\rangle_\mathrm{rms} \sim 5$. The density fluctuations and flow structure then stabilize during stage III, the nonlinear steady-state. Surveying parameter space, we have found that slow winds are typically associated with low $\beta$, low CR diffusivity (or streaming dominated) flows, fast wind with low $\beta$, high CR diffusivity flows, whereas fountain flows happen mostly for high $\beta$ flows. We will quantify these criteria and supply theoretical explanations.

We describe the flow properties of these three outcome categories in greater detail using fig.\ref{fig:three_outcomes}, which shows the density, gas pressure, CR pressure, temperature slices at $t=30 t_\mathrm{cool}$ (top 3 rows) and the time averaged projection plots\footnote{The time averaged projection plots are obtained by first averaging the slices across the $y$-axis (projection) and then time averaging over $t=31.8 - 63.6 t_\mathrm{cool}$, when the flow is well within stage III, the nonlinear steady-state.} of the density, outflow velocity and temperature (bottom row). From the slice plots, one can observe the aforementioned disk-halo structure. The central disk, spanning a height of $\sim 0.2 H \approx 2 a$ (where $a$ is the smoothing length of the gravitational field) is made up of cold gas near the temperature floor. The flow patterns of the slow-wind and fast-wind case appear collimated, with the major differences being: 1. the outflow velocity of the fast-wind case can exceed the local escape speed (bottom central panel), 2. the density is significantly lower for the fast-wind case and 3. the temperature is appreciably higher for the fast-wind case. The gas and CR pressures of the fast wind case are also greater. Some minor differences between the slow and fast wind case include smaller variability for the weak-wind (as indicated by the shaded regions in the time averaged plots). Note that once out of the central disk, the flows become isothermal\footnote{Isothermal in the sense that the temperature profile appears spatially constant, not that the equation of state is isothermal.} for both the slow and fast wind cases (from $x\approx 0.3 H$ outwards). The density and hence pressure is also relatively constant. The fountain flow is vastly different from the other two outcomes, showing more turbulent dynamics. Despite the relatively similar time averaged outflow velocity profile with the slow-wind case, both of which are sub-escape speed, the cold gas is far more extended in the fountain flow case, leading to higher average density and lower average temperature. The cold gas extending away from the disk is also low in gas pressure but high in CR pressure.

Due to mass drop-out in the atmosphere from TI, which produces the cold gas disk in the mid-plane, the ending $\alpha,\beta,\eta$ profiles could be vastly different from what it started with. For example, in fig.\ref{fig:alpha_beta_eta}, the time averaged projection plots of $\alpha,\beta,\eta$ for the slow wind, fast wind and fountain flow cases (denoted respectively by blue, red and greed solid lines) are different from the initial profiles (denoted by dotted lines) by orders of magnitude. In particular, for the slow and fast winds, the halo $\alpha$ decreases over time while the fountain flow case seems to have accumulated CR pressure. The halo $\beta$ and $\eta$ can decrease by orders of magnitude due to the substantial decrease in halo gas density (thus increasing $v_A$) and pressure.

\subsection{Energetics and dynamics of the nonlinear steady-state} \label{subsubsec:energetics_dynamics}


- \textit{Slow and fast wind case}. To understand the energetics and dynamics of the nonlinear steady-state, we plot in fig.\ref{fig:flow_properties} the time averaged projection plots of the pressures, energies and pressure gradient terms corresponding to the three displayed outcomes. For energetics, the gas energy equation (eqn.\ref{eqn:energy}) in time-steady state can be expressed as
\begin{equation}
    \frac{\gamma_g}{\gamma_g - 1}\nabla\cdot\qty(P_g \vb{v}) = \vb{v}\cdot\nabla P_g + \abs{v_A\cdot\nabla P_c} - \rho^2\Lambda + \mathcal{H}(x), \label{eqn:pg_steady_state}
\end{equation}
i.e. the gas enthalpy flux (LHS) is the sum of gas work done, CR heating, radiative cooling and residual feedback heating (RHS). Near the central disk, the density and radiative cooling rate is high, cooling some of the gas to the temperature floor. The drop in density away from the disk causes an abrupt change in the energetics. CR heating is a much weaker function of density than radiative cooling. In a streaming dominated flow with $v_x\ll v_A$ and $B$-field is constant (as in our setup), CR heating $v_A\nabla P_c\propto\rho^{1/6}$ (since $v_A \propto \rho^{-1/2}, P_c \propto \rho^{2/3}$ for constant B-fields) while cooling $\rho^2\Lambda\propto\rho^2$. Thus we can reasonably expect $v_A\nabla P_c\gg\rho^2\Lambda$ at the halo outskirts in the nonlinear steady state (thus, for the wind cases, the residual feedback heating $\mathcal{H}(x)=0$; however, it can be non-zero in the fountain flow case we later discuss). It is clear from the figure that this is indeed the case, at least for the slow and fast wind case (compare the blue and orange curves in fig.\ref{fig:flow_properties}, central column, top and middle row). Thus the halo gas is overheated. At the transition region where CR heating starts to dominate over cooling ($x\approx 0.2 H$), the velocity is low and the gas is heated to high temperatures (see the abrupt rise in temperature there, fig.\ref{fig:three_outcomes}, bottom right panel). Further out, when gas acceleration is greater, energy balance is maintained by an enthalpy flux commensurate with the overheating rate. 

We can gain intuition by noting that the steady-state gas energy equation (equation \ref{eqn:pg_steady_state}) can be rewritten as: 
\begin{equation} 
v_{x} \nabla \, {\rm ln K} = \frac{\gamma_g - 1}{t_{\rm netheat}} \Rightarrow v_x \nabla {\rm K} = \frac{(\gamma_g - 1)\rm K}{t_{\rm netheat}}
\end{equation} 
where $K \equiv P_g/\rho^{5/3}$, 
and $t_{\rm netheat} = P_g/(|v_{\rm A} \nabla P_c| - \rho^{2} \Lambda) = 1/(1/t_\mathrm{heat} - 1/t_\mathrm{cool})$ ($\approx t_\mathrm{heat}$ if CR heating dominates). This form implies that any increase in gas entropy due to heating is balanced by outward advection of entropy. It also implies that the velocity of a thermal wind driven by heating is given by 
$v_x \approx {L_{\rm K}}/{t_{\rm heat}} \approx 3/2 {L_{\rm \rho}}/{t_{\rm heat}}$
where $L_{\rm K} \equiv K/\nabla K$, i.e. $v_{\rm x} \propto t_{\rm heat}^{-1}$. Alternatively, note that 
the enthalpy flux $\gamma_g\nabla P_g v_x/(\gamma_g - 1)$ consists of two terms, first due to adiabatic expansion $\propto P_g\nabla v_x$ and the second due to work done on the gas by the flow $\propto v_x\nabla P_g$). From fig.\ref{fig:three_outcomes} (central column, top and middle row) it is apparent the enthalpy flux term is dominant over the work done term for at least a scale height above the disk (compare the green and red curves), thus the energetics there are controlled by a simple balance between CR heating and adiabatic expansion, i.e. $P_g\nabla v_x \sim -v_A\nabla P_c$. At the escape velocity $v_{\rm esc} \sim \sqrt{2 g x}$, we have $\nabla v_{\rm esc} \sim (g/x)^{1/2} \sim t_{\rm ff}^{-1}/\sqrt{2}$. This suggests: 
\begin{equation} 
v_{\rm x} \sim \zeta v_{\rm esc} \left( \frac{t_{\rm ff}}{t_{\rm heat}} \right)^{-1}
\label{eq:vx_theat_tff} 
\end{equation} 
where $\zeta \sim 0.4$ is a fudge factor which we later calibrate numerically. We investigate this scaling further in \S\ref{subsubsec:cr_heating}.
In short, the energetics of the slow and fast wind case can be described by a cool inner disk region followed abruptly by an overheated outskirt, driving a sharp rise in temperature and then a balance between CR heating and adiabatic expansion, which generates the required enthalpy flux carrying the heated gas parcels away.

For dynamics, we refer to the left column of fig.\ref{fig:flow_properties}. Although magnetic pressure dominates the the slow and fast wind case, it is unimportant in the overall dynamics of the flow due to its constancy, except for setting the Alfven/streaming speed $v_A$ (hence the CR heating rate) and collimating flows via the high magnetic tension. Since we have set $\alpha_0 = P_{c0}/P_{g0} = 1$ at the base for the displayed cases (see \S\ref{subsec:setup}), CR and gas pressures are comparable at the disk. However, CR pressure varies with density differently than the gas pressure, so they develop different profiles at the disk-halo interface, leading to different outskirt pressures. In particular, for streaming dominated flows where $v_x\ll v_A$, $P_c\propto\rho^{\gamma_c/2}$. This implies a precipitous decline in CR pressure at the disk halo interface, where there is a steep density gradient to offset the sharp change in temperature (see bottom panels of fig.\ref{fig:three_outcomes}). By contrast, the gas pressure suffers a much smaller decline in the disk, where the rise in temperature at the disk halo interface compensates for the reduced density. Thus, for streaming dominated flows, $P_c \ll P_g$ in the halo, resulting in a slow wind. 


If, instead diffusion dominates out to at least the disk halo interface, i.e. $\kappa \nabla P_c \sim \kappa P_c/a > v_A P_c \Rightarrow \kappa > v_A a$, then for $F_c \sim \kappa \nabla P_c \sim$ const (i.e., consistent with these assumptions, streaming losses $v_A \cdot \nabla P_c$ are negligible), CR suffer a linear rather than exponential decline with distance: 
\begin{equation}
    P_c \approx P_{c0} - \frac{\qty(\gamma_c - 1) F_{c0}}{\kappa} x. \label{eqn:pc_diff_dominated}
\end{equation}
Diffusion decouples CR pressure from the gas at the steep density drop, avoiding the heavy `tax' at the disk-halo interface. Since $P_c$ is higher in the halo, this allows for stronger heating at the the lower densities when radiative cooling is weak. The smaller drop in CR pressure also means that the CR pressure gradient $\nabla P_c \gg \rho g$ dominates in the more diffusive, fast wind case, while the gas pressure gradient $\nabla P_g \sim \rho g$ dominates in the streaming dominated, slow wind case (right column, Fig \ref{fig:flow_properties}). 

As the gas drops in density, heating starts to exceed cooling, and gas is abruptly heated to high temperature. The heating of the gas halts the rapid decline in gas pressure in the disk; the hot gas now has a much larger scale height $H_\mathrm{gas} \propto T$. The phase transition from cool to hot gas takes place in a very thin layer. To a rough approximation, it takes place isobarically, so that $\rho_h/\rho_c \sim T_c/T_h \ll 1$. 
Due to the low gas densities in the halo, CR transport becomes streaming dominated ($v_{\rm A} \propto \rho^{-1/2}$), with $P_c \propto \rho^{2/3}$ tracking the very gentle density decline in the halo. This change in transport is responsible for the sharp change in $\nabla P_c$ gradient at the disk-halo boundary. Note that rapid evolution in gas and CR properties typically occurs only at the disk halo interface, where gas is being heated and accelerated by CRs; fluid gradients are much gentler in the halo, where gravitational stratification is much weaker for the hot gas. The evaporative flow at the disk halo interface gives rise to a single-phase hot wind in the halo, whose velocity is given by equation \ref{eq:vx_theat_tff}.  

In summary, the slow and fast wind cases are driven by CR heating, which causes the cold gas to evaporate at the disk-halo interface, boosting the gas pressure and driving an enthalpy flux out. They differ in strength because CR heating is weaker in one case and stronger in the other. The intensity of CR heating at the halo depends on the supply of CR at the base (adjusted through $\alpha_0$ in our sims) and their transport. In particularly, for a given supply of CRs, streaming dominated flows generally lead to sharp decrease in CR energy at the interface, whereas higher diffusivity helps CRs to leak out. As we will see next, fountain flows are instead driven by CR pressure forces.

- \textit{fountain flow case}. The fountain flow case is characterized by cold, dense gas being flung out of the disk. As we shall see, this is wholly due to CR forces, rather than CR heating. When the Alfven speed $v_{\rm A}$ is small due to weak magnetic fields, the momentum input of CRs, $\nabla P_c$, is much more important than the heat input $v_{\rm A} \cdot \nabla P_c$. Due to the high density of the gas, radiative cooling is strong and the gas remains at the temperature floor. Bounded by gravity, there is a maximum height this cold gas can reach (around a scale height $H$ in the case shown in fig.\ref{fig:flow_properties}), beyond which the gas is low in density and warm. CR heating regains dominance and the system transitions into a slow wind, with CR heating balanced by the enthalpy flux.

In terms of pressure, the disk region is well supported by both the gas and CRs, but the flow becomes vastly CR dominated in the halo. The high gas density requires a high level of pressure support, most of which are provided by the CRs. 

Given the turbulent dynamics of the fountain flows, it may be more instructive to look at particular snapshots of the flow, so we refer the reader back to the slice plots shown in fig.\ref{fig:flow_properties}. There is a clear distinction in how pressure is partitioned between the gas and CR components for the cold, fountain gas and the surrounding warm gas. 
The cold gas is heavily CR dominated, whereas gas pressure is comparable to CR pressure in warm/hot gas. Outside the cold gas, gas pressure rises and CR pressure drops. Fig.\ref{fig:alpha_heat_cool_fountain} shows this more clearly: the cold gas is distinctively higher in $\alpha=P_c/P_g$ than the surrounding gas. The cold gas is also radiative cooling dominated. Thus, in contrast to the slow and fast wind case where the outflow is driven by CR heating, the cold, fountain flows here are driven by CR pressure. In particular, the presence of high levels of CR pressure extending from the disk at the cold gas indicates they are peeled off from the surface of the disk.

To supplement our discussion on the energetics and dynamics of the three flow outcomes, we plot, in fig.\ref{fig:vx_temp}, the $v_x-T$ phase plots for the three cases. In line with our expectations and observations above, faster outflow gas is generally higher in temperature. Cold gas with temperature $\lesssim 0.3 T_{0,init}$, if present, is slower and roughly equally distributed between outflows and inflows - this is particularly the case for the fountain flow, in which the cold gas is gravitationally bound and continuously circulating. 

\subsection{Effect of CR heating} \label{subsubsec:cr_heating}

\begin{figure*}
    \centering
    \includegraphics[width=\textwidth]{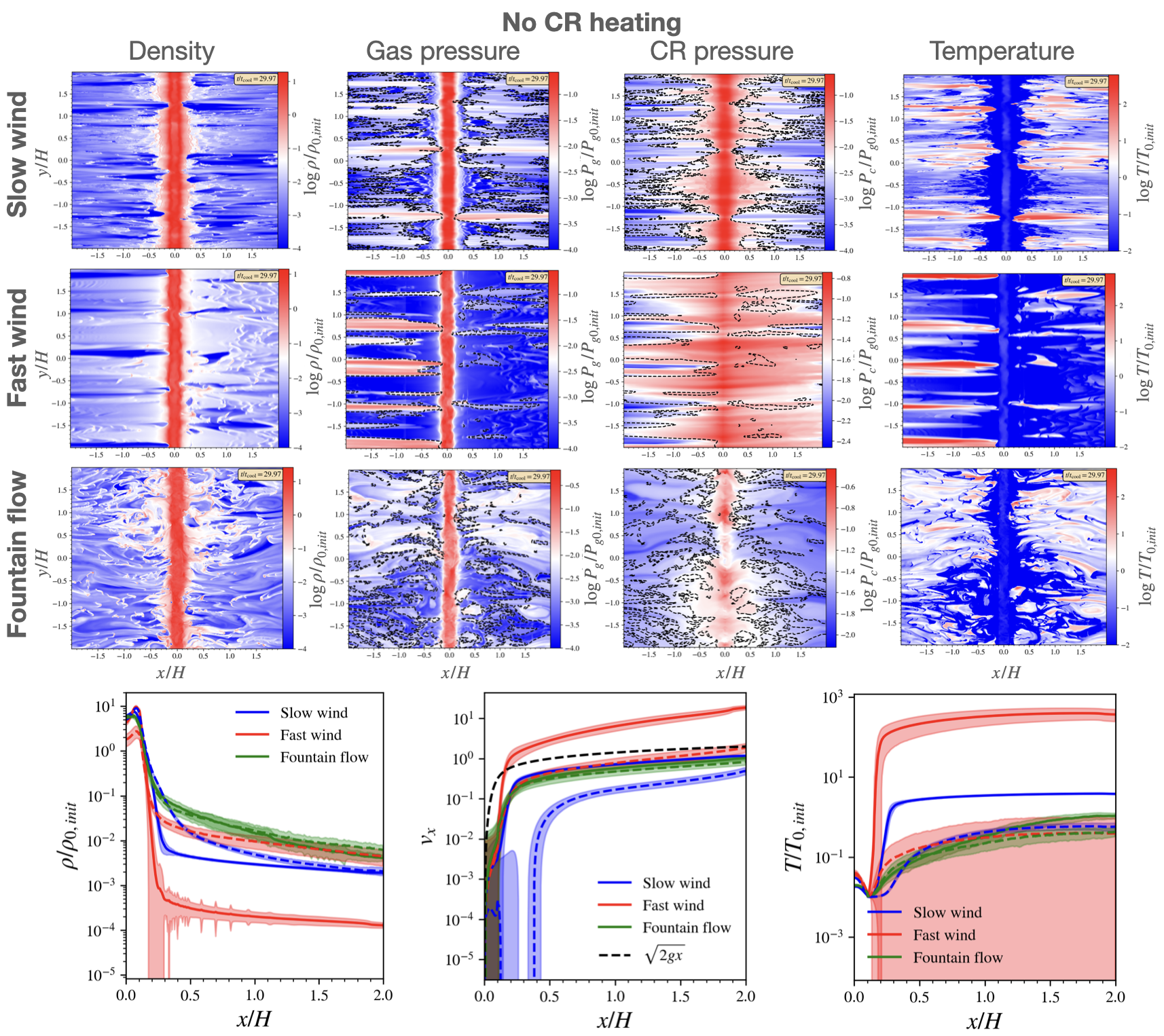}
    \caption{Nonlinear outcomes of TI with CR but without CR heating. Plots are the same as fig.\ref{fig:three_outcomes}, except the dashed lines at the bottom are time averaged projection quantities of runs without CR heating.}
    \label{fig:three_outcomes_nocrh}
\end{figure*}

\begin{figure*}
    \centering
    \includegraphics[width=\textwidth]{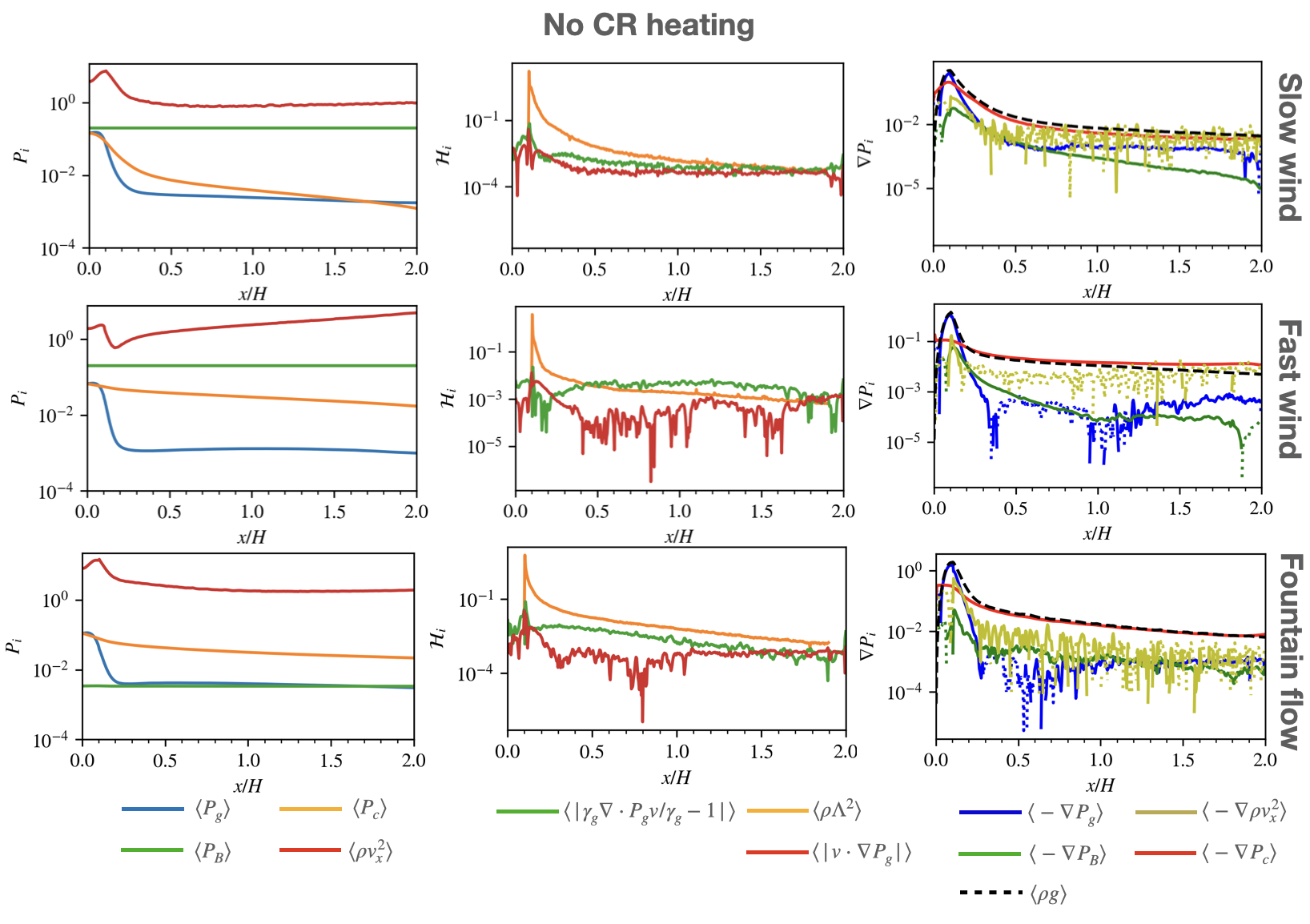}
    \caption{Same as fig.\ref{fig:flow_properties} but without CR heating.}
    \label{fig:flow_properties_nocrh}
\end{figure*}

\begin{figure*}
    \centering
    \includegraphics[width=\textwidth]{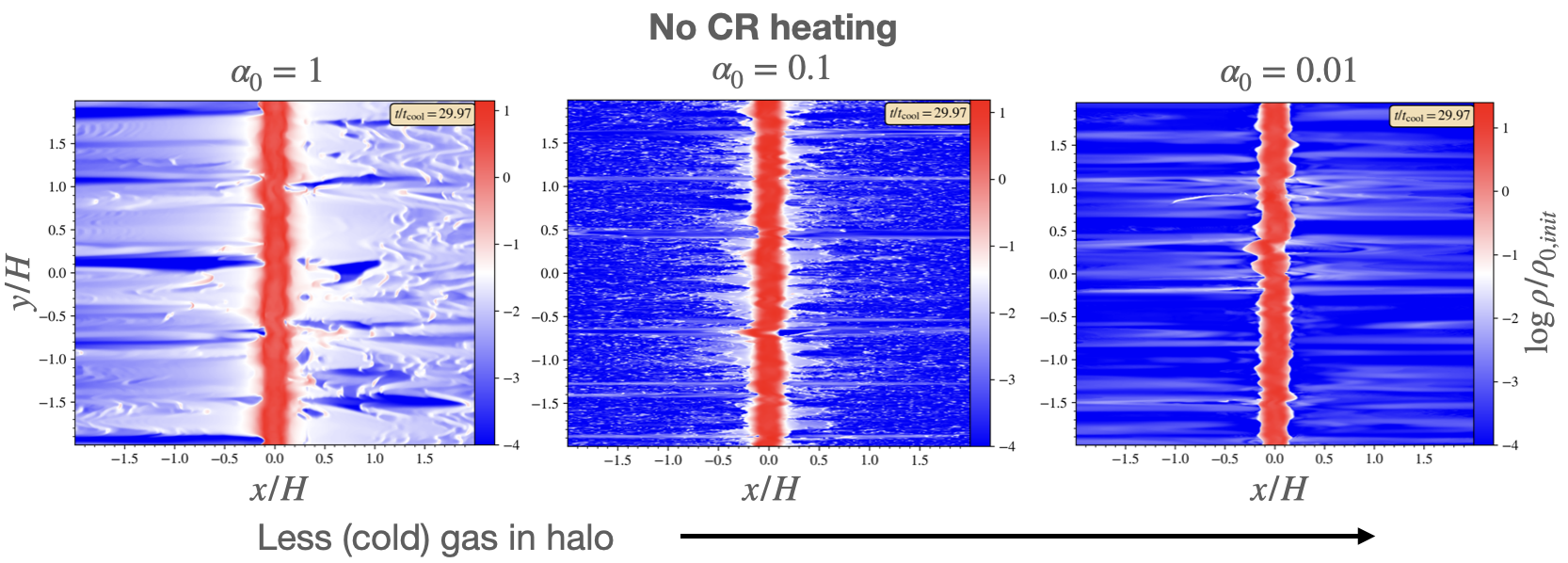}
    \caption{Density snapshots at $t=30 t_\mathrm{cool}$ for cases with different base CR pressure (as measured by $\alpha_0=P_{c0}/P_{g0}$, which in our simulation setup, is fixed) but the same starting $\beta_0$ and $\eta_H$. $t_\mathrm{cool}$ refers to the initial cooling time at $x=H$ Less cold, dense gas appears in the halo as the base CR pressure support decreases. The identifiers for the three cases displayed are (from left to right) `a1b5k1d1in.67res1024c200-nocrh', `a.1b5k1d1in.67res1024c200-nocrh' and `a.01b5k1d1in.67res1024c200-nocrh'}
    \label{fig:cold_gas_cr_pressure}
\end{figure*}

\begin{figure*}
    \centering
    \includegraphics[width=\textwidth]{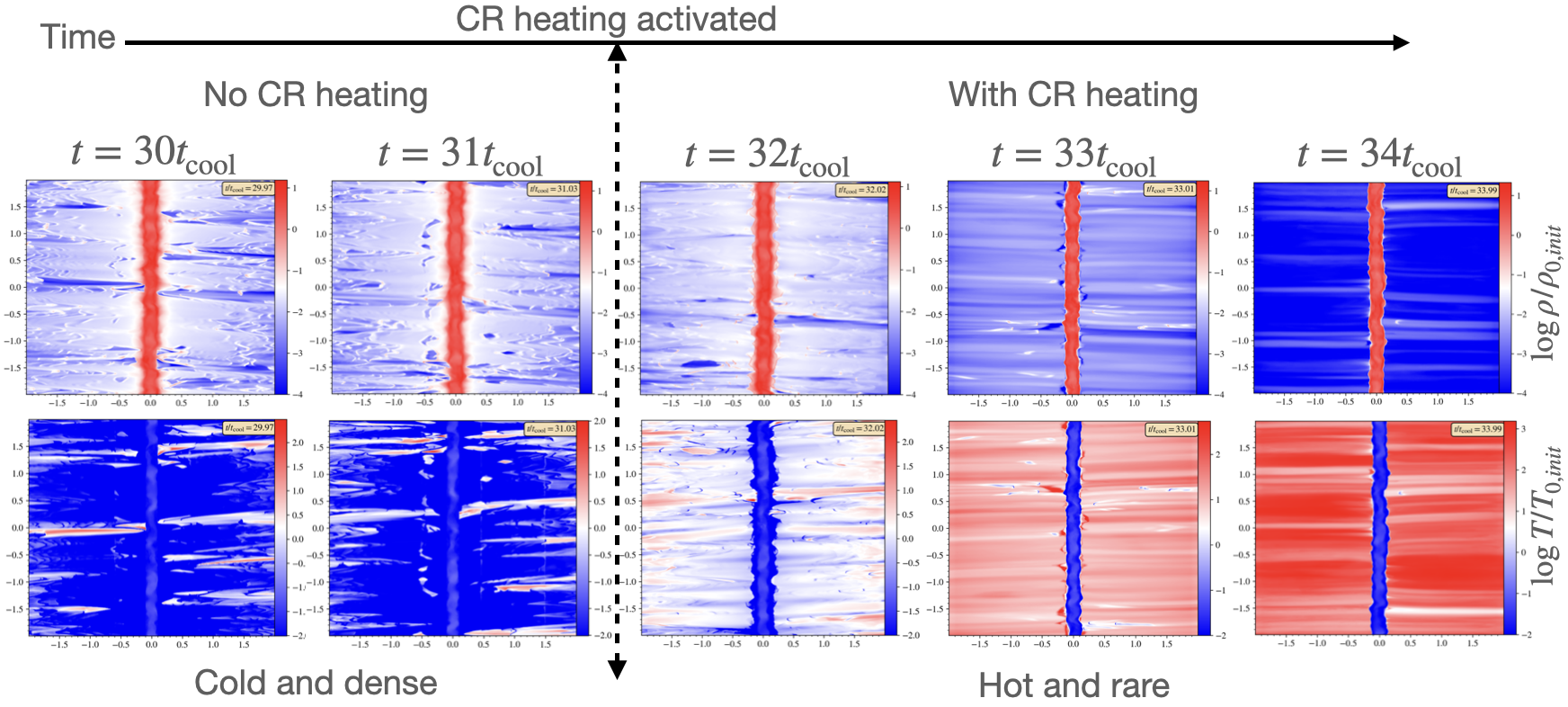}
    \caption{Transition from a cold, dense flow to a hot, fast wind case upon activation of CR heating at $t\approx 32 t_\mathrm{cool}$. $t_\mathrm{cool}$ refers to the initial cooling time at $x=H$. The case shown is a continuation of the middle row of fig.\ref{fig:three_outcomes_nocrh}, with identifier `a1b5k1d1in.67res1024c200-nocrh'.}
    \label{fig:nocrh_crh_transition}
\end{figure*}

\begin{figure}
    \centering
    \includegraphics[width=0.42\textwidth]{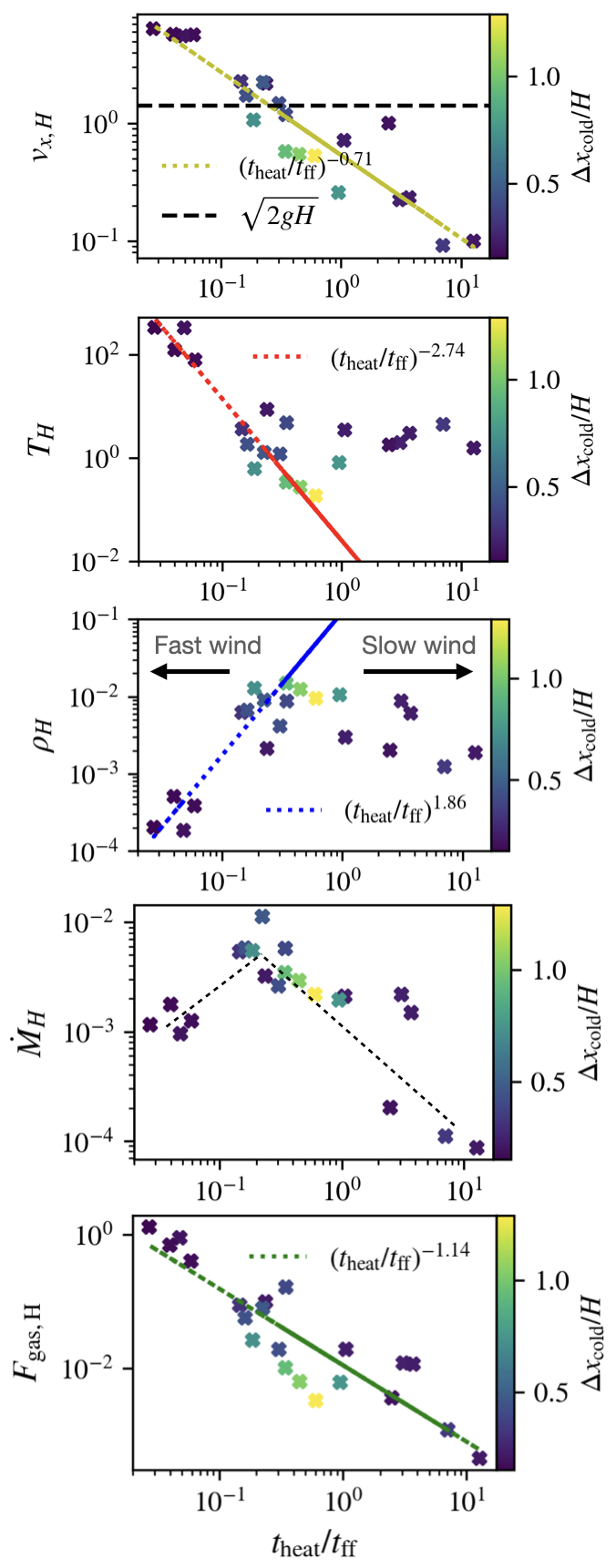} 
    \caption{
    Plot of (from top to bottom) outflow velocity $v_x$, temperature $T$, density $\rho$, mass flux $\rho v_x$ and gas energy flux $F_\mathrm{gas} = \rho v^3_x/2 + \gamma_g P_g v_x/(\gamma_g - 1)$ at a scaleheight $x=H$ against $t_\mathrm{heat}/t_\mathrm{ff}$ (also taken at a scaleheight $x=H$) for all the cases listed in table \ref{tab:cases} under \S\ref{subsec:nonlinear_outcome}. Note that $v_x, T, \rho, \rho v_x, F_\mathrm{gas}$ and $t_\mathrm{heat}/t_\mathrm{ff}$ are taken from their time averaged projection profiles. Time average projection refers to averaging from $t=31.8 t_\mathrm{cool}$ to $63.6 t_\mathrm{cool}$, when the flows have settled onto their nonlinear steady-states, and then spatial averaging over $y$. $t_\mathrm{cool}$ refers to the initial cooling time at $x=H$. The scatter points are color-coded by their cold width $\Delta x_\mathrm{cold}/H$, which is a measure of the extend of fountain flows. $\Delta x_\mathrm{cold}$ is defined in fig.\ref{fig:width_mach}. The regimes for slow and fast wind are indicated by arrows in the density plot. The dashed lines in the mass flux plot are for indicating the general trend above and below $t_\mathrm{heat}/t_\mathrm{ff}\sim 0.4$.}
    \label{fig:t_heat_t_ff_plots}
\end{figure}

In \S\ref{subsubsec:energetics_dynamics} we claimed that the slow and fast wind cases are driven by CR heating while the dynamics of the cold, fountain flow is driven by CR pressure. We demonstrate these claims further by re-running these fiducial cases, but removing CR heating to the thermal gas\footnote{In these runs, we remove the $v_A\cdot\nabla P_c$ term from the gas energy equation, yet keeping collisionless losses $v_A\cdot\nabla P_c$ in the CR energy equation.}. The results are shown in fig.\ref{fig:three_outcomes_nocrh} and \ref{fig:flow_properties_nocrh}. We can see that switching off CR heating has a significant effect on the slow and fast wind cases, but changes the fountain flow case minimally. In particular, the slice plots in fig.\ref{fig:three_outcomes_nocrh} for the slow and fast wind cases show that the density is much higher and cold gas is more prevalent. The difference 
is greatest for the `fast wind' case, where removing CR heating results in an increase in halo density by 2 orders of magnitude, a decrease of outflow speed to sub-escape speeds, and a drastic decrease in halo temperature by 3 orders of magnitude, according to the time averaged projection plots in Fig.\ref{fig:three_outcomes_nocrh}. The slow wind case also sees an increase in halo density and decrease in outflow speed and temperature, but the magnitude of the changes are considerably smaller. This reflects the importance of CR heating in driving the outflow dynamics as shown in Fig.\ref{fig:three_outcomes}. The fountain flow case continues to display fountain flow features even without CR heating, with hardly any change to the halo density, outflow velocity and temperature. This further shows that the cold, fountain flows are not a result of CR heating, but of CR forces.

In terms of energetics and dynamics, Fig.\ref{fig:flow_properties_nocrh} shows that in the absence of CR heating, gas pressure support drops, making CR pressure the dominant source of support against gravity in the halo. However, now the much higher gas densities mean that radiative cooling is important throughout the system. Note that excess radiative cooling is balanced by an artificial heating source term (equation \ref{eqn:feedback_heating}), which is not shown in Fig.\ref{fig:flow_properties_nocrh}. 
Looking again at the slice plots in fig.\ref{fig:three_outcomes_nocrh}, the presence of nearly volume-filling quantities of cold gas in the fast wind case (middle row of fig.\ref{fig:three_outcomes_nocrh}) is striking. 
The morphology of this cold gas is different from the cool clouds which typically form during thermal instability. Similar to the cold fountain flows seen in the fountain flow case, the cold halo gas here, which also has high levels of CR pressure extending from the disk, is a result of cold dense gas being flung off the disk by CR pressure. If one decreases the CR pressure at the base, e.g. by varying $\alpha_0$, as in fig.\ref{fig:cold_gas_cr_pressure}, the amount of cold gas in the halo decreases. Unlike the fountain flows seen in the fountain flow case though, the cold gas appears to be moving outwards in a monotonic wind instead of continuously recycling. The weak B-fields in the fountain case allow CRs to be alternatively trapped and released by transverse/vertical B-fields, producing outflow/infall, whereas the B-fields remain relatively straight when they are stronger. The slow wind case exhibits less cold gas in the halo. This is because the streaming-dominated CRs sustain stronger losses in the sharp density drop at the disk halo interface. The increased diffusion in the fast wind case allows CRs to leak out of the disk and act on the less dense gas, which is easier to push.

To further demonstrate the role of CR heating, we perform simulations starting without it, letting the flow settle onto a nonlinear steady state as shown in Fig.\ref{fig:three_outcomes_nocrh}, then re-activating CR heating. An example of this is shown in Fig.\ref{fig:nocrh_crh_transition} (which corresponds to a continuation of the middle row case in Fig.\ref{fig:three_outcomes_nocrh}). The cold, dense flow quickly transitions into a hot and low density wind (in just $\approx 2 t_\mathrm{cool}$ for the case shown in fig.\ref{fig:nocrh_crh_transition}). For the case shown, the low $\beta$ and high CR diffusivity generates intense heating at the halo, and results in a quick transition into a fast wind (i.e. similar to the middle row of fig.\ref{fig:three_outcomes}). 
CR heating evaporates initially cool gas leaving the disk, transforming it to a low density wind which is easy to accelerate. From equation \ref{eq:vx_theat_tff}, we see that for the flow to exceed the escape velocity, we require $t_{\rm heat}/t_{\rm ff} < 1$.



In Fig.\ref{fig:t_heat_t_ff_plots} we plot the time averaged outflow velocity, temperature and density at a scale height $x=H$ against $t_\mathrm{heat}/t_\mathrm{ff}$ (also taken at a scaleheight). The plots shows a clear transition around $t_\mathrm{heat}\sim 0.4 t_\mathrm{ff}$ when $v_x \sim v_{\rm esc} = \sqrt{2 g x}$. For $t_\mathrm{heat}/t_{\rm ff} > 0.4$, the density and temperature of the flow is roughly independent of $t_\mathrm{heat}/t_{\rm ff}$, while for $t_\mathrm{heat}/t_{\rm ff} < 0.4$, the temperature/density of the flow increase/decrease continuously as $t_\mathrm{heat}/t_{\rm ff}$ falls. By contrast, the velocity $v_{\rm x} \propto (t_{\rm heat}/t_{\rm ff})^{-0.7}$,  varies continuously with $t_\mathrm{heat}/t_{\rm ff}$, in rough accordance with equation \ref{eq:vx_theat_tff}. Surprisingly, the mass flux peaks at $t_{\rm heat}/t_{\rm ff} \sim 0.4$; it falls as heating becomes stronger. The scatter points are also color-coded by their cold width $\Delta x_\mathrm{cold}$, which measures the extent of fountain flows, and defined in fig.\ref{fig:width_mach}. Fountain flows will be discussed in more detail in the next section, but for now we simply note that while fountain flows are mostly slower, colder, denser, and have a lower gas enthalpy flux, they account for the highest mass flux among our test cases. We shall see that while the cold gas recycles in fountain flows, the warm/hot component moves monotonically outward, and because of its higher density relative to the slow/fast wind cases, it has a higher mass flux.

\subsection{Transition to fountain flows} \label{subsubsec:fountain}

\begin{figure*}
    \centering 
    \includegraphics[width=\textwidth]{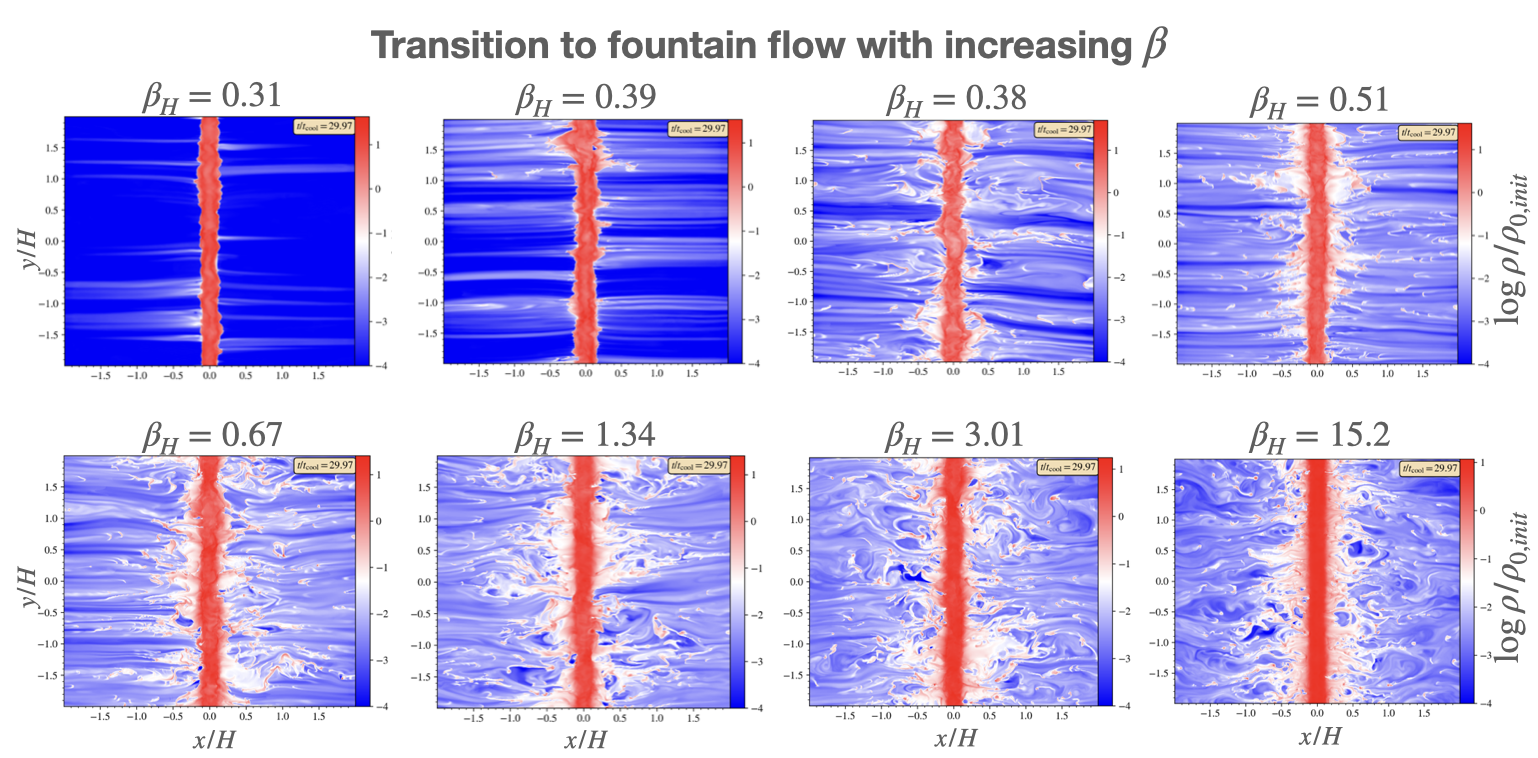}
    \caption{Density slices at $t=30 t_\mathrm{cool}$ for various cases with different $\beta$ (the time average projection of $\beta$ at $x=H$ is listed on top of each panel), showing the transition to fountain flows as the magnetic field weakens. $t_\mathrm{cool}$ refers to the initial cooling time at $x=H$. $\alpha_0=1$ and $\eta_H=1$ for these cases ($\alpha_0$ refers the initial ratio of CR to gas pressure at the base, which in our simulations, is kept fixed. $\eta_H$ is the initial ratio of CR diffusive to streaming flux at $x=H$). The initial plasma $\beta$ (at $x=0$) for these cases are: 5 (top left); 10 (top second left); 30 (top second right); 50 (top right); 100 (bottom left); 300 (bottom second left); 1000 (bottom second right); 1000 (bottom right). The respective case identifiers are: `a1b5k1d1in.67res1024c200' (top left); `a1b10k1d1in.67res1024c200' (top second left); `a1b30k1d1in.67res1024c200' (top second right); `a1b50k1d1in.67res1024c200' (top right); `a1b100k1d1in.67res1024c200' (bottom left); `a1b300k1d1in.67res1024c200' (bottom second left); `a1b1000k1d1in.67res1024c200' (bottom second right); `a1b10000k1d1in.67res1024c200' (bottom right). Note that due to reduction in gas pressure, the simulations ended up with a reduced $\beta$ compared to the initial value.}
    \label{fig:multiple_beta}
\end{figure*}

\begin{figure}
    \centering
    \includegraphics[width=0.48\textwidth]{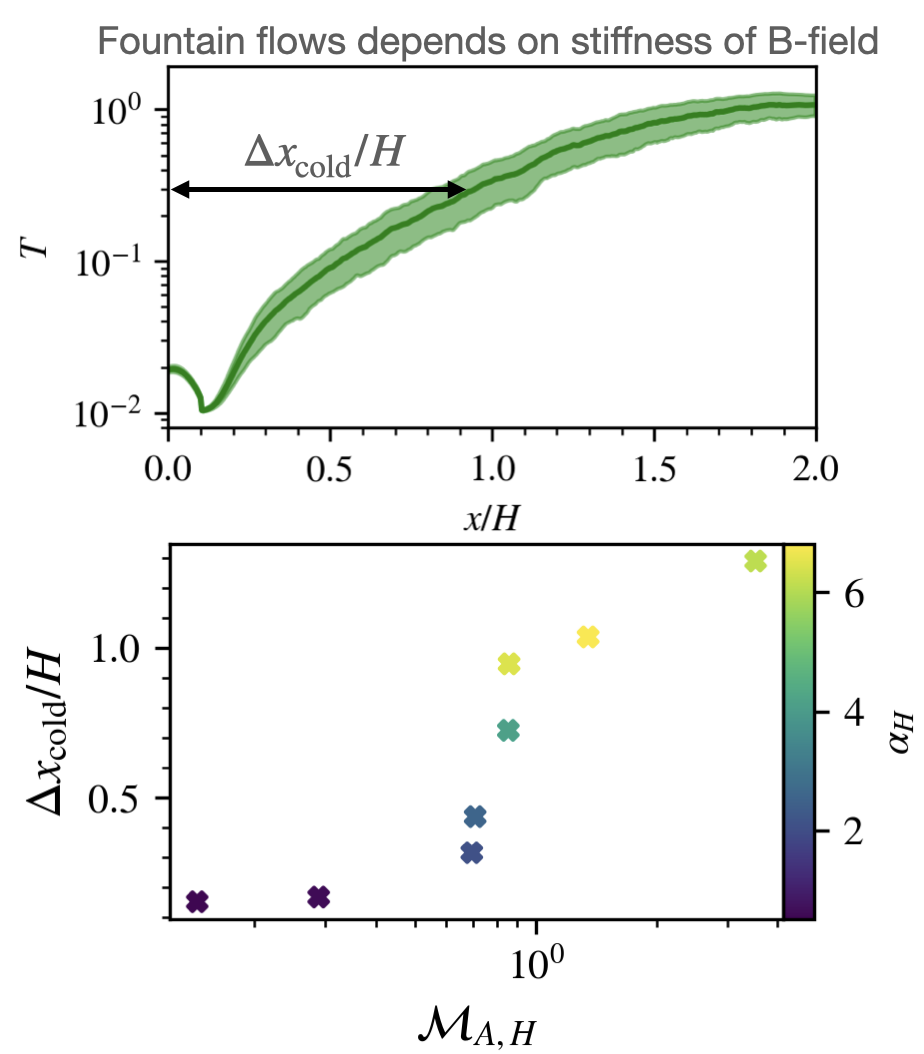}
    \caption{Bottom panel: With of cold mass $\Delta x_\mathrm{cold}$ against the Alfven Mach number $\mathcal{M}_{A,H}=v_H/v_{A,H}$. $v_H$ and $v_{A,H}$ are taken at a scaleheight from the time-averaged projection profiles of $v$ and $v_A$. The ratio of CR to gas pressure at a scaleheight $\alpha_H$ is color-coded into the scatter points. $\alpha_H$ is also taken from the time averaged projection profile of $\alpha$. The top panel is a diagram showing how the width of the cold mass is defined: from the time-averaged projection temperature profile, measure the width from the base for which the temperature is $\leq 0.3 T_{0,init}$.}
    \label{fig:width_mach}
\end{figure}

\begin{figure}
    \centering
    \includegraphics[width=0.48\textwidth]{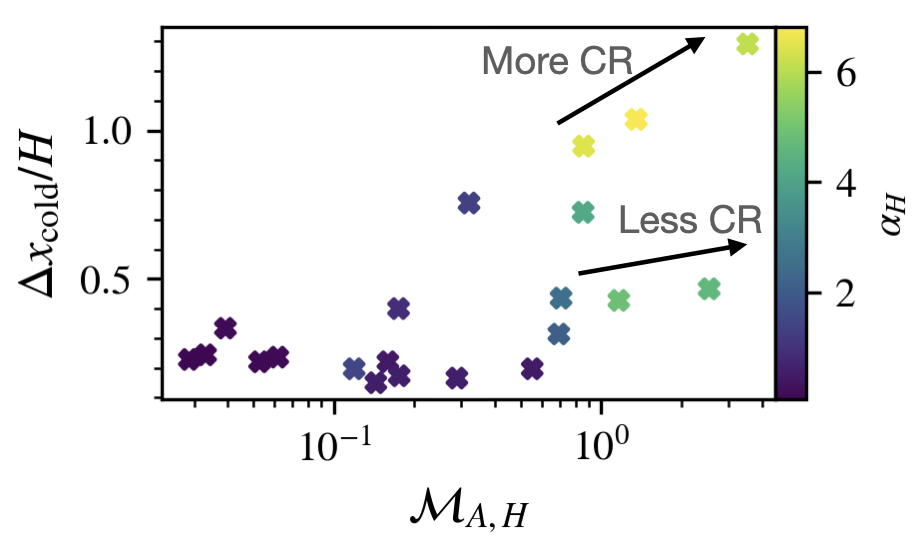}
    \caption{Same as the bottom panel of fig.\ref{fig:width_mach}, including all cases listed under \S\ref{subsec:nonlinear_outcome} in table \ref{tab:cases} with CR heating. The width of the cold region, which is a marker of fountain flows vs winds, decreases as $\alpha_{\rm H}$ is smaller. Both $\mathcal{M}_{\rm A,H} > 1$ and $\alpha_{\rm H} > 1$ are generally required for fountain flows, though the two parameters are correlated -- $\alpha_{\rm H}$ typically jumps once $M_{\rm A} > 1$.}
    \label{fig:width_mach_all}
\end{figure}

In previous sections, we focused a lot on the transition from slow winds to fast winds, and discussed how it relates to $t_\mathrm{heat}/t_\mathrm{ff}$. Here, we also want to understand the criterion for fountain flow. 
We discussed in previous sections that fountain flows are driven by CR pressure as they are a flow feature that do not vanish when CR heating is turned off. When the presence of CRs in the halo is decreased, either when the base supply of CRs is lowered or when the diffusivity is reduced, so too does the extent of fountain flows. 
The strength of the magnetic field affects fountain flows too. In the discussion and figures shown up to this point, fountain flows appear only in high $\beta$ cases. In fact, as we vary $\beta_0$ as shown in fig.\ref{fig:multiple_beta}, we could see a clear transition to a fountain flow as it increases. 

In fig.\ref{fig:width_mach}, we show how the extent of the fountain flow (as measured by the width of the cold mass $\Delta x_\mathrm{cold}$ (defined by the extent in $x$ where $T < 0.3$) depends on the Alfven Mach number $\mathcal{M}_{A}$ (measured at a scaleheight). There is a clear transition at $\mathcal{M}_A\sim 1$, below which there is generally single-phase hot gas, and above which there is cool fountain flow. This is straightforward to understand: CRs do work by direct acceleration at a rate $v \cdot \nabla P_c$, while the CR heating rate is $v_A \cdot \nabla P_c$. Thus, cool momentum driven winds arise when $\mathcal{M}_A > 1$, and hot thermally driven winds arise when $\mathcal{M}_A < 1$. 

Consistent with fig.\ref{fig:alpha_heat_cool_fountain}, the fountain cold gas is associated with CR pressure dominance, as indicated by the high ratio of CR to gas pressure $\alpha$. 
At low $\beta$, characterized by $\mathcal{M}_A\ll 1$, the magnetic field is stiff and CRs are transported monotonically outwards, producing winds. As $\beta$ increases, and $\mathcal{M}_A\gtrsim 1$, the magnetic field becomes more flexible and can wrap around cold gas, trapping CRs. The accumulated CRs build up in pressure and loft the cold, dense gas up, creating fountain flows (and significantly more turbulence). The trapping of CRs is a crucial factor in the appearance of fountain flows. In our simulations where the initial field is vertical, this realignment only happens with weak fields, though realistically it could also happen when the galactic B-field is aligned with the disk, i.e. horizontal. 

Although the mean radiative cooling rate in fountain flows is significantly larger than the mean CR heating, this does not mean the flow is exclusively a cool isothermal wind. Instead, strong gas density and CR pressure fluctuations -- seeded by the magnetic `shrink wrap' -- cause the gas to fragment into a multi-phase flow. The dense cold gas, which is gravitationally bound, is confined to low galactic heights, circulating in a fountain whose width increases with ${\mathcal M}_{\rm A}$. At higher galactic heights, the flow becomes more single phase, though some cold gas remains. Unlike the fountain flow cool gas, the hotter, lower density phase moves monotonically outward. Indeed, because the density of this phase is higher than in the hot wind case, the outward mass flux is {\it larger} for fountain flows than for hot, thermally driven winds (fig.\ref{fig:t_heat_t_ff_plots}) 

To further demonstrate the effect of CR pressure on fountain flows, in fig.\ref{fig:width_mach_all} we re-plot the $\Delta x_\mathrm{cold}$ against $\mathcal{M}_A$ graph, including all other cases listed in table \ref{tab:cases} under \S\ref{subsec:nonlinear_outcome} with CR heating. Again, there is no fountain flow for $\mathcal{M}_A \ll 1$. For $\mathcal{M}_A\gtrsim 1$, greater $\alpha_H$ (i.e., greater CR dominance) leads to greater $\Delta x_\mathrm{cold}$. Thus, both super-Alfvenic flows $\mathcal{M}_{\rm A} > 1$ and CR dominance $\alpha_{\rm H} > 1$ are required for fountain flows, although in practice the two parameters are strongly correlated, since $\alpha_{\rm H}$ increases sharply at $\mathcal{M}_{\rm A} > 1$. 

\subsection{Understanding Mass Outflow Rates; 1D Models} \label{subsubsec:unify}

\begin{figure}
    \centering
    \includegraphics[width=0.48\textwidth]{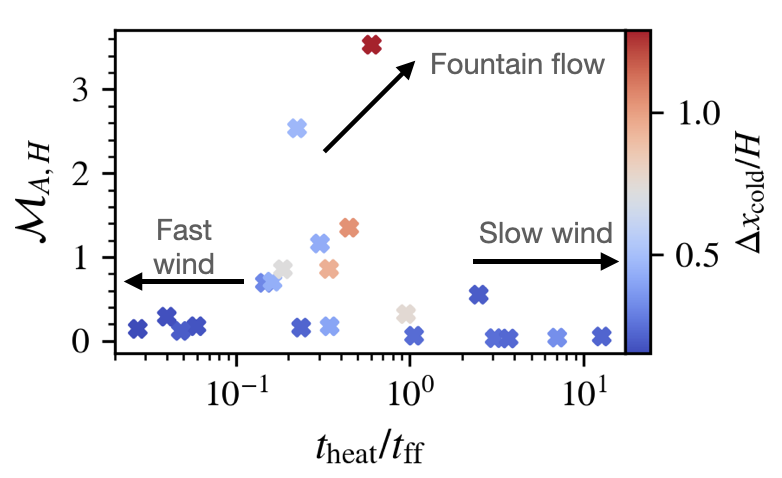}
    \caption{Alfven Mach number $\mathcal{M}_{A,H}$ against $t_\mathrm{heat}/t_\mathrm{ff}$ with $\Delta x_\mathrm{cold}/H$ color-coding. $\mathcal{M}_{A},t_\mathrm{heat}/t_\mathrm{ff}$ and $\Delta x_\mathrm{cold}/H$ are taken at a scaleheight $x=H$ from the time averaged projection profiles of $\mathcal{M}_A, t_\mathrm{heat}/t_\mathrm{ff}, \alpha$. Time average projection refers to averaging from $t=31.8 t_\mathrm{cool}$ to $63.6 t_\mathrm{cool}$, when the flows have settled onto their nonlinear steady-states, and then spatial averaging over $y$. $t_\mathrm{cool}$ refers to the initial cooling time at $x=H$. The region of parameter space typical for each nonlinear TI outcome is indicated by arrows.}
    \label{fig:ma_tratio}
\end{figure}

\begin{figure}
    \centering
    \includegraphics[width=0.48\textwidth]{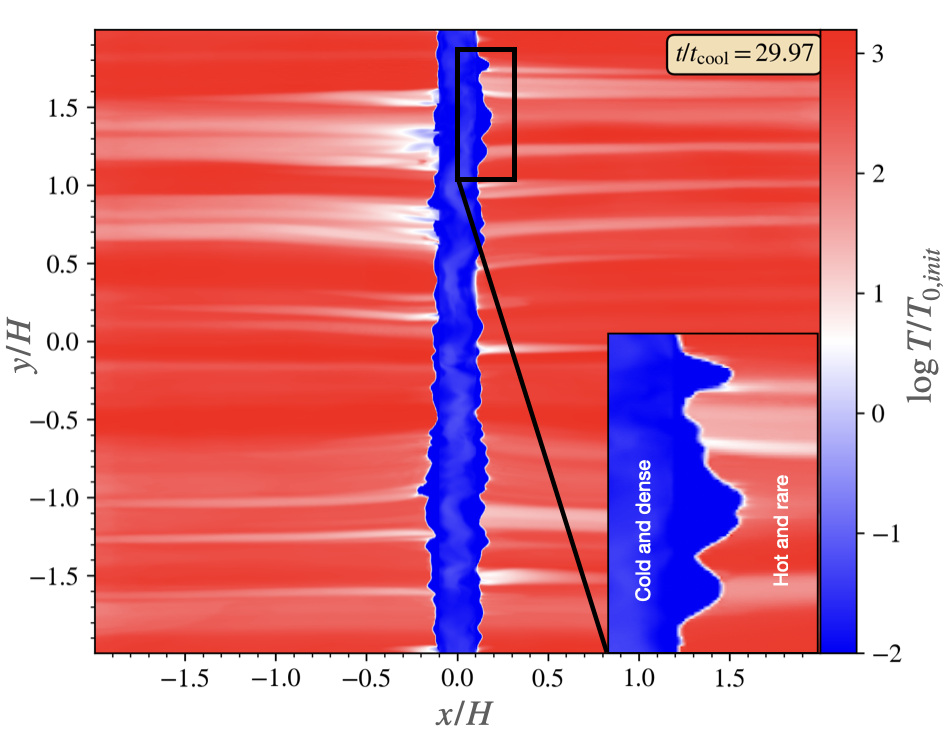}
    \caption{Zoom-in on a part of the interface region of the fast wind case, showing the multiphase structure of the interface region. The color scale shows the temperature, with blue representing cold gas and red representing hot gas.}
    \label{fig:interface}
\end{figure}

\begin{figure}
    \centering
    \includegraphics[width=0.48\textwidth]{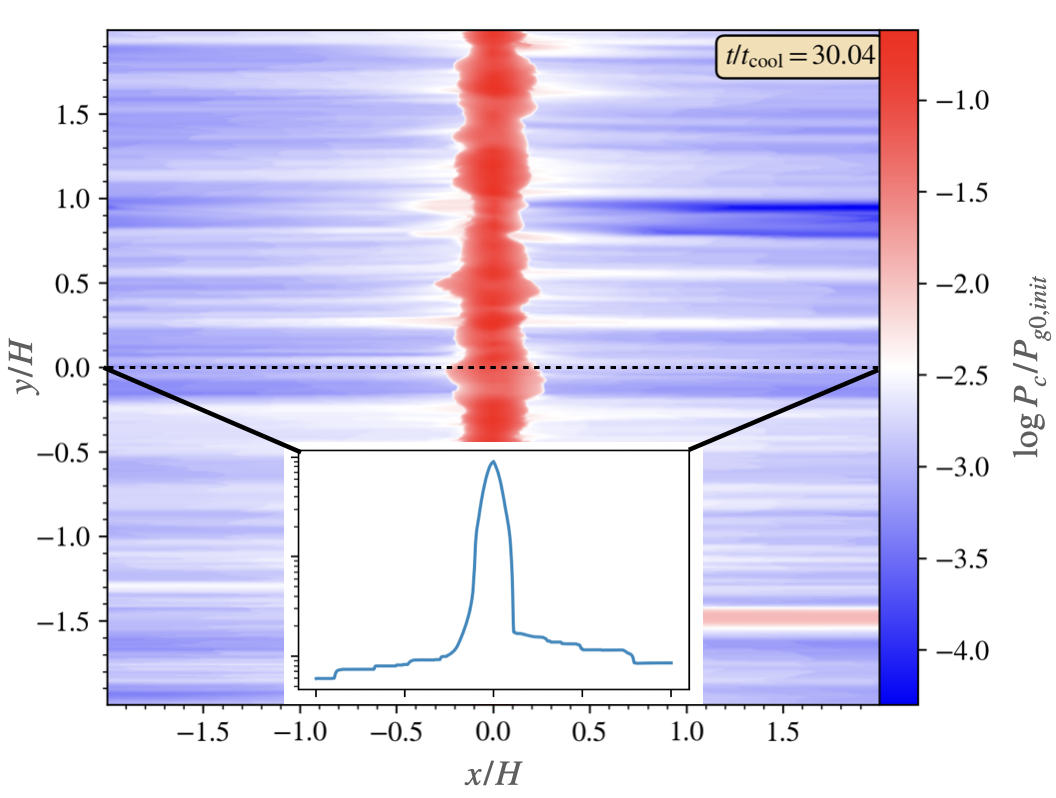}
    \caption{Top: Plot of $P_c$ at $t=30 t_\mathrm{cool}$ for the slow wind case with an inset box indicating the $P_c$ profile through the dotted line. CR staircases can be clearly seen.}
    \label{fig:cr_acoustic}
\end{figure}

\begin{figure}
    \centering
    \includegraphics[width=0.48\textwidth]{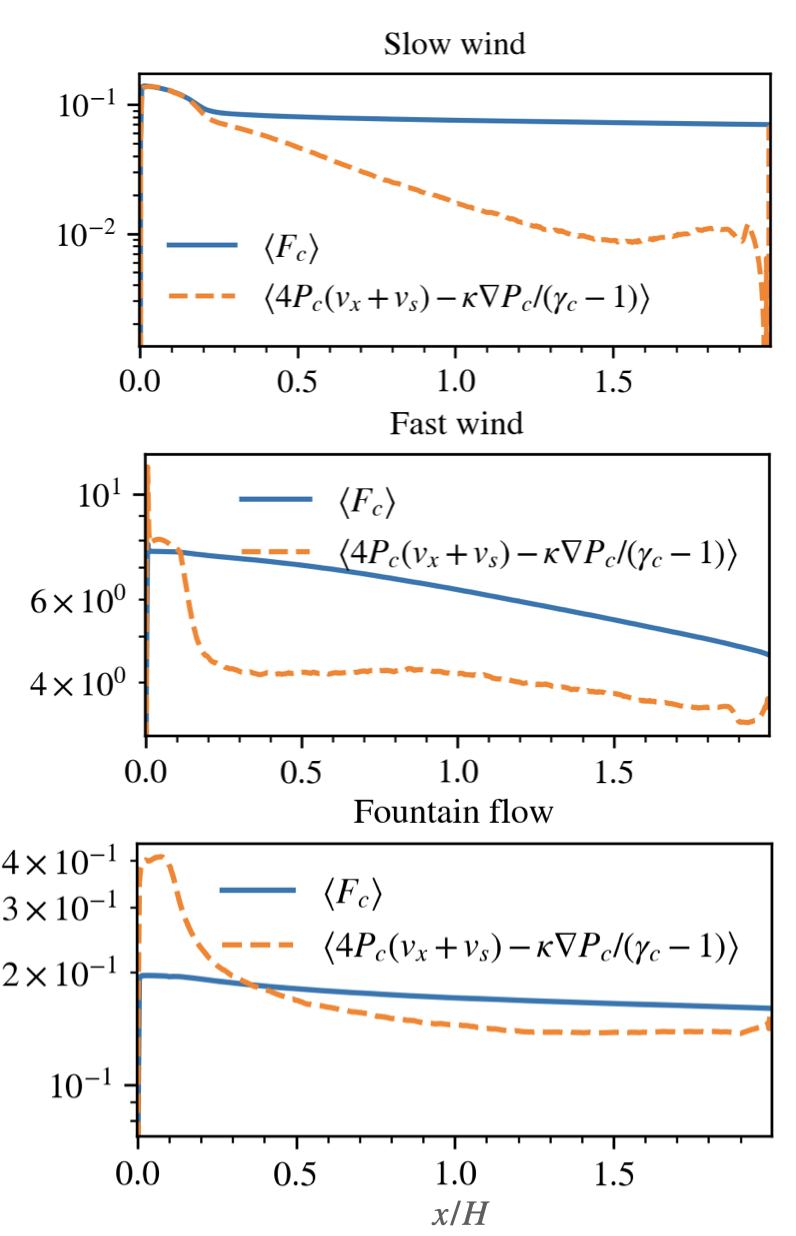}
    \caption{Comparison of the time averaged projection profile of $F_c$ and the steady state form of $F_c$ (eqn.\ref{eqn:cr_flux_steady_state}), showing there is a mismatch between the two. Time averaged projection refers to averaging from $t=31.8 t_\mathrm{cool}$ to $63.6 t_\mathrm{cool}$, when the flows have settled into their nonlinear steady-state, and then spatial averaging over $y$. $t_\mathrm{cool}$ refers to the initial cooling time at $x=H$.}
    \label{fig:flux_compare}
\end{figure}

\begin{figure}
    \centering
    \includegraphics[width=0.48\textwidth]{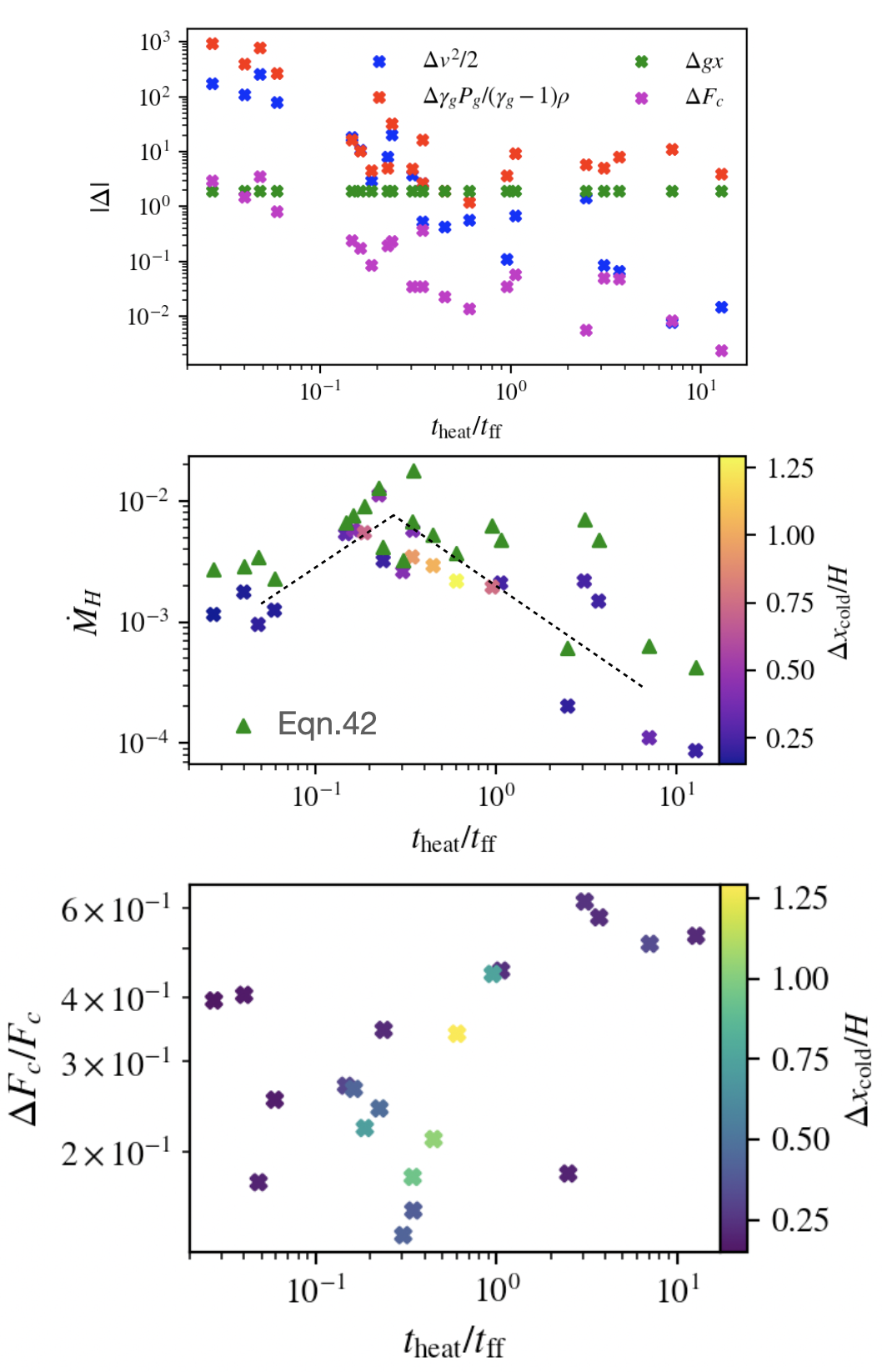} 
    \caption{Top: Breaking up of the different terms in the RHS of eqn.\ref{eqn:mdot_estimate}. Blue - $\Delta v^2/2$; green - $\Delta \phi$, where $\phi$ is estimated as $g x$; red - $\Delta \gamma_g P_g/(\gamma_g - 1)\rho$; magenta - $\Delta F_c$. Note that $\Delta F_c$, which is the numerator of eqn.\ref{eqn:mdot_estimate}, has different units form the rest. $\Delta Q$ of a generic quantity $Q$ is estimated roughly from the difference $Q(x=a) - Q(x=2 H)$ whereas $t_\mathrm{heat}/t_\mathrm{ff}$ is measured at a scaleheight $x=H$ of the time averaged projection profiles. Time average projection refers to averaging from $t=31.8 t_\mathrm{cool}$ to $63.6 t_\mathrm{cool}$, when the flows have settled onto their nonlinear steady-states, and then spatial averaging over $y$. $t_\mathrm{cool}$ refers to the initial cooling time at $x=H$. Middle: Mass flux $\dot{M}_H = \rho_H v_{x,H}$ measured at a scaleheight $x=H$ against $t_\mathrm{heat}/t_\mathrm{ff}$ for all the cases listed in table \ref{tab:cases} under \S\ref{subsec:nonlinear_outcome}. Scatter points marked with `X' and color-coded by $\Delta x_\mathrm{cold}/H$ are the true mass flux taken from the time averaged projection profiles while the green triangles are estimates of $\dot{M}_H$ using eqn.\ref{eqn:mdot_estimate} and the $\Delta$ estimates in the top panel. Bottom: Fractional change in the CR flux, $\Delta F_c/F_c$, against $t_\mathrm{heat}/t_\mathrm{ff}$ with $\Delta x_\mathrm{cold}/H$ color-coding. Note that we calculate the $\Delta$'s by taking the difference between $x=a=0.1 H$ and $x=2 H$ (the outer boundary) rather than at $x=0$ because CR heating is not added to the thermal gas in the buffer region $x=-a$ to $a$ (see \S\ref{subsec:setup})}
    \label{fig:delta_all}
\end{figure}

From our discussion above, the nonlinear outcome of TI with CR heating can be summarized with the aid of fig.\ref{fig:ma_tratio}, which shows the variation of the Alfven Mach number $\mathcal{M}_A$ against $t_\mathrm{heat}/t_\mathrm{ff}$ with $\Delta x_\mathrm{cold}/H$ (width of the cold gas) color-coding. As shown by the figure, cases with $t_\mathrm{heat}/t_\mathrm{ff}\lesssim 0.4$ result in a fast wind, as gas expansion caused by intense CR heating drives a super-escape speed flow. The halo structure is characterized by a hot, rarefied single phase where cold gas is evaporated. As $t_\mathrm{heat}/t_\mathrm{ff}$ increases beyond $\sim 0.4$, the outcome bifurcates to either a slow wind or a fountain flow. If the magnetic field is weak, such that the Alfven Mach number $\mathcal{M}_A>1$ and the easily bent magnetic field `shrink-warps' CRs (such that $\alpha > 1$), multi-phase fountain flows where cold, dense gas is flung out of the disk ensue. Otherwise, a slow wind results.

Ideally, a predictive theory should be able to tell us what the outcome is given input parameters such as $g, \kappa, B$ and boundary conditions like $\rho_0, \alpha_0, F_{c,0}$. A common approach is to solve the steady-state 1D ODEs for mass, momentum and energy conservation (i.e. 1D version of eqn.\ref{eqn:continuity} to \ref{eqn:cr_flux} omitting the time derivatives) using appropriate boundary conditions at the base, to derive the wind solution, similar to what has been done in the past (\citet{mao18_cr_isothermal_wind,quataert22_cr_wind_diffusion,quataert22_cr_wind,modak23}, except (i) the isothermal assumption has to be dropped, as in \citet{modak23}, and (ii) \textit{both} streaming and diffusion has to be incorporated, rather than considering only streaming dominated or diffusion dominated solutions, as in all of the cited works). In principle, one could then estimate what $t_\mathrm{heat}/t_\mathrm{ff}$ and $\mathcal{M}_A$ are, e.g. at a scale height, and determine using fig.\ref{fig:ma_tratio} and the conditions discussed above what the outcome would be. 

However, in practice, 1D models will likely require substantial modification; naive application of time-steady 1D fluid equations do not reproduce higher dimensional simulation results. This is obvious for the fountain flow case, where the multi-phase nature of the flow, and the effects of B-field draping which traps CRs, cannot be trivially reproduced in 1D. Surprisingly, it is also true in the slow and fast wind cases, where the gas appears mostly single phase at a given galactic height $x$, and the B-fields are relatively straight. If we compare the simulation results to the steady-state fluid equations, they do not match. 

The culprit is the disk-halo interface, where cold gas is accelerated and heated. 
In Fig \ref{fig:flux_compare}, we show that the time-averaged simulated CR flux does not assume its steady state form, as given by equation \ref{eqn:cr_flux_steady_state}. There are two reasons for this. Firstly, the interface is multi-phase, with fingers of cold gas protruding into the hot medium; accurately capturing the multi-phase character generally requires 2D or 3D simulations. Secondly, the interface can be unstable to the CR acoustic instability\footnote{The CR acoustic instability, which operates when $\beta \lesssim 0.5$, causes CRs to amplify sound waves, which grow non-linearly into weak shocks. The growth time $t_{\rm grow} \sim \kappa \beta/c_c^2$, where $c_c = (\gamma_c P_c/\rho)^{1/2}$ is the CR sound speed, is short compared to other timescales in our setup. For instance, for the fast wind case (in code-unit, $\rho\sim 2\times 10^{-4}$, $P_c\sim 0.03$, $\kappa = 2.92$, $\beta\sim 0.3$), $t_\mathrm{grow} = 1/\gamma_\mathrm{grow}\sim 4\times 10^{-3} \sim 3\times 10^{-3} t_\mathrm{cool} \sim 10^{-2} L/v_x$, where $t_\mathrm{cool}$ is the initial cooling time, $L$ is the domain size and $v_x$ is the outflow speed of the fast wind.} \citep{begelman94_acoustic_instab,tsung22-staircase}, particularly for low $\beta$ fast winds. As seen in Fig \ref{fig:interface}, the density perturbations due to these effects produce CR staircases due to the bottleneck effect \citep{tsung22-staircase}, which result in alternating regions of flat and intensely dropping $\nabla P_c$, producing haphazard regions of CR coupling to gas variables (\citealt{tsung22-staircase}; $\nabla P_c\neq 0$ is required for CR to couple with the thermal gas). The intermittent coupling causes the steady-state equation \ref{eqn:cr_flux_steady_state}, which assumes continuous coupling, to fail. It may be possible to produce models with effective coupling which can reproduce our simulation results (as has been done for multi-phase turbulent mixing layers, \citealt{tan21_radiative_mixing,tan21-lines}), but it is beyond the scope of this paper.

How can we understand the mass flux $\dot{M}$ of the flows, which is key to describing the strength of an outflow? We note that total energy conservation gives: 
\begin{equation}
\nabla \cdot \left[ \dot{M} \left( \phi + \frac{3}{2}  c_s^2 + \frac{1}{2} v^2 \right) + F_c \right] = -\rho^2 \Lambda(T). \label{eqn:total_energy_conservation}
\end{equation}
In the nonlinear regime, cooling is negligible (fig.\ref{fig:flow_properties}), thus we can estimate the mass flux, assuming $\dot{M}$ is constant, as
\begin{equation} 
\dot{M} \approx -\frac{\Delta F_c}{\Delta \left( \phi + 3 c_s^2/2 + v^2/2 \right)}, \label{eqn:mdot_estimate} 
\end{equation}
where $\Delta F_c$ is the net change in the CR flux, and $\Delta \left( \phi + 3 c_s^2/2 + v^2/2 \right)$ is the total change in the specific energy of the gas, including gravitational, thermal, and kinetic energy components ($\phi$ is approximated by $g x$ for this study). What eqn.\ref{eqn:mdot_estimate} says is that since the total energy flux is conserved, any increase in the gas energy flux comes entirely from CR, through the decrease of $F_c$. 
For a fixed $\Delta F_c$, a larger change in the specific energy requires a lower mass flux to ensure energy conservation. 

In fig \ref{fig:delta_all}, we compare simulation results against equation \ref{eqn:mdot_estimate}. The agreement is generally good, though eqn.\ref{eqn:mdot_estimate} does tend to overestimate $\dot{M}_H$ slightly as $\dot{M}$ generally is not constant near the base. \citet{modak23}, who arrive at similar estimates, assume that $\Delta F_c \sim F_{c0}$ and $\Delta (\phi + v^2/2 + 3c^2_s/) \sim v^2_\mathrm{esc,Herquist}$, where $v_\mathrm{esc,Herquist}$ is the escape speed from the base of a Herquist model gravitational potential, giving rise to an estimated mass flux of $\dot{M}_\mathrm{Modak}\approx F_{c0}/v^2_\mathrm{esc,Herquist}$. In our simulation, we can see that this estimate is justified for $t_\mathrm{heat}/t_\mathrm{ff} > 1$, corresponding to slow wind cases, as $\Delta\phi$ does contribute significantly to the change in the specific energy (top panel of fig.\ref{fig:delta_all}) and $\Delta F_c$ is of order (but not exactly) $F_c$ (bottom panel of fig.\ref{fig:delta_all}). This is indeed the case that \citet{modak23} simulated. As one transitions to the $t_\mathrm{heat} < t_\mathrm{ff}$ regime, however, this estimation is no longer valid, as the change in the specific energy is now dominated by the kinetic and thermal energy terms. Accurate estimation of $\Delta v^2$ and $\Delta T$ are needed. This requires knowledge of the wind solutions, which from our discussion above, is left for future work. Observe also that $\Delta F_c/F_c$ tends to be larger for the slow wind cases ($t_\mathrm{heat}/t_\mathrm{ff} > 1$). One reason for this is that the slow wind cases generally have smaller $\eta$ (CR diffusivity). The CRs are therefore more strongly coupled to the gas, and lose most of their energy. Overall, $\Delta F_c/F_c \sim 0.1-0.6$, i.e. the halo is at best marginally optically thick. Echoing our discussion in \S\ref{subsubsec:cr_heating}, the trends in $\dot{M}_H$ as shown in fig.\ref{fig:t_heat_t_ff_plots} (and in the middle panel fig.\ref{fig:delta_all}) can be explained as follows: For $t_\mathrm{heat}/t_\mathrm{ff} > 1$, the tight coupling between CRs and the thermal gas implies greater CR losses by proportion, with $\Delta F_c\sim 0.6 F_c$. The change in the specific energy is of order $\Delta\phi$, which in our simulations is fixed. Reducing $t_\mathrm{heat}/t_\mathrm{ff}$, for example by increasing the CR supply at the base, increases $\Delta F_c$ and therefore $\dot{M}$. As $t_\mathrm{heat}/t_\mathrm{ff}$ is reduced below $0.4$, the opposite trend occurs. Due to increased $\eta$ (CR diffusivity) for the fast wind cases, CR losses decrease by proportion ($\Delta F_c/F_c$ decreases). 
Furthermore, the gas specific energy is no longer fixed by $\Delta \phi$, but is dominated by the (much larger) kinetic and thermal energy, which leads to a drop in $\dot{M}$. Thus, the maximum $\dot{M}$ occurs at the transition $t_\mathrm{heat}/t_\mathrm{ff}\sim 0.4-1$.

\section{Discussion} \label{sec:discussion}

\subsection{Translating from Code to Physical Units}
\label{sec:discussion-code} 
 
The fluid equations we solve are scale-free, and our results are characterized essentially by dimensionless ratios. The only constraint is that the cooling index we used, $-2/3$, necessarily requires the initially condensing gas to be between $10^{5}-10^{6}\ \mathrm{K}$. With this constraint, our results can be dimensionalized if the reference quantities $\rho_0, T_0, g_0$ in physical units are given (they are all set to 1 in our simulations). 
If we set the reference gravitational acceleration, temperature and density to be $g_0 = 10^{-8}\ \mathrm{cm}\ \mathrm{s}^{-2}$ (as appropriate for the Milky Way disk; \citealt{benjamin97}), $T_0 = 10^{6}\ \mathrm{K}$ and $\rho_0 = 10^{-26}\ \mathrm{g}\ \mathrm{cm}^{-1}$ ($n_0 \sim 10^{-2} {\rm cm^{-3}}$) respectively, the other reference quantities would then scale as: length $H = k_B T_0/m_u g_0 = 2.7\ \mathrm{kpc}$, pressure $P_0 = \rho_0 k_B T_0/m_u = 8.3\times 10^{-13}\ \mathrm{erg}\ \mathrm{cm}^{-3}$, velocity $v_0 = (k_B T_0/m_u)^{1/2} = 91\ \mathrm{km}\ \mathrm{s}^{-1}$, and the flux $F_0 = P_0 v_0 = 7.5\times 10^{-6}\ \mathrm{erg}\ \mathrm{s}^{-1}\ \mathrm{cm}^{-2}$, where $k_B$ and $m_u$ are the Boltzmann constant and the atomic mass unit\footnote{Note that all of these quantities are equal to 1 in code-units. By expressing the ideal gas law as $P_g = \rho T$ in the code, we have absorbed factors of $k_B$ into $T$ and $m_u$ into $\rho$.}. From fig.\ref{fig:three_outcomes} we can see that at $x=H$ the fast wind can acquire velocity $\sim 600\ \mathrm{km}\ \mathrm{s}^{-1}$ (or generally hundreds of $\mathrm{km}\ \mathrm{s}^{-1}$) whereas the slow wind is around $\sim 60\ \mathrm{km}\ \mathrm{s}^{-1}$ (or generally tens of $\mathrm{km}\ \mathrm{s}^{-1}$). The halo density can get to as low as $10^{-30}\ \mathrm{g}\ \mathrm{cm}^{-3}$ ($n_0 \sim 10^{-6} {\rm cm^{-3}}$) and $10^{-28}\ \mathrm{g}\ \mathrm{cm}^{-3}$ ($n_0 \sim 10^{-4} {\rm cm^{-3}}$) for the fast and slow wind respectively while the temperature remains $T \sim 10^{6}\ \mathrm{K}$ for the slow wind but can reach up to $T \sim 10^{8}\ \mathrm{K}$ for the fast wind. Scaling the CR diffusivity by $\kappa_\mathrm{ref} = H v_0 = 7.6\times 10^{28}\ \mathrm{cm}^2\ \mathrm{s}^{-1}$, the slow, fast wind and fountain case diffusion coefficients $\kappa$ are $2.2\times 10^{27}\ \mathrm{cm}^2\ \mathrm{s}^{-1}$, $2.2\times 10^{29}\ \mathrm{cm}^2\ \mathrm{s}^{-1}$ and $2.9\times 10^{28}\ \mathrm{cm}^2\ \mathrm{s}^{-1}$ respectively.

\begin{figure}
    \centering
    \includegraphics[width=0.48\textwidth]{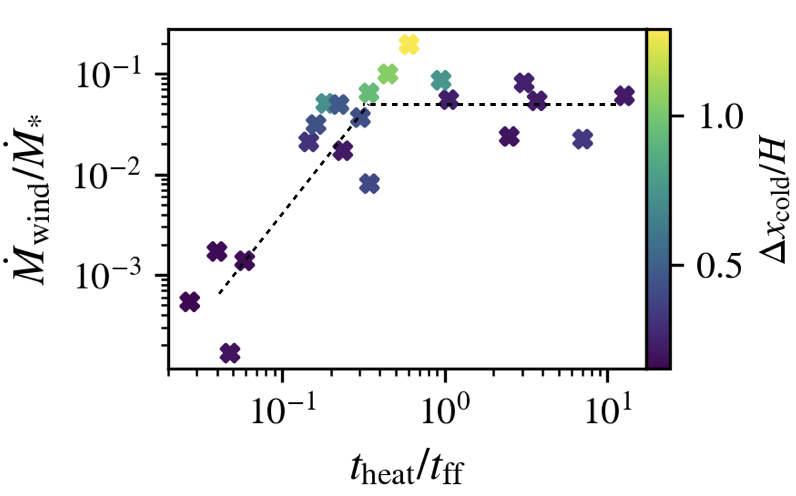} \\
    \includegraphics[width=0.48\textwidth]{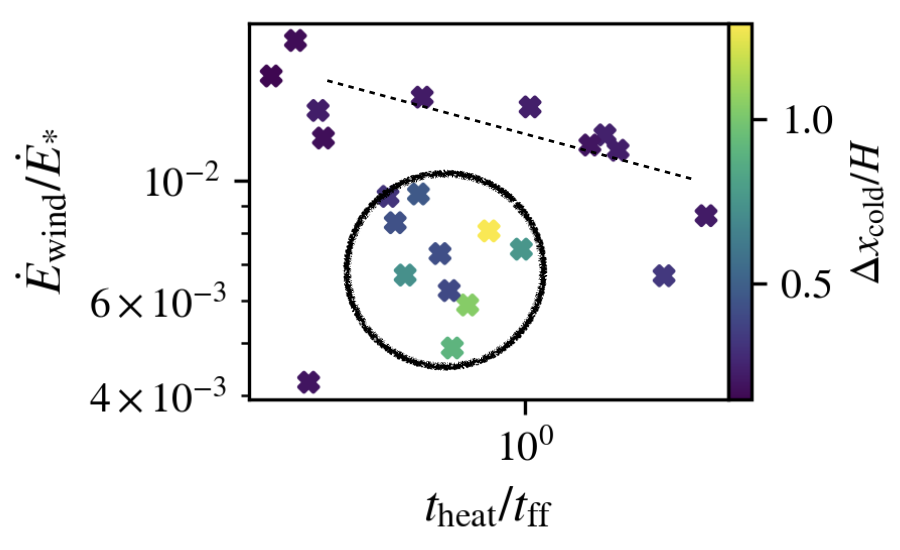}
    \caption{Top: Mass loading factor $\dot{M}_\mathrm{wind}/\dot{M}_*$ against $t_\mathrm{heat}/t_\mathrm{ff}$ with $\Delta x_\mathrm{cold}/H$ color-coding. Bottom: Energy loading factor $\dot{E}_\mathrm{wind}/\dot{E}_*$ against $t_\mathrm{heat}/t_\mathrm{ff}$ with $\Delta x_\mathrm{cold}/H$ color-coding. In both plots, the dotted lines are there to highlight the general trend. In the bottom plot (energy loading), fountain flow cases are circled and the trend for energy loading to increase with stronger heating (indicated by the dashed line) applies only for the wind cases.}
    \label{fig:mass_loading}
\end{figure}

A key physical quantity for winds is the mass loading factor, which expresses the mass outflow rate per unit star formation rate (SFR) ($\dot{M}_\mathrm{wind}/\dot{M}_*$). In physical units, the mass loss rate $\dot{M}_\mathrm{wind}$ can be expressed as $\rho v A$, where $A$ is the cross-sectional area the wind passes through. To get the SFR, we make several simplifying assumptions to connect the CR flux at the base $F_{c0}$ to $\dot{M}_*$. First, assuming all of the CRs originate from supernovae (SN), we can express $F_{c0}\approx\epsilon_\mathrm{SN} E_\mathrm{SN} \dot{N}_\mathrm{SN}/A$, where $\epsilon_\mathrm{SN}\approx 0.1$ is the CR acceleration efficiency by SN, $E_\mathrm{SN}\sim 10^{51}\ \mathrm{erg}$ is the energy released from each SN event, $\dot{N}_\mathrm{SN}$ is the SN rate. If we make the further assumption that a fraction $f_\mathrm{SN}$ of the stars formed becomes SN, i.e. $\dot{N}_\mathrm{SN} = f_\mathrm{SN} (\dot{M}_*/\bar{M})$, where $\bar{M}$ is the mean stellar mass, then $\dot{M}_*\approx F_{c0} A\bar{M}/\epsilon_\mathrm{SN} f_\mathrm{SN} E_\mathrm{SN}$ and the mass loading factor $\dot{M}_\mathrm{wind}/\dot{M}_* = (\rho v/F_{c0})(\epsilon_\mathrm{SN} f_\mathrm{SN} E_\mathrm{SN}/\bar{M})$. Again, crudely estimating $f_\mathrm{SN}$ and $\bar{M}$ using the initial-mass-function $\phi$ (IMF): $f_\mathrm{SN}\approx\int^{20 M_\odot}_{8 M_\odot}\phi\dd{M}/\int^{20 M_\odot}_{0.1 M_\odot}\phi\dd{M}$ and $\bar{M}\approx\int^{20 M_\odot}_{0.1 M_\odot}M\phi\dd{M}/\int^{20 M_\odot}_{0.1 M_\odot}\phi\dd{M}$, we get, if we adopt a power law IMF with index $-2.35$ \citep{salpeter55_imf}, $f_\mathrm{SN} \approx 0.002$, $\bar{M}\approx 0.33 M_\odot$. At the top of fig.\ref{fig:mass_loading} we plot the mass loading factor against $t_\mathrm{heat}/t_\mathrm{ff}$ using these conversion and scaling factors. As discussed in \S\ref{subsubsec:cr_heating}, despite the high outflow velocity of fast winds ($t_\mathrm{heat}/t_\mathrm{ff}\lesssim 0.4$), they are inefficient in carrying mass out. Slow winds ($t_\mathrm{heat}/t_\mathrm{ff}\gtrsim 0.4$) appear more efficient, and the mass loading factor seems to be roughly independent of $t_\mathrm{heat}/t_\mathrm{ff}$. The result for slow winds is in agreement with the literature that shows that the mass loading factor depends generally on the escape velocity, which is held fixed in our study. We note that one should take the numerical values for the mass loading in fig.\ref{fig:mass_loading} with a grain of salt, as it involved some simplifying assumptions and depends quite heavily on the reference values we used to map our code-units to physical units (e.g. if the reference temperature $T_0$ is decreased by a factor of 10 to $10^{5}\ \mathrm{K}$, as one might imagine for less massive galaxies with lower virial temperature, the mass loading would be boosted by a factor of 10). Also, the mass loading calculated here takes into account only the effect of CRs; the total mass loading is likely a culmination of many factors (direct mechanical injection from SNe, radiation, and from multiphase clouds). With just CRs alone, the mass loading factor of fast winds is low compared to that observed (e.g. $\dot{M}_{\rm wind}/\dot{M}_{*} \sim 0.3$ for $M_*\approx 10^{11} M_\odot$ galaxies and $\dot{M}_{\rm wind}/\dot{M}_{*} \sim 3$ for $M_*\approx 10^{9} M_\odot$ galaxies, see \citet{chisholm17_mass_loading}), whereas the slow wind regime mass loading appears higher and more consistent. In any case, the trend shows that fast winds have appreciably lower mass loading than the slow wind. 

Similarly, the energy loading factor $\dot{E}_\mathrm{wind}/\dot{E}_*$, compares thermal and kinetic power in the wind to the rate of energy input from stellar feedback. Here, we only consider CR feedback, and write $\dot{E}_\mathrm{wind} = F_\mathrm{wind} A$ and $F_{\rm c0} = \epsilon_{\rm SN} \dot{E}_{*}/A$ to obtain $\dot{E}_\mathrm{gas}/\dot{E}_* = \epsilon_\mathrm{SN} F_\mathrm{wind}/F_{c0}$. 
We plot the energy loading factor $\dot{E}_\mathrm{gas}/\dot{E}_* = \epsilon_\mathrm{SN} F_\mathrm{wind}/F_{c0}$ at the bottom panel of fig.\ref{fig:mass_loading} (the numerator $F_\mathrm{wind}$, which varies in the flow, is evaluated at $x=H$). We again caution the reader to take the numerical values with a grain of salt, but the trend is clear: the energy loading increases gently from slow to fast winds, while the fountain flow cases (circled in the plot) show distinctively low energy loading. This echos existing studies showing that hot, fast outflows are generally more efficient in energy loading but less so in mass loading, while the opposite is true for colder flows \citep{li-bryan20,fielding22_multiphase_wind}. Given the drastic change in fluid properties of the fast wind cases from the slow wind case, it might seem surprising that the rise in energy loading is so gentle across the two regimes. The energy loading factor is given by
\begin{gather}
    \frac{\dot{E}_\mathrm{wind}}{\dot{E}_*} = \epsilon_\mathrm{SN}\frac{F_\mathrm{gas}}{F_\mathrm{c0}} \approx \epsilon_\mathrm{SN}\frac{\Delta\qty(F_\mathrm{gas} + F_\mathrm{grav})}{F_\mathrm{c0}}\frac{\Delta F_\mathrm{gas}}{\Delta\qty(F_\mathrm{gas} + F_\mathrm{grav})} \nonumber \\
    \quad \approx \epsilon_\mathrm{SN}\frac{\Delta F_c}{F_\mathrm{c0}}\frac{\Delta\qty(v^2/2 + c^2_s/(\gamma_g - 1))}{\Delta\qty(\phi + v^2/2 + c^2_s/(\gamma_g - 1))} \label{eqn:energy_loading},
\end{gather}
where we have approximated $F_\mathrm{gas}$ by $\Delta F_\mathrm{gas}$ in the first step as the gas energy flux near the base is negligible, and used the fact that $\Delta (F_\mathrm{gas} + F_\mathrm{cr} + F_\mathrm{grav}) = 0$ (for negligible radiative cooling) in the second step, where $\Delta F_\mathrm{grav}$ is the work done against gravity. In the slow wind regime, the change in gravitational potential dominates (fig.\ref{fig:delta_all}), so the last term in eqn.\ref{eqn:energy_loading} can be approximated by $\Delta (v^2/2 + c^2_s/(\gamma_g - 1))/\Delta\phi$; it decreases as $t_\mathrm{heat}/t_\mathrm{ff}$ increases. However, this term approaches 1 in the fast wind regime, when the flow is dominated by kinetic and thermal energy. On the other hand, faster winds are usually associated with higher CR diffusivity in our study (since we fix the base CR pressure, greater CR heating in the halo is brought on by either increasing the magnetic field or allowing CRs to diffuse faster out of the disk by increasing $\kappa$), thus the coupling between the thermal gas and CR is generally weaker for fast wind in our study, leading to a smaller $\Delta F_c/F_{c0}$. The overall increase in the energy loading factor across the slow wind regime to the fast wind regime is thus gentle. We note that $t_{\rm heat}/t_{\rm ff}$ can be changed in a number of ways, so the steepness of this scaling may change in other scenarios with different boundary conditions (for instance, fixing $F_{\rm c}$ rather than $P_{\rm c}$ in the central disk). 


One of the most interesting outcomes of this study is the presence of fountain flows when $\mathcal{M}_A > 1$, characterized by circulation of cold gas. The extent of the cold gas, from fig.\ref{fig:width_mach}, can reach up to $H$ (or more). From the right-most panel of fig.\ref{fig:vx_temp}, we can see that cold gas with $T < 0.1 T_{\rm 0,init}$ is distributed roughly equally between outflows and inflows, indicative of a circulation, with velocity of order $\pm 1$ (in code-units). Using the scaling factors as discussed above, $H\sim 2.7\ \mathrm{kpc}$ and the cold gas circulation speed $\sim 90\ \mathrm{km}\ \mathrm{s}^{-1}$. These values are consistent with the observed intermediate-velocity-clouds (IVC, \citealt{marasco22_clouds_ii}), which appear also to be circulating above the galactic disk at a height of $\sim \mathrm{kpc}$ with velocity $\lesssim 90\ \mathrm{km}\ \mathrm{s}^{-1}$. 

\subsection{Comparisons against larger scale simulations}

An obvious question is whether any of the phenomenon we discuss can be seen in existing CR wind simulations, particularly those which are larger scale and less idealized. We have already discussed how the slow heated wind has been studied in 1D by \citet{modak23}. We mention two potential cases where previous simulations are in the right parameter regime for CR fountain flows and heated winds. Of course, a positive identification requires further detailed analysis. We merely show that existing simulation runs often lie in the parameter regimes we describe, and show similar behavior. They can be reanalysed with these considerations in mind.   

A number of FIRE simulations, with CR physics incorporated in a two-moment formalism \citep{chan19_fire,ji20_cr_halo,hopkins21_cr_outflows}, are potentially in the fountain flow regime. Their simulated CGM has high plasma $\beta \gg 1$ and therefore has a low CR heating rate and high Alfven Mach number, as required for fountain flows. Indeed, cold gas which circulates in a fountain is seen \citep{chan22_disk_halo_interface,ji21_virial_shocks}. Their disk-halo interface has significant gas motions near the disk, and is more hydrostatic further out (with gravity balanced primarily by CR pressure), as in our fountain flow picture (see fig.\ref{fig:flow_properties}). They found that CR feedback causes a greater uplift of gas, with gas of all phases (both cold and hot) more abundant above the disk. The hot gas moves further out, while the cold, $T \sim 10^4\ \mathrm{K}$ gas is generally more confined. Our simulation set-ups are of course very different, but this broad-brush agreement is encouraging. 

AREP0 simulations also include CR physics in a two-moment formalism \citep{thomas22_cr_wind}. In contrast to the FIRE simulations, their CR-driven wind is strongly magnetized ($\beta < 1$), with magnetic fields relatively collimated and vertical in the inner parts of the wind, and low in Alfven Mach number. In the innermost CGM, B-field lines are vertical, as in our setup. The authors describe a wind launched by CR pressure at the disk-halo interface. However, their wind also exhibits properties consistent with a CR-heated wind. From their Fig 3, we can see that $t_{\rm heat}/t_{\rm cool} \ll 1$ at the wind launch region at $z\gtrsim 0.6$   kpc, where (from the slice plots in their Fig 1) it is heated to high temperatures and low densities. Thus, a phase transition mediated by CR heating, similar to what we see, seems plausible. The heating of the disk gas is crucial: it reduces its density so that $\nabla P_c \gg \rho g$; without CR heating, the wind is much weaker (\S\ref{subsubsec:cr_heating}).  The inner disk wind ($R_{\rm disk} < 1$ kpc) shows wind velocities comparable to the escape velocity $v \sim 200 \, {\rm km \, s^{-1}}$, and indeed crude estimates\footnote{We estimate $t_{\rm heat} \sim P_g/\Gamma \sim 0.5 \, {\rm Myr}$, from $P_g \sim 10^{-2}\ \mathrm{eV}\ \mathrm{cm}^{-3}$, $\Gamma \sim 10^{-27} \, {\rm erg \, s^{-1} \, cm^{-3}}$, while $t_{\rm ff} \sim r/v_{\rm esc} \sim 3 \, {\rm Myr}$.} from their figures show that $t_{\rm heat}/t_{\rm ff} \sim 0.2$. This is consistent with our `fast wind regime' where $v \sim v_{\rm esc}$ for $t_{\rm heat}/t_{\rm ff} \sim 0.2$ in Fig \ref{fig:t_heat_t_ff_plots}. Similar to our fast wind simulations (Fig \ref{fig:flow_properties}), in their simulations CR pressure gradients dominate in the hot wind above the disk. Regions of weak coupling between the CRs and the thermal gas (which the authors called `dark Alfven regions') are found within the wind, and the transport speed of the CRs is frequently found to lag the Alfven speed. This echos our claim that the CR flux is, for the most part, not given by the steady-state form (eqn.\ref{eqn:cr_flux_steady_state}, and see fig.\ref{fig:flux_compare}). The authors attribute the `dark Alfven regions' to field lines perpendicular to the CR gradient, so that there is no streaming instability. We note that the low plasma $\beta$ of the gas implies that the CR acoustic instability \citep{tsung22-staircase} can potentially also play a role, particularly further out in the CGM. This can also cause regions of flat CR pressure where the CR streaming instability does not develop, and subsequent onset of the bottleneck effect causes a slowing of the CRs transport speed \citep{tsung22-staircase}, which is limited by the Alfven speed at bottlenecks.

\section{Conclusions} \label{sec:conclusions}

In this study we explored the effect of CR heating on TI, both in the linear and nonlinear phase, using a gravitationally stratified setup with vertical magnetic fields and streaming CRs. In the linear phase, we found that in accordance to linear theory \citep{kempski20_thermal_instability}, CR heating can cause gas entropy modes to propagate at some velocity proportional the Alfven velocity up the $P_c$ gradient. The propagation of the modes is a result of differential CR heating on different parts of the gas perturbation, resulting in a net phase velocity. We verified with simulations that the modes propagate at the expected velocity (see \S\ref{subsec:propagation} and fig.\ref{fig:xshow}). This propagation is subdued when we increase CR diffusivity (fig.\ref{fig:xshow_diff}), as CRs  diffuse out of the perturbation before they have time to heat it. 

Mode propagation, under the action of CR heating, in a streaming dominated flow could in principle suppress TI. The idea is that if the time it takes for the modes to propagate across a cooling radius $t_\mathrm{cross}$ is less than the time it takes the modes to grow, which scales as $t_\mathrm{cool}$, the perturbations would not reach nonlinear amplitudes and become cold clouds. 
However, since both the crossing time $t_{\rm cross} \sim L/v_{\rm A}$ and the heating time $ t_{\rm heat} \sim L/v_{\rm A}$ are closely related, there is only a small range of $\theta=t_\mathrm{cool}/t_\mathrm{heat}$ near unity where suppression by propagation effects operates, before the gas becomes over-heated. 

Our most interesting results do not relate to the linear thermal instability, but rather the non-linear outcome of our simulations. There are two important things to note. Firstly, thermal instability causes substantial mass dropout and evolution in CGM properties. A cold disk containing most of the mass forms in the mid-plane, while the density and pressure of extraplanar gas is significantly reduced. Thus, initial values of $\alpha=P_c/P_g$, $\beta = P_g/P_B$ evolve substantially during the non-linear stage\footnote{For instance, in our `fountain flow' simulations, $\beta_{H,i} =300$ becomes $\beta_{\rm H,f} \approx 1$, due to the reduction in gas pressure. See Fig \ref{fig:multiple_beta} for other examples.}. The strong phase transition from cool disk gas to hot atmosphere has important consequences for CR wind properties, different from calculations which assume single-phase isothermal winds. Secondly, the dual role of CRs in pushing ($\nabla P_c$) and heating ($v_A \cdot \nabla P_c$) the gas implies that global thermal and hydrostatic equilibrium is not possible without fine-tuning. Indeed, the gas generally loses force and/or thermal balance at the disk halo interface, leading to the development of winds. The sharp reduction in gas density at the disk halo interface $\rho_{\rm h} \sim \rho_c (T_c/T_h)$ reduces radiative cooling ($\propto \rho^2$) and gravitational forces ($-\rho g$). The resulting loss of dynamical and thermal equilibrium can cause gas to accelerate outwards and heat up. CR winds and fountain flows are most efficient if diffusion is sufficiently strong such that transport is diffusion dominated until the flow reaches low densities (i.e., in the halo). Otherwise, since $P_{\rm c} \propto \rho^{2/3}$ for streaming dominated flows (assuming $B \approx$const), the sharp density gradient means that CRs suffer strong losses at high densities, when radiative cooling is still efficient. Since $v_{\rm A}$ increases strongly as the atmosphere thins due to mass dropout, in our simulations flows are typically streaming dominated in the halo, even if they are diffusion dominated near the disk. 

We find two general classes of solutions, depending on whether the momentum or heat imparted by CRs dominates. 
CRs do work by direct acceleration at a rate $v \cdot \nabla P_c$, while the CR heating rate is $v_A \cdot \nabla P_c$. Thus, momentum driven winds arise when $M_A > 1$, and thermally driven winds arise when $M_A < 1$. This typically means that momentum or energy driven winds arise for high and low $\beta$ atmospheres respectively. 
\begin{itemize}
\item{{\it Momentum-Driven Winds; Fountain Flows} ($M_A>1$; high $\beta$; typically $\alpha = P_c/P_g > 1$). Cool disk gas is accelerated directly via CR forces -- indeed, winds with almost unchanged characteristics are launched if CR heating is turned off. Since the flow is super-Alfvenic, magnetic fields are easily warped; as they wrap around rising gas they can trap CRs, with a consequent strong jump in CR dominance $\alpha = P_c/P_g$ once $M_{\rm A} > 1$. Density fluctuations result in a multi-phase fountain flow at low galactic heights, with the cold dense gas being lifted off the disk and falling back, while hotter gas flows outwards. At larger distances, when CR heating exceeds radiative cooling, the gas becomes hotter and mostly single phase. The extent of the fountain region increases with ${\mathcal M}_{\rm A}$ and CR dominance $\alpha_{\rm H}$. }

\item{{\it Energy-Driven Winds} ($M_A < 1$; low $\beta$; typically $\alpha = P_c/P_g < 1$). Cool disk gas is strongly heated and evaporated at the disk halo interface, resulting in a hot wind powered by CR heating. In steady state, the divergence of the enthalpy flux of the hot gas balances CR heating. The sharp transition to a hot phase leads to a strong drop in gas density at the disk halo interface, with $\rho_{\rm h} \sim \rho_{\rm c} (T_{\rm c}/T_{\rm h})$. This low wind gas density means that the mass flux of CR heated winds is relatively low; they are an inefficient form of feedback compared to cool, denser momentum-driven winds, even though a large fraction of the latter circulates in a fountain flow. The velocity of the wind is $v \sim 0.4 v_{\rm esc} (t_{\rm heat}/t_{\rm ff})^{-1}$, i.e. the flow exceeds the escape velocity once $t_{\rm heat} \sim t_{\rm ff}$. The flow becomes even hotter and lower density for these fast winds, leading to very low mass fluxes. The strong magnetic tension in these sub-Alfvenic flows means there is little warping of field lines, and the single-phase wind flows monotonically outward.} 
\end{itemize} 

There are numerous potential avenues for future work. A key issue is geometry. We have simulated a plane parallel Cartesian setup. This is appropriate close to the disk; thus, our simulation domain of $\sim 2H$ is relatively small. As the flow opens up, a spherical geometry becomes appropriate further away. The flow properties become quite different, as do the density, velocity, B-field and hence Alfven speed profiles. Thus, our simulations do not address the asymptotic properties of the flow far out in the CGM; also, Parker-type sonic points do not develop in plane-parallel flows. Our work is complementary to 1D models which use spherical geometry \citep{ipavich75,mao18,quataert21-diffusion,quataert21-streaming,modak23}, but make other idealizations which we relax. Other possible extensions include: (i) using more physically motivated diffusion coefficients which depend on plasma conditions -- e.g., from quasi-linear self-confinement theory  \citep{wiener_13_cr_stream_clusters}, or using a model for field-line wandering \citep{sampson22}; (ii) incorporating other sources of thermal and momentum driving besides CRs, and understanding their mutual interaction; (iii) considering more complex B-field geometry (e.g., tangled fields due to turbulence). One interesting avenue for future work include formulating an `effective' 1D model which takes the effect of multi-phase structure and CR bottlenecks into account, to match our time-averaged multi-dimensional flows (similar to effective 1D models for turbulent mixing layers; \citealt{tan21_radiative_mixing,tan21-lines,chen22}). Another would be to make predictions for the nature of CR outflows for different galaxies lying on the SFR-$M_*$ relation (which, in our language, correspond to different values of $F_{\rm c0}$ and gravity $g$ respectively, leading to different values of $t_{\rm heat}/t_{\rm ff}$). It would also be interesting to make observational predictions (e.g., in gamma-ray emission) for CR dominated fountain flows. Of course, the biggest unknowns are still the strength of magnetic fields in the CGM, and the nature of CR transport, particularly the relative importance of streaming and diffusive transport.

\section*{Acknowledgements}

We thank Lucia Armillotta, Eliot Quataert, Philipp Kempski, Suoqing Ji, Yan-Fei Jiang, Eve Ostriker, Ellen Zweibel for helpful discussions. We acknowledge NSF grant AST-1911198 and NASA grant 19-ATP19-0205 for support. CB was supported by the National Science Foundation under Grant No. NSF PHY-1748958 and by the Gordon and Betty Moore Foundation through Grant No. GBMF7392. Additionally, this work made considerable use of the Stampede2 supercomputer through allocation PHY-210004 from the Advanced Cyberinfrastructure Coordination Ecosystem: Services \& Support (ACCESS) program, which is supported by National Science Foundation grants \#2138259, \#2138286, \#2138307, \#2137603, and \#2138296.
\section*{Data Availability}

The data underlying this article will be shared on reasonable request to the corresponding author. The reader can view videos pertaining to the discussion in \S\ref{subsec:nonlinear_outcome} at the following link: \url{https://www.youtube.com/playlist?list=PLQqhpX30dsYq2cD51L4M2pNQAlm0GSle9}.



\bibliographystyle{mnras}
\bibliography{main,master_references} 




\appendix

\section{1D Linearized Equations in Uniform Medium} \label{app:linear}

\citet{kempski20_thermal_instability} showed that 1D calculations can approximately capture the behavior of CR-modified thermal modes in the linear regime as perturbations perpendicular to the magnetic field is usually small. In the 1D, fully-coupled limit, eqn.\ref{eqn:continuity}-\ref{eqn:cr_flux} reduces to
\begin{gather}
    \pdv{\rho}{t} + v\pdv{\rho}{x} + \rho\pdv{v}{x} = 0, \label{eqn:continuity_1d} \\
    \pdv{v}{t} + v\pdv{v}{x} = -\frac{1}{\rho}\pdv{P_g}{x} - \frac{1}{\rho}\pdv{P_c}{x} - g, \label{eqn:momentum_1d} \\
    \pdv{P_g}{t} + v\pdv{P_g}{x} + \gamma_g P_g \pdv{v}{x} = -\qty(\gamma_g - 1) v_s\pdv{P_c}{x} + \qty(\gamma_g - 1)\mathcal{L}, \label{eqn:therm_energy_1d} \\
    \pdv{P_c}{t} \qty(\gamma_c - 1)\pdv{F_c}{x} = \qty(\gamma_c - 1)\qty(v + v_s)\pdv{P_c}{x} + \qty(\gamma_c - 1)\mathcal{Q}, \label{eqn:cr_energy_1d} \\
    F_c = \frac{\gamma_c}{\gamma_c - 1} P_c\qty(v + v_s) - \frac{\kappa_\parallel}{\gamma_c - 1}\pdv{P_c}{x}, \label{eqn:cr_flux_1d}
\end{gather}
where we have used the fact that the magnetic field $\vb{B}$ is constant in 1D. The streaming velocity $v_s = -v_A\sign{\pdv*{P_c}{x}}$. Ignoring contributions from CR sources $\mathcal{Q}$ and assuming the background is nearly uniform such that we can ignore derivatives of the background but CRs remain coupled to gas, the linearized equations are
\begin{gather}
    -\frac{\omega}{\omega_s}\frac{\delta\rho}{\rho} + \frac{\delta v}{c_s} = 0, \label{eqn:lin_continuity} \\
    -\frac{\omega}{\omega_s}\frac{\delta v}{c_s} + \frac{1}{\gamma_g}\frac{\delta P_g}{P_g} + \frac{1}{\gamma_g}\frac{\delta P_c}{P_g} = 0, \label{eqn:lin_momentum} \\
    \qty[-i\qty(\gamma_g - 1)\frac{\omega_c}{\omega_s}\Lambda_T - \frac{\omega}{\omega_s}]\frac{\delta P_g}{P_g} - i\qty(\gamma_g - 1)\frac{\omega_c}{\omega_s}\qty(2 - \Lambda_T)\frac{\delta\rho}{\rho} \nonumber \\ \quad + \gamma_g\frac{\delta v}{c_s} + \qty(\gamma_g - 1)\frac{\omega_A}{\omega_s}\frac{\delta P_c}{P_g} = 0, \label{eqn:lin_energy} \\
    \gamma_c\alpha\frac{\delta v}{c_s} - \frac{\gamma_c\alpha\omega_A}{2\omega_s}\frac{\delta\rho}{\rho} + \qty(\frac{\omega_A}{\omega_s} - i\frac{\omega_d}{\omega_s} - \frac{\omega}{\omega_s})\frac{\delta P_c}{P_g} = 0, \label{eqn:lin_cr_energy}
\end{gather}
where $c_s = \sqrt{\gamma_g P_g/\rho}$ is the adiabatic sound speed, $\omega_s = k c_s$, $\omega_A = k v_A$, $\omega_d = k^2\kappa_\parallel$, $\omega_c = \rho^2\Lambda/P_g$, $\alpha = P_c/P_g$ and we have assumed the background flow is static. Substituting $\delta v$ in eqn.\ref{eqn:lin_continuity} into eqn.\ref{eqn:lin_momentum}-\ref{eqn:lin_cr_energy}, the equations simplify to
\begin{gather}
    \frac{\delta P_g}{P_g} + \frac{\delta P_c}{P_c} = \gamma_g\qty(\frac{\omega}{\omega_s})^2\frac{\delta\rho}{\rho}, \label{eqn:simplify_momentum} \\
    i\omega\qty(\gamma_g\frac{\delta\rho}{\rho} - \frac{\delta P_g}{P_g}) = -\qty(\gamma_g - 1)\omega_c\qty[\qty(2 - \Lambda_T)\frac{\delta\rho}{\rho} + \Lambda_T\frac{\delta P_g}{P_g}] \nonumber \\
    \quad - i\qty(\gamma_g - 1)\omega_A\frac{\delta P_c}{P_g}, \label{eqn:simplify_energy} \\
    \frac{\delta P_c}{P_g}\qty(\omega - \omega_A + i \omega_d) = \gamma_c\alpha\qty(\omega - \frac{\omega_A}{2})\frac{\delta\rho}{\rho}. \label{eqn:simplify_cr_energy}
\end{gather}
For thermally unstable modes, we expect $\omega\sim\omega_c$. If $\omega_c\gg\omega_s$, i.e. the cooling time is shorter than the sound crossing time, $\delta\rho/\rho\ll\delta P_g/P_g$ from eqn.\ref{eqn:simplify_momentum}, i.e. the mode is isochoric. If $\omega_c\ll\omega_s$, the otherwise is true and the mode is pressure balanced. 

The ratio of perturbed CR heating to cooling is
\begin{equation}
    \frac{\delta\qty(\mathrm{CR Heating})}{\delta\qty(\mathrm{Cooling})} = \frac{i\omega_A\qty(\delta P_c/P_g)}{\omega_c\qty[\qty(2 - \Lambda_T)\qty(\delta \rho/\rho) + \Lambda_T\qty(\delta P_g/P_g)]}. \label{eqn:ratio_full}
\end{equation}
Eqn.\ref{eqn:ratio_full} reduces to eqn.\ref{eqn:ratio_cr_heat_cool} in the isochoric and isobaric limits. The ratio $(\delta P_c/P_g)/(\delta\rho/\rho)$ is given directly by eqn.\ref{eqn:simplify_cr_energy}
\begin{equation}
    \frac{\qty(\delta P_c/P_g)}{\qty(\delta \rho/\rho)} = \gamma_c\alpha\frac{\omega - \omega_A/2}{\omega - \omega_A + i\omega_d}. \label{eqn:ratio_dpc_drho}
\end{equation}
For $\omega_c\ll\omega_A$ (small scale modes) and $\omega_d\gg\omega_A$, the RHS is purely imaginary, i.e. $\delta P_c$ is $\pi/2$ out of phase with $\delta\rho$. This phase shift allows CR heating to counteract cooling.

\section{Hydrostatic Boundary Conditions for Eulerian Grid Codes} \label{app:hydro_boundary_implement}

We employ hydrostatic boundary conditions in the $x$-direction, which requires eqn.\ref{eqn:bond_hydro} to be satisfied at the boundaries. In a grid code, cell-center values are indicated with subscripts $i,j,k$, all of them integers, representing cells in the $x,y,z$-directions respectively. Below we will shorten the notation to $i$ to reduce clutter, the relations derived in the following are implied for all $j,k$. We shall use $is$ and $ie$ to denote the first and last active zones in the $x$-directions. The cell-center ghost zones are expressed by $is-n$ and $ie+n$, where $n = 1,\ldots, n_g$, $n_g$ is the number of ghost zones (typically 2 for piecewise linear method (PLM)). We want eqn.\ref{eqn:bond_hydro} to hold at the boundary cell face, for example at the outer-$x$ boundary
\begin{equation}
    \eval{\dv{P_g}{x}}_{ie+n-1/2} + \eval{\dv{P_c}{x}}_{ie+n-1/2} = -\eval{\rho}_{ie+n-1/2}\eval{g}_{ie+n-1/2}, \label{eqn:bound_face_out}
\end{equation}
where the fractional index indicates cell faces. The cell-faced values are approximated linearly as
\begin{multline}
    \frac{P_{g,ie+n} - P_{g,ie+n-1}}{\Delta x} + \frac{P_{c,ie+n} - P_{c,ie+n-1}}{\Delta x} \\= -\frac{1}{2}\qty(\rho_{ie+n} + \rho_{ie+n-1}) g_{ie+n-1/2}, \label{eqn:bond_apprx_out}
\end{multline}
we do not need to approximate $g$ because it is a given function. Using the ideal gas law $P_g = \rho T$ and since our initial profiles are isothermal and $P_c\propto\rho^{\gamma_c/2}$ for streaming dominated flows,
\begin{equation}
    r - 1 + \alpha\qty(r^{\gamma_c/2} - 1) = -\xi\qty(r + 1), \label{eqn:bond_step1_out}
\end{equation}
where $\alpha,\xi, r$ are defined by
\begin{equation}
    \alpha = \frac{P_{c,ie+n-1}}{P_{g,ie+n-1}}, \quad \xi = \frac{\rho_{ie+n-1} g_{ie+n-1/2}\Delta x}{2 P_{g,ie+n-1}}, \quad r = \frac{\rho_{ie+n}}{\rho_{ie+n-1}}. \label{eqn:bond_axir_out} 
\end{equation}
Rearranging,
\begin{equation}
    \alpha r^{\gamma_c/2} + \qty(1 + \xi) r - \qty(1 + \alpha - \xi) = 0. \label{eqn:bond_step2_out}
\end{equation}
Solving for $r$ (e.g. using a non-linear root finder) gives us the value for $\rho_{ie+n}$ in terms of quantities in the $ie+n-1$ cell-centers, which we can use to determine $P_{g,ie+n}$ and $P_{c,ie+n}$. $F_{c,ie+n}$ can be obtained from the CR equation by imposing time steadiness, i.e.
\begin{equation}
    \eval{\dv{F_c}{x}}_\mathrm{bond} = \eval{v_A}_\mathrm{bond}\eval{\dv{P_c}{x}}_\mathrm{bond} 
\end{equation}
Performing a linear approximation,
\begin{multline}
    F_{c,ie+n} = F_{c,ie+n-1} \\+ \frac{1}{2}\qty(v_{A,ie+n} + v_{A,ie+n-1})\qty(P_{c,ie+n} - P_{c,ie+n-1}). \label{eqn:bond_cr_step1_out}
\end{multline}
We copy the velocity of the last active zone to the ghost zones if the flow is outbound and set them to zero otherwise.

The same can be performed for the inner-$x$ boundary, an equation similar to \ref{eqn:bond_step2_out} arises,
\begin{gather}
    \alpha r^{\gamma_c/2} + \qty(1 - \xi) r - \qty(1 + \alpha + \xi) = 0, \label{eqn:bond_step2_in}
\end{gather}
where now $\alpha,\xi, r$ are defined by
\begin{equation}
    \alpha = \frac{P_{c,is-n+1}}{P_{g,is-n+1}}, \quad \xi = \frac{\rho_{is-n+1} g_{is-n+1/2}\Delta x}{2 P_{g,is-n+1}}, \quad r = \frac{\rho_{is-n}}{\rho_{is-n+1}}. \label{eqn:bond_axir_in} 
\end{equation}
$F_{c,is-n}$ is given by
\begin{multline}
    F_{c,is-n} = F_{c,is-n+1} \\+ \frac{1}{2}\qty(v_{A,is-n} + v_{A,is-n+1})\qty(P_{c,is-n} - P_{c,is-n+1}). \label{eqn:eqn:bond_cr_step1_in}
\end{multline}
The magnetic field is set to the value of the nearest $x$-layer. 
Note that we assume an isothermal background. If this assumption is relaxed, the energy eqn.\ref{eqn:energy} will have to be invoked.

\begin{figure}
    \centering
    \includegraphics[width=0.35\textwidth]{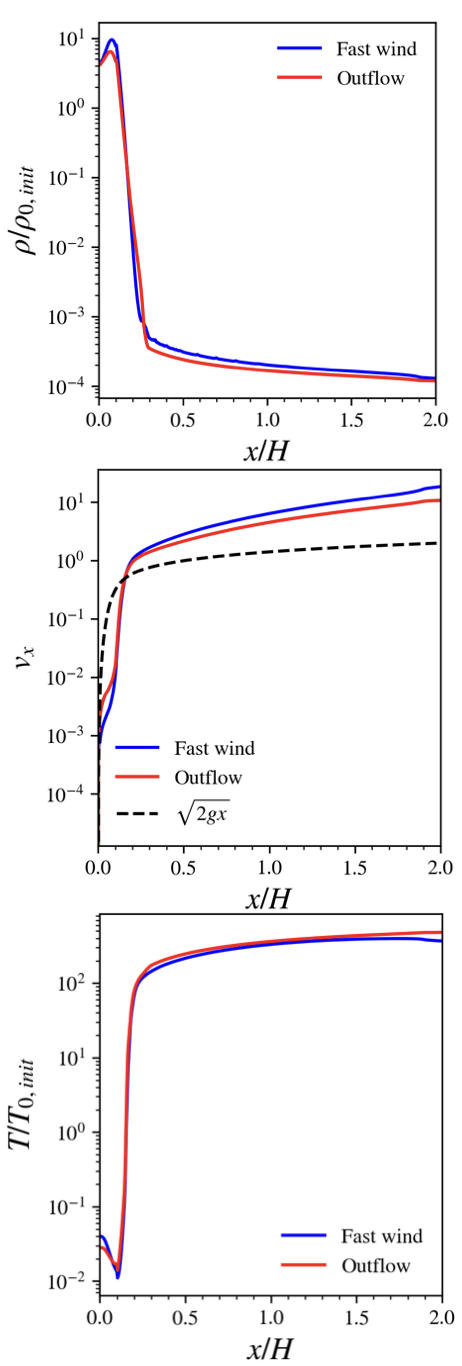}
    \caption{Comparison of the fast wind case using different boundary conditions: hydrostatic (blue) and outflowing (red), as described in the text, showing no significant difference is found.}
    \label{fig:boundary_compare}
\end{figure}

In \S\ref{subsec:nonlinear_outcome} we see that the nonlinear evolution of thermal instability with CR heating can lead to winds, bringing the system out of hydrostatic equilibrium. Using the fast wind case as an example, in fig.\ref{fig:boundary_compare} we show that there is no significant difference when one uses an outflow type condition along the $x$-boundaries. We copy the density and gas pressure of the last active zone to the ghost zones and adopted a diode condition for the $x$-velocity (i.e. the $x$-velocity of the last active zone is copied to the ghost zone if the gas is outflowing, and zero otherwise). The $y$-velocity at the last active zone is copied to the ghost zones. As for the CR pressure and $x$-flux, we set the ghost zone values to be 0.99 times the value at the previous cell. We note that simply copying the CR pressure and $x$-flux at the last active zone to the ghost zones would lead to unphysical confinement of CRs as there would be no CR gradient to transport the CRs out. The CR $y$-flux at the last active zone is copied to the ghost zones.

\section{Simulations fixing the base CR flux instead of pressure} \label{app:fix_fc}

As discussed in \S\ref{subsec:nonlinear_outcome}, the nonlinear outcome of TI depends heavily on the CR heating time $t_\mathrm{heat}$ at the halo, which depends on the halo gas, CR pressure and Alfven speed. In our simulations, we control these quantities through specifying $\alpha_0$, which sets the CR pressure at the base, $\beta_0$, which sets the magnetic field, and $\eta_H$, which sets the CR diffusion coefficient and therefore how much of the base CRs will leak into the halo. 
Moreover, these quantities evolve as mass dropout proceeds in the halo, and the gas pressure $P_g$ and Alfven speed evolve. Nonetheless, it is important to understand the sensitivity of our results to boundary conditions at the disk. 
Instead of specifying $\alpha_0$, which sets $P_c$ at the base, we could alternatively specify $F_{c0}$, the base CR flux. The combination of $F_{c0},\beta_0,\eta_H$ will self-consistently determine what the base $P_c$ will be, and is therefore just a different way of specifying $\alpha_0,\beta_0,\eta_H$. The advantage of setting $\alpha_0,\beta_0,\eta_H$ instead of $F_{c0},\beta_0,\eta_H$ is that the former involves only dimensionless parameters whereas the latter requires physical units\footnote{One could specify a parameter like $P_{\rm c,0}/(P_{\rm g,0} v_{\rm esc})$, but this is similar to our definition of $\alpha_0$.}. As we shall demonstrate, our conclusions remain unchanged whether you choose to fix $F_{c0}$ or $\alpha_0$. For example, increasing $F_{c0}$ for a fixed CR transport model will result in greater presence of CRs in the halo, and is equivalent to increasing $\alpha_0$.

What value of $F_{c0}$ should we set? Let's assume, in galaxies, all of the CRs are generated from supernovae. Each supernova releases about $10^{51}\ \mathrm{ergs}$ of energy, for which around $10\%$ goes into accelerating CRs \citep{caprioli14_acc_eff}. Supernovae occur around once per 100 years, so if we assume all of the CRs produced eventually make it out of the disk, the rate of CRs released into the halo, would on averaged be $\dot{E}_c\sim 10^{51}\ \mathrm{ergs}*0.1/(100\ \mathrm{yr})\sim 3\times 10^{40}\ \mathrm{ergs}\ \mathrm{s}^{-1}$. Assuming CRs escape mostly perpendicular to the disk, we can relate the CR flux $F_c$ with $\dot{E}_c$ through $F_c A\approx \dot{E}_c$, where $A$ is the galactic disk face area. For a disk with radius $10\ \mathrm{kpc}$, the CR flux would then be $F_c\sim 10^{-5}\ \mathrm{ergs}\ \mathrm{s}^{-1}\ \mathrm{cm}^{-2}$. Let's convert this to code units. In our simulations we set $g_0,T_0,\rho_0$ all to unity (see \S\ref{subsubsec:initial}). If these variables scale, in real units as $g_0=10^{-8}\ \mathrm{cm}\ \mathrm{s}^{-2}$, $T_0=10^{6}\ \mathrm{K}$ and $\rho_0=10^{-26}\ \mathrm{g}\ \mathrm{cm}^{-3}$, as typically in galactic environments, then the scale-height $H= k_B T_0/m_u g_0 =2.7\ \mathrm{kpc}$, pressure $P_0=\rho_0 k_B T_0/m_u = 8.3\times 10^{-13}\ \mathrm{erg}\ \mathrm{cm}^{-3}$, velocity $v_0= (k_B T_0/m_u)=1=91\ \mathrm{km}\ \mathrm{s}^{-1}$, and the CR flux $F_{c0}=P_0 v_0=1=7.5\times 10^{-6}\ \mathrm{ergs}\ \mathrm{cm}^{-2}\ \mathrm{s}^{-1}$. In code units, the CR flux $F_c\sim 10^{-5}\ \mathrm{ergs}\ \mathrm{s}^{-1}\ \mathrm{cm}^{-2}$ would then be $\sim 1.3$. In reality galaxies could, depending on e.g. the star formation rate, size and structure, be supplying CRs at different rates, thus we also explore different values of $F_{c0}$.

\begin{figure*}
    \centering
    \includegraphics[width=\textwidth]{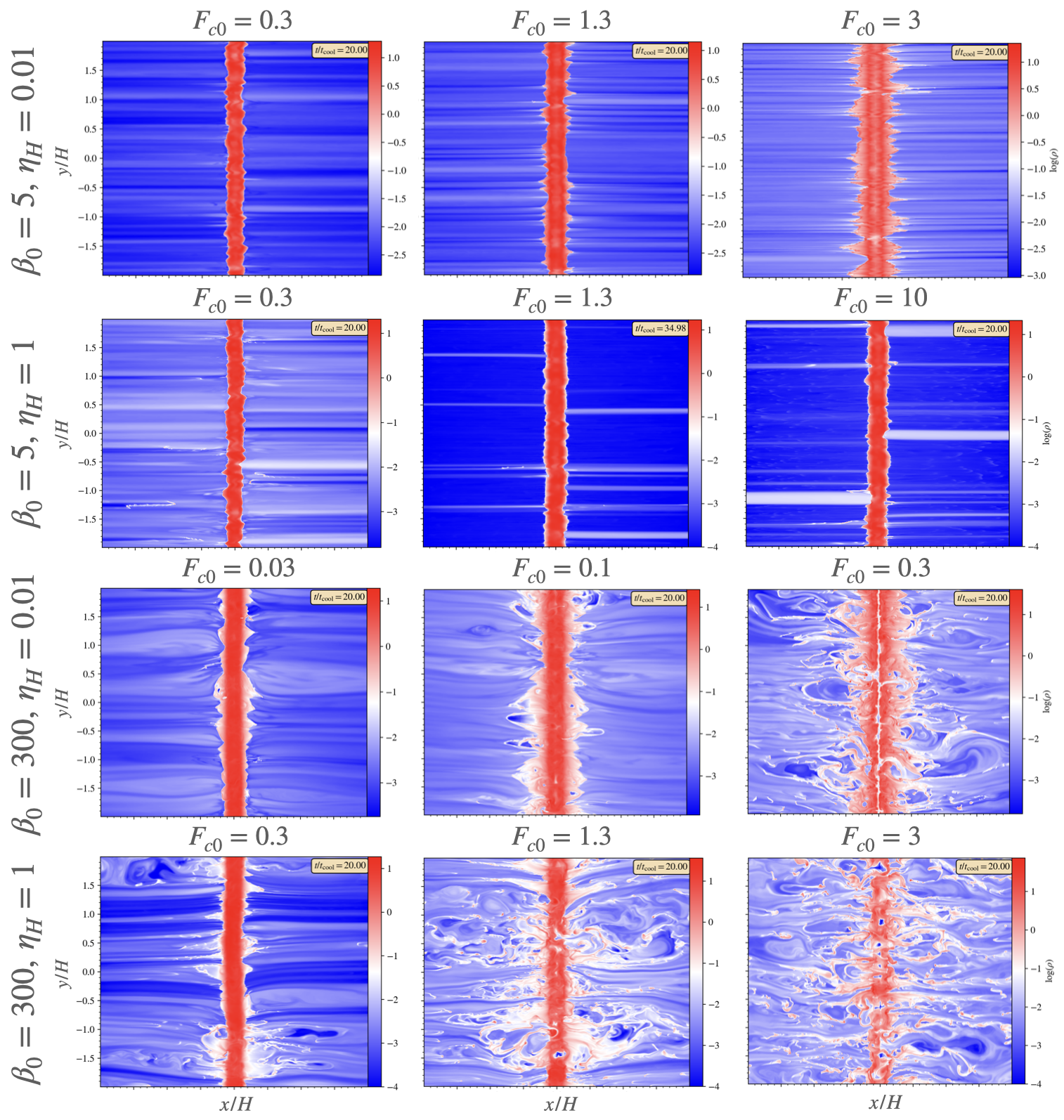}
    \caption{Density slice plots at $t=20 t_\mathrm{cool}$ for a variety of test cases fixing the base CR flux $F_{c0}$. $F_{c0}$ is given in code-units, but conversion to real units for scenario specific flows can be found in Appendix \ref{app:fix_fc}. The nonlinear outcomes of TI are explored for varying $F_{c0},\beta_0,\eta_H$. Specifically, the first row displays flows that are low in $\beta$ and streaming dominated, the second row for low $\beta$ and higher diffusivity. The third and forth row are replica of the first and second row at higher $\beta$.}
    \label{fig:fix_fc}
\end{figure*}

In fig.\ref{fig:fix_fc} we display the nonlinear outcome of TI for various combinations of $F_{c0},\beta_0,\eta_H$. Once again we observe the three outcomes discussed in \S\ref{subsec:nonlinear_outcome}: slow wind, fast wind and fountain flows. The fast wind is again marked by a single phase, rarefied halo (e.g. middle and right panels of the second row) while fountain flows are marked by filamentary cold flows (e.g. right panel of the third row and the middle and right panels of the bottom row). By varying $F_{c0}$,  we can observe the transition into different outcomes clearly. For example, from a slow wind to a fast wind in the second row and into a fountain flow in the third and bottom row. All these are the result of greater supply of CRs to the halo, which increases both CR pressure support and heating. 

\begin{figure}
    \centering
    \includegraphics[width=0.48\textwidth]{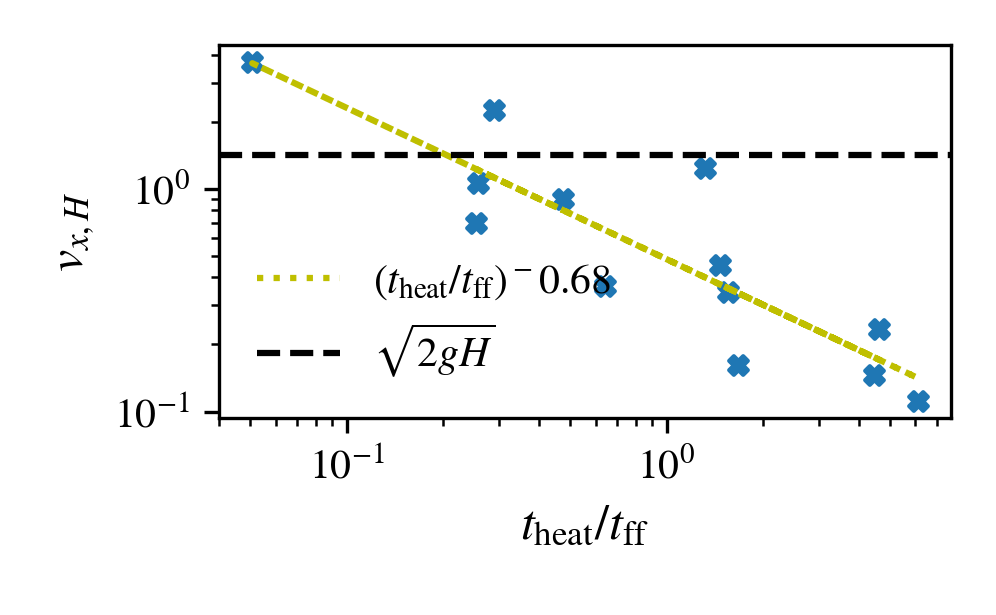}
    \caption{$v_x$ against $t_\mathrm{heat}/t_\mathrm{ff}$ at a scaleheight for the cases shown in fig.\ref{fig:fix_fc}, which the base CR flux is fixed instead of CR pressure.}
    \label{fig:vx_tratio_fixfc}
\end{figure}

In fig.\ref{fig:vx_tratio_fixfc} we again plot the outflow velocity $v_x$ against $t_\mathrm{heat}/t_\mathrm{ff}$ (taken at a scale height), recovering the same trend as in fig.\ref{fig:t_heat_t_ff_plots} that the flow transitions to a fast wind when $t_\mathrm{heat}\ll t_\mathrm{ff}$. In short, there is no fundamental difference whether one fixes $F_{c0}$ or $\alpha_0$ at the base. All that matters is $t_\mathrm{heat}$ in the halo.

\section{Resolution and 3D} \label{app:res_3d}

\begin{figure}
    \centering
    \includegraphics[width=0.45\textwidth]{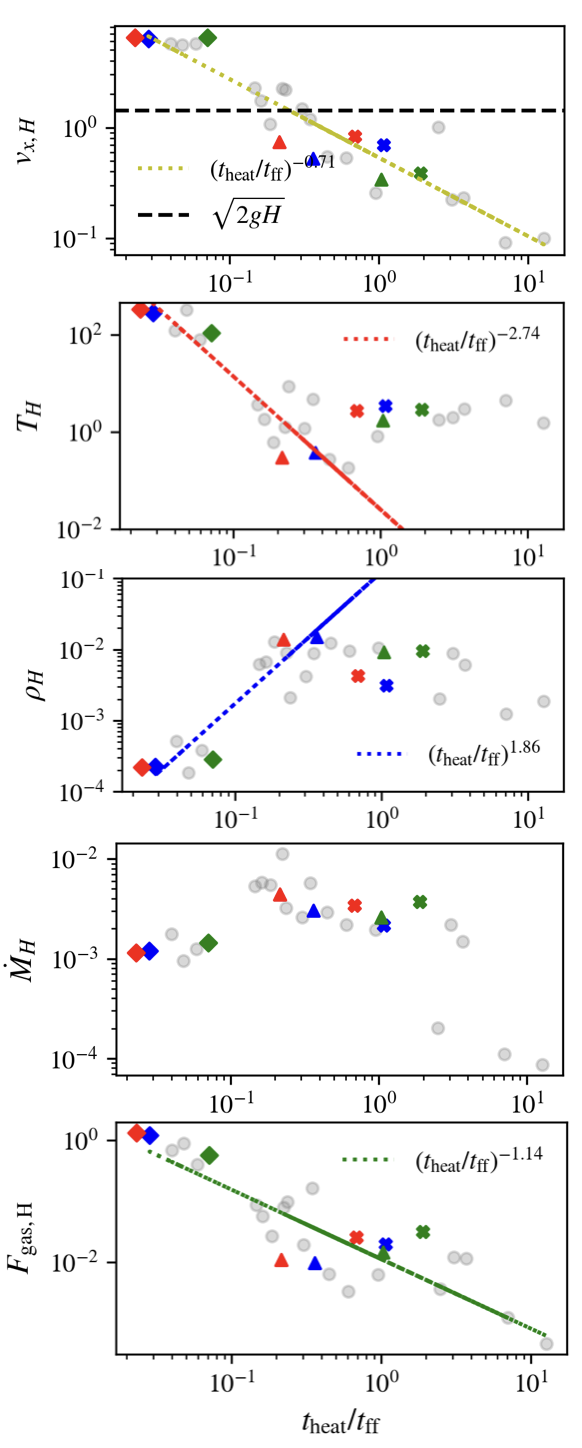}
    \caption{Same as fig.\ref{fig:t_heat_t_ff_plots} including cases 2D standard resolution (blue) with 2D higher resolution (red) and lower-resolution 3D (green markings). The markers indicate `slow wind' (crosses) `fast wind' (diamond) and `fountain flow' (triangular) profile parameters respectively. Test cases with flow parameters other than the three mentioned are indicated by grey circles and are shown for reference.}
    \label{fig:t_heat_tff_plots_res3D}
\end{figure}


\begin{figure*}
    \centering
    \includegraphics[width=\textwidth]{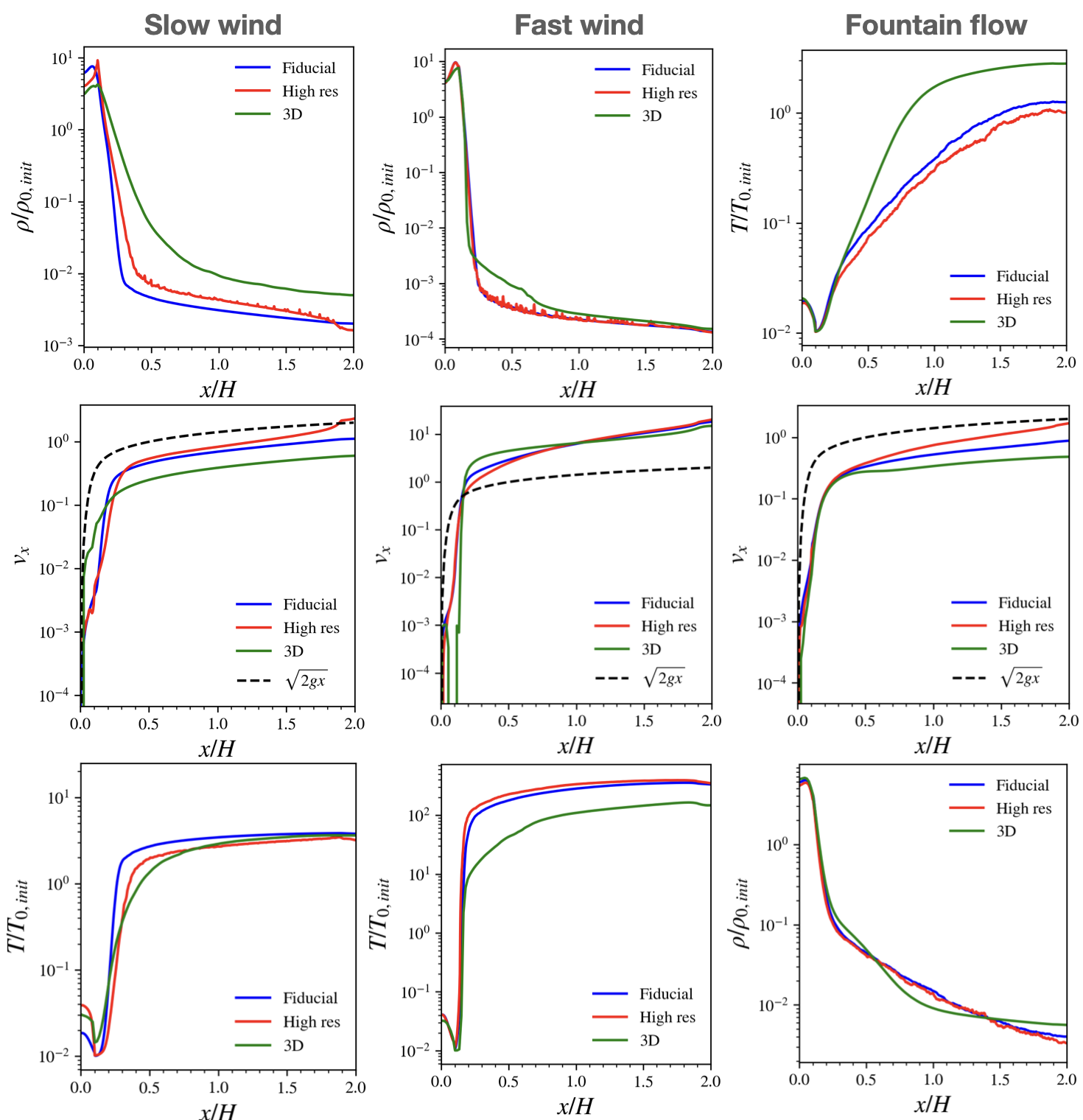}
    \caption{Time averaged projection plots of the density, velocity and temperature for the `slow wind', `fast wind' and `fountain flow' cases. In each case, the fiducial profile (blue) is compared against the higher resolution profile (red) and the 3D profile (green).}
    \label{fig:profile_res3D}
\end{figure*}

We rerun the `slow wind', `fast wind' and `fountain flow' cases in \S\ref{subsec:nonlinear_outcome} with higher resolution and in 3D to check that our results hold. For increased resolution, we resolve the simulation domain ($-2 H < x < 2 H$) by $2048\times 512$ grids (doubling the $x$-resolution) whereas for 3D simulations, the grid resolution is reduced to $256\times 128\times 128$ (again higher resolution along the $x$-axis) to save computational time. Further details regarding the setup are listed in table \ref{tab:cases}. As shown in fig.\ref{fig:t_heat_tff_plots_res3D}, the fluid properties of the higher resolution and 3D runs are all in line with the trends given by simulations with fiducial resolution. Our conclusions remain unchanged. While there is some scatter as resolution and dimensionality change, the scatter lies along the trends we have already found. 
In fig.\ref{fig:profile_res3D} we compare the time averaged projection plots of the density, velocity and temperature for all three solution outcomes. The fiducial and the high resolution 2D profiles are very similar. The 3D profiles also give very similar outcomes. The time-averaged profiles do deviate somewhat more: for instance, up to a factor of $\sim 2$ in asymptotic temperatures or densities, though at least part of this is due to the much lower resolution in our 3D sims. 


\bsp	
\label{lastpage}
\end{document}